# Numerical Simulations of the Two-phase flow and Fluid-Structure Interaction Problems with Adaptive Mesh Refinement

A DISSERTATION
SUBMITTED TO THE FACULTY OF THE GRADUATE SCHOOL
OF THE UNIVERSITY OF MINNESOTA
BY

Yadong Zeng

IN PARTIAL FULFILLMENT OF THE REQUIREMENTS
FOR THE DEGREE OF
DOCTOR OF PHILOSOPHY

Lian Shen

Mar, 2022

# Acknowledgements

I would like to thank my esteemed supervisor Prof. Lian Shen for his invaluable supervision, support, and tutelage during the course of my Ph.D. degree. He led me to an interesting topic of Adaptive Mesh Refinement (AMR). I learned a lot from him about conducting research of high quality, writing papers, writing proposals, and giving presentations.

Additionally, I would like to express gratitude to Dr. Ann Almgren, Dr. Dan Martin, Dr. Johannes Blaschke, Dr. Andy Nonaka, Dr. Zhi Yao, and Dr. Revathi Jambunathan for their treasured support of my internships in the Berkeley Lab.

I really enjoy working with Prof. Amneet Bhalla and I had many wonderful discussions with him during my Ph.D. study. He helps me a lot in many aspects, including deriving formulas, writing the codes, and giving insightful feedback. I couldn't finish my Ph.D. study without his help and support.

I would like to thank my preliminary and final committee members, Prof. Jeff Tithof, Prof. Sungyon Lee, and Prof. Li Wang for their valuable feedback and support.

I am grateful to all members of the Fluid Mechanics Lab, especially Dr. Anqing Xuan, Dr. Qiang Cao, Dr. Sida He, Dr. Tianyi Li, and Dr. Han Liu for their help, support, and insightful discussion of my research.

Finally, I would like to give specifically thanks to my parents, my friend Yujia Ke, and my girlfriend Ruby for their continuous support and encouragement.

# Dedication

This thesis is dedicated to my parents.

# Abstract

Numerical simulations of two-phase flow and fluid structure interaction problems are of great interest in many environmental problems and engineering applications. To capture the complex physical processes involved in these problems, a high grid resolution is usually needed. However, one does not need or maybe cannot afford a fine grid of uniformly high resolution across the whole domain. The need to resolve local fine features can be addressed by the adaptive mesh refinement (AMR) method, which increases the grid resolution in regions of interest as needed during the simulation while leaving general estimates in other regions.

In this work, we propose a block-structured adaptive mesh refinement (BSAMR) framework to simulate two-phase flows using the level set (LS) function with both the subcycling and non-subcycling methods on a collocated grid. To the best of our knowledge, this is the first framework that unifies the subcycling and non-subcycling methods to simulate two-phase flows. The use of the collocated grid is also the first among the two-phase BSAMR framework, which significantly simplifies the implementation of multi-level differential operators and interpolation schemes. We design the synchronization operations, including the averaging, refluxing, and synchronization projection, which ensures that the flow field is divergence-free on the multi-level grid. It is shown that the present multi-level scheme can accurately resolve the interfaces of the two-phase flows with gravitational and surface tension effects while having good momentum and energy conservation.

We then develop another consistent scheme, in which the conservative momentum equations and the mass equation are solved in the aforementioned BSAMR framework. This consistent mass and momentum transport treatment greatly improves the accuracy and robustness for simulating two-phase flows with a high density ratio and high Reynolds number. We demonstrate that the consistent scheme results in a numerically stable solution in flows with high density ratios (up to $10^6$) and high Reynolds numbers (up to $10^6$), while the inconsistent scheme exhibits nonphysical fluid behaviors in these tests.

For solving single- and multiphase fluid-structure interaction (FSI) problems, we present an adaptive implementation of the distributed Lagrange multiplier (DLM) immersed boundary (IB) method on multilevel collocated grids. We also developed a force-averaging algorithm to maintain the consistency of Eulerian immersed boundary (IB) forces across multiple levels. The efficacy of the force averaging algorithm is validated using the lid-driven cavity with a submerged cylinder problem. We demonstrate the versatility of the present multilevel framework by simulating problems with various types of kinematic constraints imposed by structures on fluids, such as imposing a prescribed motion, free motion, and time-evolving shape of a solid body. The accuracy and robustness of the codes are validated using several canonical test problems.

# Contents

# List of Tables

# List of Figures

xiv

xvi

xvii

# Chapter 1

# Introduction

## 1.1 Background

Numerical simulation of air–water interactions is of great interest in many environmental problems and engineering applications, such as wind over breaking waves (Iafrati et al., 2019; Yang et al., 2018), ship hydrodynamics (Bertram, 2011), bubbly flows (Hua and Lou, 2007; Deike et al., 2018), and liquid jets (Ménard et al., 2007). To capture the complex physical processes involved in the two-phase flow problems, a high grid resolution is needed around the free surface to resolve small flow structures (Prosperetti and Tryggvason, 2009). However, those fine structures may not be present everywhere, therefore one does not need or maybe cannot afford a fine grid of uniformly high resolution across the whole domain. The need to resolve local fine features can be addressed by the adaptive mesh refinement (AMR) method. AMR increases the grid resolution in regions of interest as needed during the simulation while leaving general estimates in other regions. Since proposed in the 1980s (Berger and Oliger, 1984; Berger and Colella, 1989), various classes of AMR have been developed and applied to a wide range of physical problems, including the wave energy converters (Yu and Li, 2013; Zeng and Shen, 2020), shallow water flows (Popinet, 2015), marine ice sheet (Cornford et al., 2013), supersonic flows (Al-Marouf and Samtaney, 2017), surfactant driven flows (de Langavant et al., 2017), and stratified oceanic flows (Chalamalla et al., 2017).

Based on the grid hierarchy and data structure, AMR methods can generally be classified into two groups, the quadtree/octree-based AMR (TBAMR) (Guittet et al., 2015;

Mirzadeh et al., 2016; Popinet, 2003) and block-structured AMR (BSAMR) (Berger and Oliger, 1984; Berger and Colella, 1989; Almgren et al., 1998; Martin and Colella, 2000; Martin et al., 2008; Minion, 1996; Zeng and Shen, 2019). In the TBAMR, each cell can be split into four cells in two dimensions or eight cells in three dimensions and the hierarchy of the grid cells is organized using a tree structure (Burstedde et al., 2011). Although the tree-based structure is an intuitive representation of the grid hierarchy for the multi-level grid and simplifies the management of the grid refinement and coarsening, it is relatively difficult to implement the data structure (Burstedde et al., 2011; Isaac et al., 2015). Moreover, the connectivity between adjacent cells and the refinement history need to be stored at every time step (Williamschen and Groth, 2013; Min and Gibou, 2007). The BSAMR, on the other hand, builds the mesh as nested Cartesian grids (Berger and Oliger, 1984; Berger and Colella, 1989; Almgren et al., 1998; Martin and Colella, 2000; Martin et al., 2008; Minion, 1996; Vanella et al., 2010). It is relatively easy to use the domain decomposition method for parallelization (Gunney and Anderson, 2016). Equations on the nested grid can also be solved efficiently utilizing the multigrid (MG) solver (Minion, 1996; Almgren et al., 1998).

For both TBAMR and BSAMR, the choice of the grid layout can affect the complexity of the discretization scheme and multi-level algorithm on the adaptive grid (Burstedde et al., 2014; Griffith et al., 2007). Most existing grid layouts fall into three categories, the collocated grid, the staggered grid, and the semi-staggered grid. On a semi-staggered grid (Almgren et al., 1998; Sussman et al., 1999), the velocities are defined at the cell center whereas the pressure is defined at the nodal center. As a result, different interpolation schemes are required for the velocities and pressure on the multi-level grid. Moreover, two types of implicit solvers are needed for the velocity and pressure equations. The staggered grid (or MAC layout), while facilitating a divergence-free velocity field, still needs different interpolation schemes for the velocities and pressure (Bhalla et al., 2013; Nangia et al., 2019a; Griffith et al., 2007; Pivello et al., 2014; Vanella et al., 2010). Furthermore, producing a compact, accurate implicit solver for the viscous terms is not straightforward in the context of the MAC layout (Guittet et al., 2015). The collocated grid is attractive for non-orthogonal grids. Variables on different levels are coupled through the coarse-fine boundaries and are solved simultaneously. However, because of this coupling, the time step is restricted by the finest grid spacing for numerical stability. On the other hand, the level-by-level advancement method decouples the time advancement among different levels (Berger and Colella, 1989; Almgren et al., 1998). This method can be further divided into the non-subcycling method and subcycling methods (Martin and Colella, 2000; Minion, 1996). The non-subcycling method uses a uniform time step for all levels. Thus, the time step is also restricted by the finest level for numerical stability. The subcycling method, where variables on different levels are advanced with different time steps, can reduce the number of advancement steps and save the simulation time. However, the variables across different levels need to be interpolated in time for the subcycling method, which is not required in the non-subcycling method. Furthermore, the subcycling method needs recursive advancement steps between different levels, making it relatively difficult to implement (Griffith et al., 2007).

In the past several decades, many researchers have combined AMR with the interface-tracking techniques, such as the front-tracking method (Unverdi and Tryggvason, 1992), and the interface-capturing techniques, such as the volume-of-fluid (VOF) method (Hirt and Nichols, 1981; Renardy et al., 2001; Pilliod Jr and Puckett, 2004) and level set (LS) method (Osher and Sethian, 1988; Adalsteinsson and Sethian, 1994; Sethian, 1996), to simulate two-phase flow problems. Pivello et al. (2014) presented a BSAMR-based adaptive front tracking method to simulate bubbly flows. This method represents the interface precisely with a Lagrangian mesh but needs frequent re-meshing when the interface deforms significantly, e.g., in a violent two-phase flow. Sussman et al. (1999) proposed an adaptive LS approach for the incompressible two-phase flow within the BSAMR framework. The LS method is also developed for the unstructured TBAMR by Kohno and Tanahashi (2004). Antepara et al. (2019) embedded a conservative LS method into the TBAMR framework. Popinet (2009) combined the VOF method with the TBAMR for surface-tension-driven interfacial flows and de Langavant et al. (2017) presented a LS method on the non-graded tree-based adaptive grids for surfactant driven flows. In the works cited above, the two-phase flow solutions are updated in time using the composite advancement method, i.e., discretized equations of velocity and pressure are constructed and solved for multiple levels simultaneously. To the best of our knowledge, there has been no implementation of the AMR framework that can utilize the level-by-level advancement method, especially the subcycling method, for the simulation of two-phase flows. Considering that complex flow structures are often present in two-phase flows, advancement using the subcycling method is desired for reducing the computational cost.

A challenging problem in the multiphase flow community is the treatment of flows with high density ratios (Rudman, 1998; Bussmann et al., 2002; Raessi, 2008) and high Reynolds numbers (Nangia et al., 2019a; Yang et al., 2021b; Gao et al., 2021a), for which special numerical techniques are needed to ensure numerical stability. Recent studies have begun to address the numerical instabilities associated with flows with density ratios on the order of $10^3 - 10^4$ and greater. A variety of numerical approaches have been reported, including the LS method (Nangia et al., 2019a), VOF method (Rudman, 1998; Bussmann et al., 2002), lattice Boltzmann method (Bussmann et al., 2002; Inamuro et al., 2004), and diffused interface method (Ding et al., 2007). For example, Sussman et al. (2007) proposed a sharp-interface coupled level-set and volume-of-fluid (CLSVOF) method to solve the liquid and gas system separately. Because the extrapolated velocity from one side is used to advect the LS function, the VOF function, and the velocity, this method can handle flows with a high density ratio. Ding et al. (2007) formulated a divergence-free staggered-velocity field from conservation laws, which works well for two-phase flows with a high density ratio in the context of the diffused interface method. Rudman (1998) and Bussmann et al. (2002) found that using a scheme to consistently transport the momentum equations and VOF scalar can reduce the problem of numerical instability. In the LS context, Raessi (2008) and Raessi and Pitsch (2012) first introduced geometric mass flux transport for tightly coupling the mass and momentum. However, their methods are limited to one and two dimensions owing to the inherent difficulty of reconstructing the interface using the LS function in three dimensions. Desjardins and Moureau (2010) designed a consistent transport scheme for the 3D staggered grid in the context of the LS method, and Ghods and Herrmann (2013) extended this scheme to the collocated unstructured grid. However, only the first-order upwind scheme was used for the density and velocity advection in (Desjardins and Moureau, 2010; Ghods and Herrmann, 2013), and this scheme smears both interface and velocities. An improved third-order Cubic Upwind Interpolation (CUI) interpolation scheme was proposed by Patel and Natarajan (2015) in the hybrid staggered/nonstaggered framework for the consistent transport of mass and momentum. The CUI interpolation scheme was also used in (Nangia et al., 2019a,b; Bhalla et al., 2020; Yang et al., 2021b) in the context of the staggered grid. In the present work, we employ the CUI scheme on a collocated grid to convect mass and momentum for robust simulations of flows with high density ratios. Using the collocated grid significantly simplifies the implementation of the interpolation schemes and the differential operators when multiple levels of grids are involved.

In addition, many engineering fields address problems that involve complex interactions between fluids and solids. Examples include biomedical engineers who address heart valves (Kamensky et al., 2015) and drug particles (Kolahdouz et al., 2020), control and mechatronic engineers who address swimming fish (Bhalla et al., 2013; Deng et al., 2013), flying drones (Floreano and Wood, 2015), and tiny insects (Santhanakrishnan et al., 2014; Miller and Peskin, 2005), marine engineers who address offshore platforms (Faltinsen, 1993) and risers (Trim et al., 2005; Song et al., 2016), and mechanical and energy engineers who address membrane distillation (Lou et al., 2020), wind turbines (Natarajan et al., 2019b) and wave energy converters (WECs) (Yu and Li, 2013; Khedkar et al., 2021). Broadly speaking, FSI problems can be simulated using interface conforming arbitrary Lagrangian-Eulerian (ALE) methods or interface nonconforming fictitious domain/immersed boundary methods. Although an ALE-like method sharply resolves the fluid boundary layer and imposes exact boundary conditions on the fluid-structure (Hu et al., 2001) or the fluid-fluid (Ramaswamy and Kawahara, 1987) interface, for a solid structure with complex geometry and large displacements or deformations, the ALE method poses several numerical and implementation challenges because the computational domain needs to be frequently remeshed to adhere to the evolving structure or the interface (Sotiropoulos and Yang, 2014; Tian et al., 2014). In recent years, the IB method has been widely employed for simulating complex FSI problems that involve large deformations of a structure or geometrical changes of the fluid-fluid interface due to the motion of the structure. In contrast to an ALE approach, in the IB method, the background (Cartesian) fluid grid does not deform due to a moving structure. Instead, the moving structure is accounted for through IB forces that are applied near the structure surface to satisfy the velocity matching boundary condition on the fluid-solid interface (Mittal and Iaccarino, 2005). Depending on the forcing technique, IB methods can be further classified as a diffuse interface or a sharp interface IB method (Sotiropoulos and Yang, 2014; Balaras, 2004). In the diffuse-interface IB method, the surface force calculated on the (Lagrangian) immersed boundary is distributed to several adjacent fluid grid nodes by a regularized integral kernel (Peskin, 1972, 2002; Fadlun et al., 2000; Yang et al., 2009; He et al., 2022). One of the most efficient ways to compute the rigid body IB force is through the distributed Lagrange multiplier (DLM) technique of Patankar et al. (2000). The sharp-interface IB method, on the other hand, directly imposes the velocity of the moving solid boundary by fitting a spatial polynomial through the solid interface and nearby fluid nodes (Udaykumar et al., 1996; Kim et al., 2001; Yang and Balaras, 2006; Gilmanov et al., 2015; Cui et al., 2018; Roman et al., 2009; Kang, 2015; Zeng et al., 2021). In addition to the IB method, the cut-cell method (Ye et al., 1999; Almgren et al., 1997; Udaykumar et al., 1996) and the Brinkman penalization method (Liu and Vasilyev, 2007) have also been proposed to simulate complex FSI problems.

To reduce the execution times of FSI simulations, many researchers have applied AMR frameworks to implement some of the aforementioned FSI schemes. Popinet (2003) implemented the volume of fluid (VOF) method on a TBAMR framework for simulating incompressible Euler flows (no flow penetration through the solid boundary, but tangential slip is present) in complex stationary domains. The solid boundaries in (Popinet, 2003) are represented by an Eulerian level set (LS) function that is embedded on the finest grid level. Guittet et al. (2015) developed a LS-based method for the incompressible Navier-Stokes equations adaptive quad/octrees grids, which can address arbitrary refinement/coarsening ratios between adjacent cells. Roma et al. (1999) implemented the diffuse-interface IB method on a multilevel staggered grid within the BSAMR framework. However, this method is only suitable for low Reynolds numbers ($Re = 10 - 100$) because of the explicit treatment of the nonlinear convective term with a nonconservative central differencing scheme and the lack of limiters. A similar formulation was later implemented by Griffith et al. (2007) on both multilevel collocated grids and staggered grids. Vanella and Balaras (2009) and Vanella et al. (2010) implemented a direct-forcing immersed boundary method within the BSAMR framework for laminar and turbulent flows, in which the governing equations are integrated simultaneously and iteratively using a strong coupling scheme. Bhalla et al. (2013) and Nangia et al. (2019b) combined the DLM method with BSAMR to capture thin boundary layers and vortical structures present in high-density-ratio, wave-structure interaction (WSI) problems. An advantage of the DLM/IB method is that fluid-structure coupling is implicit, which implies that there is no need to explicitly evaluate hydrodynamic stresses on the solid surface or to iterate between fluid domain integrators and solid domain integrators. In this study, we combine the DLM method with a collocated BSAMR framework for simulating both single and multiphase FSI problems.

## 1.2   Thesis overview

The motivation of this thesis is to numerically simulate the two-phase flow and fluid-structure interaction problems with adaptive mesh refinement. The thesis is organized as follows.

Chapter 2 introduces some important concepts of the BSAMR and describes the definitions of the variables and operators for a multilevel adaptive grid. A general multi-level time advancement framework is also given, which includes both the subcycling and non-subcycling methods. The synchronization step is also introduced to synchronize the composite solution across different levels.

Chapter 3 develops a unified adaptive level set (LS) framework using the multi-level collocated grid for incompressible two-phase flows. This framework allows us to advance all variables level by level using either the subcycling or the non-subcycling method such that the data advancement on each level is fully decoupled. A multi-level re-initialization method for the LS function is also proposed, which greatly improves the mass conservation of the two-phase flows. The collocated grid allows the use of a single set of differential schemes and interpolation operations for all variables, which greatly simplifies the numerical implementation. The capability and robustness of the computational framework are validated by a variety of canonical problems, including the inviscid shear layer, gravity wave, rising bubble, and Rayleigh-Taylor instability.

In chapter 4, we develop a consistent adaptive framework in a multilevel collocated grid layout for simulating two-phase flows with adaptive mesh refinement (AMR). The conservative momentum equations and the mass equation are solved in the present consistent framework. This consistent mass and momentum transport treatment greatly improves the accuracy and robustness for simulating two-phase flows with a high density ratio and high Reynolds number. The interface capturing level set method is coupled with the conservative form of the Navier–Stokes equations, and the multilevel reinitialization technique is applied for mass conservation. The accuracy and robustness of the framework are validated by a variety of canonical two-phase flow problems. We demonstrate that the consistent scheme results in a numerically stable solution in flows with high density ratios (up to $10^6$) and high Reynolds numbers (up to $10^6$), while the inconsistent scheme exhibits nonphysical fluid behaviors in these tests.

In chapter 5, we present an adaptive implementation of the distributed Lagrange multiplier (DLM) immersed boundary (IB) method on multilevel collocated grids for solving single- and multiphase fluid-structure interaction (FSI) problems. Both a non-subcycling time advancement scheme and a subcycling time advancement scheme, which are applied to time-march the composite grid variables on a level-by-level basis, are presented; these schemes use the same time step size and a different time step size, respectively, on different levels. This is in contrast to the existing adaptive versions of the IB method in the literature, in which coarse- and fine-level variables are simultaneously solved and advanced in a coupled fashion. A force-averaging technique and a series of synchronization operations are constructed to achieve excellent momentum and mass conservation across multiple levels of grid hierarchy. We demonstrate the versatility of the present multilevel framework by simulating problems with various types of kinematic constraints imposed by structures on fluids, such as imposing a prescribed motion, free motion, and time-evolving shape of a solid body. The DLM method is also coupled to a robust level set method-based two-phase fluid solver to simulate challenging multiphase flow problems, including wave energy harvesting using a mechanical oscillator. The capabilities and robustness of the computational framework are validated against a variety of benchmarking single-phase and multiphase FSI problems from the literature, which include a three-dimensional swimming eel model to demonstrate the significant speedup and efficiency that result from employing the present multilevel subcycling FSI scheme.

The summary and discussion can be found in chapter 6.

# Chapter 2

# Adaptive Mesh Refinement

## 2.1 Variables and operators on multi-level adaptive grids

### 2.1.1 Concepts and definitions in BSAMR

This section introduces some important concepts of the BSAMR. In this thesis, the coarsest level of the grid in the entire computational domain $\Gamma$ is referred to as level 0. The finest level that the grid can be refined to is denoted as level $l_{max}$. In other words, the total number of levels is $l_{max} + 1$. Fig. 2.1 illustrates a three-level adaptive grid with $l_{max} = 2$ as an example.

Grid cells can be dynamically tagged and refined following certain criteria (Berger and Colella, 1989). In BSAMR, although the tagging is done on individual cells, we do not refine or de-refine the cells individually. Instead, these tagged cells are grouped to form a series of rectangular patches for two-dimensional grids or cuboid patches for three-dimensional grids. There can be more than one patch on a specific level and these patches are refined simultaneously to the next level. For example, in Fig. 2.1, level 2 consists of two patches. Because BSAMR uses a nesting hierarchy of rectangular patches, the union patches on level $l+1$ must be contained in the union patches on level $l$ for all $0 \le l < l_{max}$, i.e. $\Gamma^{l+1} \subset \Gamma^l$, where $\Gamma^l$ denotes the union of the patches on level $l$. Because of this nesting property, three types of boundaries exist on the adaptive grid:

○ Physical boundary: the boundary that encloses the computational domain, illustrated using the dashed lines in Fig. 2.1.

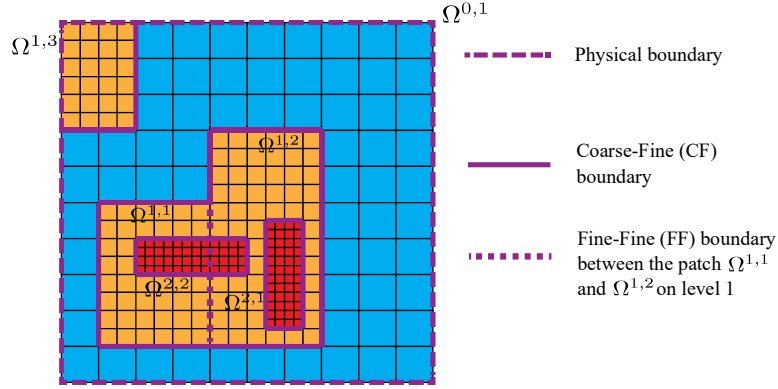

Figure 2.1: Diagram of a three-level adaptive grid with three types of boundaries. $\Omega^{i,j}$ represents the patch $j$ on level $i$ for all $i \geq 0$, $j \geq 1$.

○ Coarse-fine (CF) boundary: the boundary between the grid cells of different levels. These boundaries are illustrated using the thick solid lines in Fig. 2.1.

○ Fine-fine (FF) boundary: the boundary between two patches at the same level, marked using the dotted lines in Fig. 2.1.

Ghost cells are defined at all boundaries and their values are assigned to represent the boundary effects. The ghost cells at the physical boundaries are filled based on the physical boundary conditions. At the CF boundaries, we adopt a conservative interpolation ($\mathcal{I}_{cons}$) method (Almgren et al., 1998; Zhang et al., 2019) to fill the ghost cells of the fine level. To be specific, we reconstruct a continuous functional form, $f(x)$, on the ghost cells by combining the values of the coarse level and the values of the fine level. Besides satisfying the continuity across the CF boundary, the function $f(x)$ is also subject to the requirement that the average of $f(x)$ over the area of a coarse cell is equal to the original coarse cell value. The conservative interpolation scheme maintains the second-order accuracy of the proposed multi-level algorithms. At the FF boundaries, the ghost cell values are copied from neighboring patches.

### 2.1.2 Definitions of variables and operators

We first define the following types of regions for a specific level $l$.

○ Valid region ($\Gamma_{valid}^{l}$): grid cells on level $l$ that are not covered by finer patches.

○ Invalid region ($\Gamma^l_{invalid}$): grid cells on level $l$ that are covered by finer patches.

We note that on level $l_{max}$, $\Gamma^{l_{max}}_{valid} = \Gamma^{l_{max}}$ and $\Gamma^{l_{max}}_{invalid} = \varnothing$. On level $l < l_{max}$, $\Gamma^l_{valid} = \Gamma^l \backslash \Gamma^{l+1}$ and $\Gamma^l_{invalid} = \Gamma^{l+1}$. Fig. 2.2 shows an example, where the green and orange cells represent the valid regions of level $l$ and $l+1$, respectively. The orange area is also the invalid region of level $l$.

In the present work, all variables, including the velocity $\boldsymbol{u}$, pressure $p$, and LS function $\phi$, are defined at cell centers, i.e., the collocated grid is used. Because our multi-level scheme uses a level-by-level advancement method, variables need to be available in both the valid and invalid regions. Depending on which regions we use to describe the flow field, we have the following two sets of variables.

○ Composite variables: variables defined in the valid regions across a multi-level grid.

○ Level variables: variables defined on the whole level, including both the valid and invalid regions.

All simulation results in this paper are presented using the composite variables unless specified otherwise. Corresponding to the composite and level variables, there are also two sets of operators in the adaptive grid algorithms.

○ Composite operators: operators defined in the valid regions across multiple levels using cell values on different levels.

○ Level operators: operators extended from the composite operators to all regions on a single level.

We use a 2D LS function $\phi$ to show the definitions of the above two types of operators, as illustrated in Fig. 2.2. Although only 2D operators are presented in this section, 3D operators can be defined in a straightforward way. For any point $(i,j) \subset \Gamma^l_{valid}$ on level $l$, the 2D composite gradient operator $\boldsymbol{G}^{cc,\mathrm{comp},l}$ is defined as

$$
\begin{aligned}
\left( G^{cc,\mathrm{comp},l}\phi \right)^x_{i,j} &= \frac{\phi_{i+1,j} - \phi_{i-1,j}}{2\Delta x^l}, \\
\left( G^{cc,\mathrm{comp},l}\phi \right)^y_{i,j} &= \frac{\phi_{i,j+1} - \phi_{i,j-1}}{2\Delta y^l},
\end{aligned}
\tag{2.1}
$$

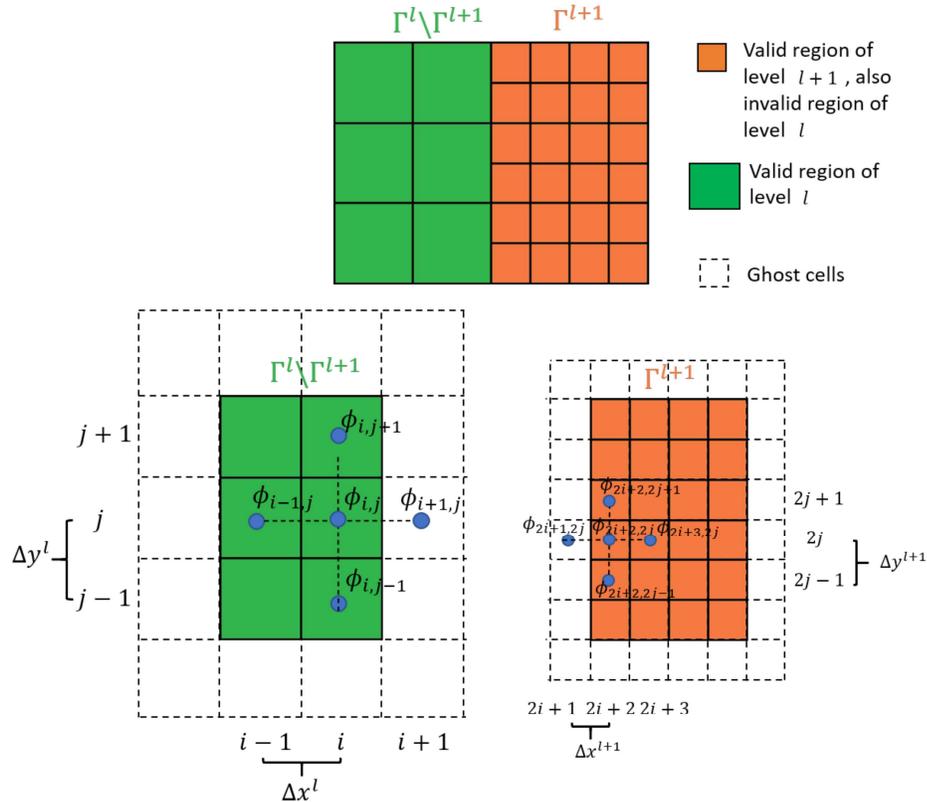

Figure 2.2: Diagram of the variable definitions on a multi-level grid and the stencil of the discretization operators.

Here, the superscript $^{cc}$ means that the operator applies to the cell-centered variables; the superscripts $^{comp}$ and $^l$ indicate that the composite operator $G$ is evaluated in the valid regions of level $l$; the superscripts $x$ and $y$ denote the $x$- and $y$-components of the gradient, respectively; $\Delta x^l$ and $\Delta y^l$ are the grid spacings in the $x$ and $y$ directions on level $l$, respectively. If the grid cell is away from the CF boundary, no ghost cell values are needed. If the grid cell is adjacent to the CF boundary, the values in the ghost cells are used for the evaluation of the composite gradient. For example, for the stencil shown in Fig. 2.2, the value of ghost cell $\phi_{i+1,j}$ at the CF boundary is averaged from level $l + 1$.

Similarly, for all $\phi_{2i+2,2j} \subset \Gamma_{valid}^{l+1}$ on level $l + 1$, the composite gradient operator

$\boldsymbol{G}^{cc,\mathrm{comp},l+1}$ is defined as

$$
\begin{aligned}
\left(G^{cc,\mathrm{comp},l+1}\phi\right)^x_{2i+2,2j} &= \frac{\phi_{2i+3,2j} - \phi_{2i+1,2j}}{2\Delta x^{l+1}}, \\
\left(G^{cc,\mathrm{comp},l+1}\phi\right)^y_{2i+2,2j} &= \frac{\phi_{2i+2,2j+1} - \phi_{2i+2,2j-1}}{2\Delta y^{l+1}}.
\end{aligned}
\tag{2.2}
$$

The value of ghost cell $\phi_{2i+1,2j}$ at the CF boundary is filled using the aforementioned conservative interpolation by combining the data on both level $l$ and level $l+1$.

Similar to the composite gradient operator $\boldsymbol{G}^{cc,\mathrm{comp},l}$, the composite divergence operator $D^{cc,\mathrm{comp},l}$ and the composite Laplacian operator $L^{cc,\mathrm{comp},l}$ are defined as

$$
\left(D^{cc,\mathrm{comp},l}\phi\right)_{i,j} = \frac{\phi_{i+1,j} - \phi_{i-1,j}}{2\Delta x^l} + \frac{\phi_{i,j+1} - \phi_{i,j-1}}{2\Delta y^l},
\tag{2.3}
$$

$$
\left(L^{cc,\mathrm{comp},l}\phi\right)_{i,j} = \frac{\phi_{i+1,j} - 2\phi_{i,j} + \phi_{i-1,j}}{(\Delta x^l)^2} + \frac{\phi_{i,j+1} - 2\phi_{i,j} + \phi_{i,j-1}}{(\Delta y^l)^2}.
\tag{2.4}
$$

To simplify the notations in the following sections, we denote the union set of the composite operators as

$$
\begin{aligned}
G^{cc,\mathrm{comp}} &= \bigcup_{i=0}^{l_{max}} G^{cc,\mathrm{comp},l}, \\
D^{cc,\mathrm{comp}} &= \bigcup_{i=0}^{l_{max}} D^{cc,\mathrm{comp},l}, \\
L^{cc,\mathrm{comp}} &= \bigcup_{i=0}^{l_{max}} L^{cc,\mathrm{comp},l}.
\end{aligned}
\tag{2.5}
$$

For a specified level $l$, the expressions of the level divergence operator $D^{cc,\mathrm{level},l}$, the level gradient operator $\boldsymbol{G}^{cc,\mathrm{level},l}$, and the level Laplacian operator $L^{cc,\mathrm{level},l}$ are the same as Eqs. (2.1), (2.3), and (2.4), respectively. However, these level operators also apply to the invalid regions of level $l$, i.e., $\phi_{i,j} \subset (\Gamma^l_{valid} \cup \Gamma^l_{invalid})$. For example, when evaluating the level gradient of $\phi^x_{i,j}$ in Eq. (2.1), $\phi_{i+1,j}$ is the value from the invalid region of level $l$, instead of the averaged value from level $l+1$. For conciseness, we replace the $D^{cc,\mathrm{level},l}$, $\boldsymbol{G}^{cc,\mathrm{level},l}$, and $L^{cc,\mathrm{level},l}$ with the conventional mathematical expressions $\nabla$, $\nabla\cdot$, $\nabla^2$ hereafter.

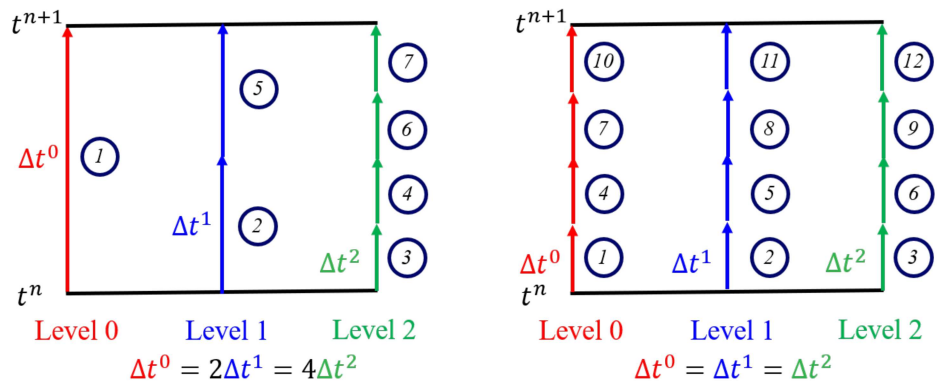

Figure 2.3: Schematic of the sub-steps in the level-by-level advancement method for a three-level grid ($l_{max} = 2$). Left: the subcycling method. Right: the non-subcycling method. The sub-steps are represented by the circled numbers.

## 2.2 Multi-level time advancement and Synchronization

### 2.2.1 Subcycling and non-subcycling methods

We consider two cycling methods, subcycling and non-subcycling, to advance variables on a multi-level grid. For the subcycling method, the solutions on different levels are advanced with different time steps. The larger grid spacings on the coarser levels allow a larger time step if the CFL number is kept the same on different levels. For example, if the refining ratio between two neighboring levels is two and if the velocities are approximately the same on both levels, the time step on the coarser level, $\Delta t^l$, and on the finer level, $\Delta t^{l+1}$, has the relation $\Delta t^l = 2\Delta t^{l+1}$. In the non-subcycling method, the variables on different levels advance with the same time step, restricted by the finest level $l_{max}$ to maintain the numerical stability. In this scenario, all levels are always at the same time instant.

Fig. 2.3 schematically shows how the individual levels are advanced with the subcycling and non-subcycling methods for a multi-level grid with $l_{max} = 2$. As shown in the sketch, seven sub-steps are needed to advance all levels from $t^n$ to $t^n + \Delta t^0$ using the subcycling method. For each sub-step, the single-level advancement algorithm is used for time advancement. By comparison, it takes 12 sub-steps for the non-subcycling method. Although the non-subcycling method has more steps, the subcycling method needs the time interpolation because of the mismatch of the time among different levels. The values at the middle time instant are obtained using a mid-point averaging, i.e. $f(t^n + \Delta t^l/2) = [f(t^n) + f(t^n + \Delta t^l)]/2$, which gives a second-order time accuracy. This interpolation is avoided in the non-subcycling method because all levels are at the same time instant.

### 2.2.2 Synchronization

Synchronization is the process of modifying the data on multiple levels to make them consistent and to better represent the composite solution. The synchronization step is needed for both the subcycling and the non-subcycling methods (Almgren et al., 1998; Martin and Colella, 2000; Martin et al., 2008). There are three sub-steps of the synchronization step.

Sub-step 1. Average

Because variables on the finer levels are considered more accurate, the velocity $\boldsymbol{u}$, pressure $p$, and LS function $\phi$ are replaced by those on the finer levels after the averaging process. In this way, the composite solution can be obtained in all valid regions of the multi-level grid. Because the collocated grid is used, the same averaging operator, i.e. the conservative interpolation operator introduced in section 2.1.2, can be used for all flow variables.

Sub-step 2. MAC synchronization and refluxing

When calculating the advection terms, the Marker and Cell (MAC) projection is only applied level by level. As a result, the advection velocity $\boldsymbol{u}^{adv}$ is only divergence-free on the specific level where it is calculated but not across all levels (Almgren et al., 1998; Martin et al., 2008). For example, at the CF boundary, the advection velocity $\boldsymbol{u}^{adv,l}$ on the coarser level $l$ is not equal to the edge average of the advection velocity $\boldsymbol{u}^{adv,l+1}$ on the finer level, leading to an imbalance of the momentum fluxes. As a result, the freestream preservation is violated while advancing the variables level by level.

To remedy this problem, the differences between $\boldsymbol{u}^{adv}$ on the coarser level and the finer level are quantified during the single-level advancement. These velocity differences, together with flux differences, form the registers to make the corrections for each level. Specifically, the velocity registers that hold the difference of the edge centered advection velocity are given by

$$\delta \boldsymbol{u}^l = -A^l \boldsymbol{u}^{adv,l} + \frac{1}{2} \sum_{k=1}^{2} \sum_{faces} A^{l+1} \boldsymbol{u}^{adv,k,l+1}. \tag{2.6}$$

In the above equation, the subscript $k$ represents the sub-steps of the finer level $l+1$ since it takes two sub-steps for level $l+1$ to catch up with level $l$ (Fig. 2.3), $\sum_{faces}$ is the sum over the cell faces, and $A$ is the area of each face. The velocity flux registers, including both the advective flux register $\delta \boldsymbol{f}_{\boldsymbol{u}}^{adv,l}$ and the viscous flux register $\delta \boldsymbol{f}_{\boldsymbol{u}}^{visc,l}$, are defined in a similar way as

$$\delta \boldsymbol{f}_{\boldsymbol{u}}^{adv,l} = \Delta t^l \left( A^l \boldsymbol{f}_{\boldsymbol{u}}^{adv,l} + \frac{1}{2} \sum_{k=1}^{2} \sum_{faces} A^{l+1} \boldsymbol{f}_{\boldsymbol{u}}^{adv,k,l+1} \right), \tag{2.7}$$

$$\delta \boldsymbol{f}_{\boldsymbol{u}}^{visc,l} = \Delta t^l \left( A^l \boldsymbol{f}_{\boldsymbol{u}}^{visc,l} + \frac{1}{2} \sum_{k=1}^{2} \sum_{faces} A^{l+1} \boldsymbol{f}_{\boldsymbol{u}}^{visc,k,l+1} \right). \tag{2.8}$$

The LS function has the advective flux register $\delta \boldsymbol{f}_{\phi}^{adv,l}$ only, which is calculated as

$$\delta \boldsymbol{f}_{\phi}^{adv,l} = \Delta t^l \left( A^l \boldsymbol{f}_{\phi}^{adv,l} + \frac{1}{2} \sum_{k=1}^{2} \sum_{faces} A^{l+1} \boldsymbol{f}_{\phi}^{adv,k,l+1} \right), \tag{2.9}$$

In Eqs. (2.7–2.9), $\boldsymbol{f}_{\boldsymbol{u}}^{adv,l}$, $\boldsymbol{f}_{\phi}^{adv,l}$, and $\boldsymbol{f}_{\boldsymbol{u}}^{visc,l}$ are given by

$$\boldsymbol{f}_{\boldsymbol{u}}^{adv,l} = \boldsymbol{u}^{adv} \phi^{n+1/2}, \tag{2.10}$$

$$\boldsymbol{f}_{\phi}^{adv,l} = \boldsymbol{u}^{adv} \phi^{n+1/2}, \tag{2.11}$$

$$\boldsymbol{f}_{\boldsymbol{u}}^{visc,l} = \frac{1}{2Re} \left( \mu(\phi^{n,l}) \nabla \boldsymbol{u^{n,l}} + \mu(\phi^{n+1,l}) \nabla \boldsymbol{u^{*,n+1,l}} \right). \tag{2.12}$$

The mismatch of the velocity register in Eq. (2.6) forms the right hand side of a MAC

solve for the correction $\delta e^l$ on level $l$,

$$\nabla \cdot \left( \frac{A^l}{\rho^{n+1/2,l}} \nabla(\delta e^l) \right) = \nabla \cdot \delta \boldsymbol{u}^l. \tag{2.13}$$

After solving Eq. (2.13), a velocity correction $\boldsymbol{u}_{corr}^l$ is obtained by

$$\boldsymbol{u}_{corr}^l = \frac{-\nabla(\delta e^l)}{\rho^{n+1/2,l}}. \tag{2.14}$$

The flux corrections associated with the above velocity correction are

$$\boldsymbol{f}_\phi^{corr,l} = \boldsymbol{u}_{corr}^l \phi^{n+1/2,l}, \tag{2.15}$$

$$\boldsymbol{f}_{\boldsymbol{u}}^{corr,l} = \boldsymbol{u}_{corr}^l \boldsymbol{u}^{n+1/2,l}. \tag{2.16}$$

The final correction to the LS function on level $l$, $\phi_{sync}^l$, is determined by the flux correction $\boldsymbol{f}_\phi^{corr,l}$ in Eq. (2.15) and the advective flux register $\delta \boldsymbol{f}_\phi^{adv,l}$ in Eq. (2.9) as

$$\phi_{sync}^l = -\nabla \cdot \boldsymbol{f}_\phi^{corr,l} - \frac{\delta \boldsymbol{f}_\phi^{adv,l}}{\Delta t \cdot Vol^l}, \tag{2.17}$$

where $Vol^l$ is volume of the grid cell on level $l$, i.e., $Vol^l = \Delta x^l \Delta y^l$ for the 2D case and $Vol^l = \Delta x^l \Delta y^l \Delta z^l$ for the 3D case. The LS function on level $l$ is then updated as

$$\phi^{n+1,l} := \phi^{n+1,l} + \Delta t^l \phi_{sync}^l, \tag{2.18}$$

The flux correction about the velocity $\boldsymbol{f}_{\boldsymbol{u}}^{corr,l}$ in Eq. (2.16), together with its advective flux register $\delta \boldsymbol{f}_{\boldsymbol{u}}^{adv,l}$ in Eq. (2.7) and viscous flux register $\delta \boldsymbol{f}_{\boldsymbol{u}}^{adv,l}$ in Eq. (2.8), form a subsequent parabolic equation,

$$\boldsymbol{u}_{sync}^l - \frac{\Delta t}{2\rho^{n+1/2,l} Re} \nabla \cdot \left( \mu(\phi^{n+1}) \nabla \boldsymbol{u}_{sync}^l \right) = -\nabla \cdot \boldsymbol{f}_{\boldsymbol{u}}^{corr,l} - \frac{1}{\Delta t \cdot Vol^l} \left( \delta \boldsymbol{f}_{\boldsymbol{u}}^{adv,l} + \frac{1}{\rho^{n+1/2,l}} \delta \boldsymbol{f}_{\boldsymbol{u}}^{adv,l} \right), \tag{2.19}$$

which gives the final correction of the velocity $\boldsymbol{u}_{sync}^l$ on level $l$. The updated velocity on level $l$ is then given by

$$\boldsymbol{u}^{n+1,l} := \boldsymbol{u}^{n+1,l} + \Delta t^l \boldsymbol{u}_{sync}^l. \tag{2.20}$$

The corrections also need to propagate to all the finer levels $q$ as

$$\phi^{n+1,q} := \phi^{n+1,q} + \Delta t^l \mathcal{I}_{cons}(\phi_{sync}^l), \tag{2.21}$$

and

$$\boldsymbol{u}^{n+1,q} := \boldsymbol{u}^{n+1,q} + \Delta t^l \mathcal{I}_{cons}(\boldsymbol{u}_{sync}^l) \tag{2.22}$$

for all $l < q \leq l_{max}$. Here, the conservative interpolation $\mathcal{I}_{cons}$ is used.

At last, for any level $l > 0$, the velocity registers and flux registers on the coarser level $l-1$ are affected by the above correction and thus need to be updated as follows,

$$\delta \boldsymbol{u}^{l-1} := \delta \boldsymbol{u}^{l-1} + \frac{1}{2} \sum_{faces} (A^l \boldsymbol{u}_{corr}^l), \tag{2.23}$$

$$\delta \boldsymbol{f}_{\boldsymbol{u}}^{adv,l-1} := \delta \boldsymbol{f}_{\boldsymbol{u}}^{adv,l-1} + \frac{1}{2} \Delta t^{l-1} \sum_{faces} (A^l \boldsymbol{f}_{\boldsymbol{u}}^{corr,l}), \tag{2.24}$$

$$\delta \boldsymbol{f}_{\boldsymbol{u}}^{visc,l-1} := \delta \boldsymbol{f}_{\boldsymbol{u}}^{visc,l-1} + \frac{1}{2} \Delta t^l \sum_{faces} \left( \frac{1}{2} A^l \mu(\phi^{n+1}) \nabla \boldsymbol{V}_{sync}^l \right), \tag{2.25}$$

$$\delta \boldsymbol{f}_{\phi}^{adv,l-1} := \delta \boldsymbol{f}_{\phi}^{adv,l-1} + \frac{1}{2} \sum_{faces} (A^l \boldsymbol{f}_{\phi}^{corr,l}). \tag{2.26}$$

As a reminder, the above MAC synchronization and refluxing sub-step is used to maintain the conservation of momentum and scalar in the whole flow field. To validate the efficacy of this sub-step, we assess the conservation of a passive scalar in the inviscid flow of a counter-rotating vortex pair (Martin and Colella, 2000). The initial azimuthal velocity $u_\theta(r)$ of one vortex is given by

$$u_\theta(r) = \begin{cases} \tilde{\Gamma} \left( \frac{8}{3R^3} r^4 - \frac{5}{R^4} r^3 + \frac{10}{3R^2} r \right), & r < R, \\ \tilde{\Gamma} \left( \frac{1}{r} \right), & r \geq R, \end{cases} \tag{2.27}$$

where $R$ is the radius of the vortex core, $r$ is the distance from the vortex center $(x_c, y_c)$, and $\tilde{\Gamma}$ is the vortex strength. One vortex is centered at $(x_c, y_c) = (0.3, 0.35)$ with $\tilde{\Gamma} = -0.35$, $R = 0.15$ and the other is at $(x_c, y_c) = (0.3, 0.65)$ with $\tilde{\Gamma} = 0.35, R = 0.15$. The computational domain is $\Omega : [0,1] \times [0,1]$. The grid size on level 0 is $100 \times 100$. A passive scalar advected by the above vortex pair is simulated. The initial scalar field is

Table 2.1: Parameters for cases of the counter vortex problem.

| Case No. | Mesh refinement type | Is refluxing performed? |
|----------|----------------------|-------------------------|
| 1 | Static | No |
| 2 | Static | Yes |
| 3 | Dynamic | No |
| 4 | Dynamic | Yes |

set as

$$s(x,y) = \begin{cases} 2.0, & \text{if } x \in [0.2, 0.8] \text{ and } y \in [0.2, 0.8], \\ 1.0, & \text{otherwise}. \end{cases} \tag{2.28}$$

As the LS function is essentially a passive scalar governed by the advection equation, the simulation of $s$ is carried out using Eq. (3.3) presented in the next chapter. A total of four subcycling cases are considered, varying in the mesh refinement and whether the refluxing step is performed, as listed in Table 2.1. For the mesh refinement, we consider the static and dynamic refinement. For the static refinement, grid cells are refined to $l_{max} = 1$ in the rectangular region $x \in [0.2, 0.8]$ and $y \in [0.2, 0.8]$. For the dynamic refinement, the vorticity magnitude, $|\omega_z| > 0.75|\omega_z^{max}|$, is used as the refinement criterion.

The vorticity field at $t = 0.36$ for the dynamic refinement case with refluxing (case 4) is shown in the left part of Fig. 2.4, which is in good agreement with (Martin and Colella, 2000). The right part of Fig. 2.4 shows the scalar concentration and grid hierarchy at $t = 0.36$ for case 4. Because of the advection by the vortices, a high concentration of the scalar crosses the CF boundary, which can lead to errors in the conservation of the scalar if the MAC synchronization and refluxing operations are not considered. To quantify this error, the relative change of the total amount of scalar compared to the initial time is evaluated as

$$e(t) = \frac{\int_\Omega \left( s|_t - s|_{t=0} \right) dx}{\int_\Omega s|_{t=0} dx}. \tag{2.29}$$

The results for the above four cases are plotted in Fig. 2.5. When the refluxing is used (cases 2 and 4), the relative error is within $10^{-16}$ for both the static and dynamic refinement, while noticeable errors are present in simulations without refluxing (cases 1 and 3). This test shows that the MAC synchronization and refluxing operations are necessary and can help the conservation of the scalar.

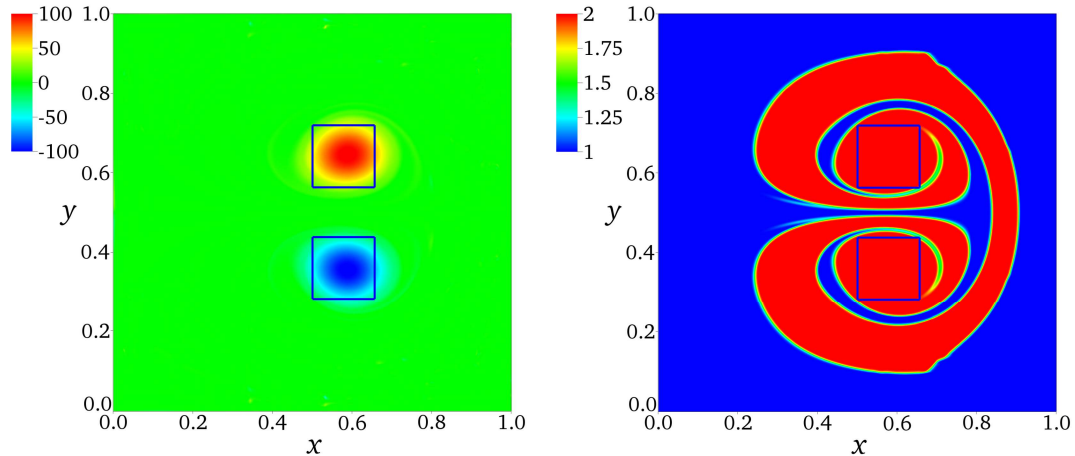

Figure 2.4: Left: Vorticity contours and grid hierarchy at $t = 0.36$ of the two-level subcycling case with the dynamic mesh refinement with refluxing (case 4) for the counter vortex problem. Right: Concentration of the passive scalar $s$ and grid hierarchy at $t = 0.36$ for case 4. The blue lines are the grid patches on level 1.

### Sub-step 3. Synchronization projection

Because the level projection is only applied level by level, it does not guarantee that the velocity is divergence-free across all levels. The synchronization projection, as the last sub-step of the synchronization, is applied to enforce this constraint (Almgren et al., 1998; Martin and Colella, 2000). Using the composite operators defined in section 2.1.1, we first solve a correction field $e_s$ by projecting velocities on all levels (Minion, 1996),

$$L^{cc,\text{comp}}_{\rho^{n+1}} e_s = \frac{1}{\Delta t^{sync}} D^{cc,\text{comp}} \boldsymbol{u}^{n+1}, \qquad (2.30)$$

where $\Delta t^{sync}$ is equal to the time step of level 0, i.e. $\Delta t^{sync} = \Delta t^0$; $L^{cc,\text{comp}}_{\rho^{n+1}} e_s$ is a density-weighted approximation to $\nabla \cdot (1/\rho^{n+1} \nabla e_s)$ on all levels. We note that the ghost cell values of $e_s$ need to be appropriately specified for the evaluation of the composite operators $L^{cc,\text{comp}}$, $D^{cc,\text{comp}}$, and $G^{cc,\text{comp}}$ in Eq. (2.5). On level 0, the ghost cell values

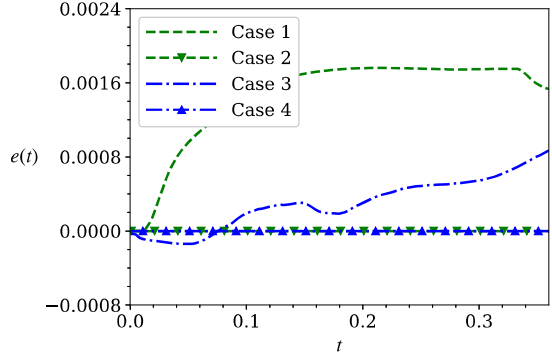

Figure 2.5: Comparison of the conservation errors of the passive scalar among the four cases of the counter rotating vortex. Case 1 and case 2 use the static refinement while case 3 and case 4 use the dynamic refinement. Case 2 and case 4 consider the refluxing while case 1 and case 3 do not.

of $e_s$ are determined by the physical boundary conditions. On level $l$ $(l > 0)$,

$$e_s^l = \mathcal{I}_{cons}(e_s^l, e_s^{l-1}), \tag{2.31}$$

which means that the ghost cell values $e_s^l$ on level of $l$ at the CF boundaries are computed using the conservative interpolation (section 2.1.1) by combining the values of $e_s$ on both level $l$ and $l-1$ (Martin and Colella, 2000; Martin et al., 2008). We remark that a single conservative interpolation scheme can be used with all composite operators to fill the ghost cells of all cell-centered variables at the CF boundaries (Howell and Bell, 1997; Almgren et al., 1998). Finally, the multi-level velocity field is updated as

$$\boldsymbol{u}^{n+1} := \boldsymbol{u}^{n+1} - \Delta t^{sync} \boldsymbol{G}^{cc,comp} e_s, \tag{2.32}$$

after which $\boldsymbol{u}^{n+1}$ becomes divergence-free in a multi-level sense.

The stability of the projection operation is an important issue. For example, for the non-graded TBAMR, the stability of the discrete projection can be affected by the presence of the large size ratio of the adjacent cells, for which a different projection formulation was proposed to enforce the orthogonality of the projection (Min and Gibou, 2006; Gibou et al., 2007). The stability of the synchronization projection method used in the present work has been demonstrated by Martin and Colella (Martin and Colella, 2000) using a three-vortex problem. The robustness and the stability of the approximate projection are related to the careful interpolations of the ghost cell values (Guy and Fogelson, 2005; Brown et al., 2001). Before doing the approximate projection, we use the constant extrapolation for the ghost cell values at the physical boundaries (Guy and Fogelson, 2005). We also apply the conservative interpolation for those on the finer levels at the CF boundaries (Almgren et al., 1998). Another factor contributing to the stability of the algorithm is the form of the projection. The projection is applied to the updated velocity at the new time ($\boldsymbol{u}^{n+1}$) rather than the increment velocity ($\boldsymbol{u}^{n+1} - \boldsymbol{u}^n$), which helps to stabilize the synchronization projection as found by Martin and Colella (2000). Furthermore, the nesting property of the BSAMR guarantees the regularity of the multi-level grid and probably eases the burden on the convergence of the multi-level multigrid solver compared to the non-graded TBAMR. In case a higher refining ratio is required in BSAMR, one may need to carefully implement the interpolation scheme to ensure the stability of the projection. Later in section 3.3.1, both the level projection and the synchronization projection methods are tested with a sample problem (Min and Gibou, 2006), which further confirms the stability of the approximate projection method.

As the final remark of this section, we discuss some differences of our synchronization algorithm from other implementations in the literature. For the averaging step, all variables can share the same averaging operator because of our adoption of the collocated layout. We believe this averaging process is simpler than that in Almgren et al. (1998) because the latter, with a semi-staggered layout, requires different averaging operators for velocity and pressure. Secondly, in our algorithm for the MAC synchronization and refluxing, the errors are collected from the instantaneous field as velocity and flux registers, which are then used to correct the multi-level solution. In Martin and Colella (2000) and Martin et al. (2008), the volume discrepancy method is used to maintain the freestream preservation, where an auxiliary scalar is used as the indicator of the errors and is advected in time along with the flow field. Finally, in our algorithm, the synchronization projection is performed only when all the finer levels catch up with level 0. In other words, for every time step from $t^n$ to $t^{n+1}$, only one synchronization projection step is conducted. Compared to Almgren et al. (1998), where the synchronization projection is performed iteratively on pairs of two consecutive levels, our algorithm has fewer projection steps but is still effective as demonstrated.

# Chapter 3

# Inconsistent scheme for
# two-phase flow

## 3.1 Mathematical formulation

In this section, we apply the inconsistent scheme to simulate two-phase incompressible flows with gravity and surface tension effects. As illustrated in the left part of Fig. 3.1, the densities of the two phases are denoted by $\rho_1$ and $\rho_2$, respectively; and the dynamic viscosities are $\mu_1$ and $\mu_2$, respectively. The Navier–Stokes equations for the fluid flow with variable density and viscosity read

$$\frac{\partial \boldsymbol{u}}{\partial t} + \nabla \cdot (\boldsymbol{u}\boldsymbol{u}) = \frac{1}{\rho(\phi)} \left[ -\nabla p + \frac{1}{Re} \nabla \cdot 2\mu(\phi)\tilde{\boldsymbol{D}} + \rho(\phi)\frac{\hat{\boldsymbol{g}}}{Fr^2} - \frac{1}{We}\kappa(\phi)\delta(\phi)\boldsymbol{n} \right], \quad (3.1)$$

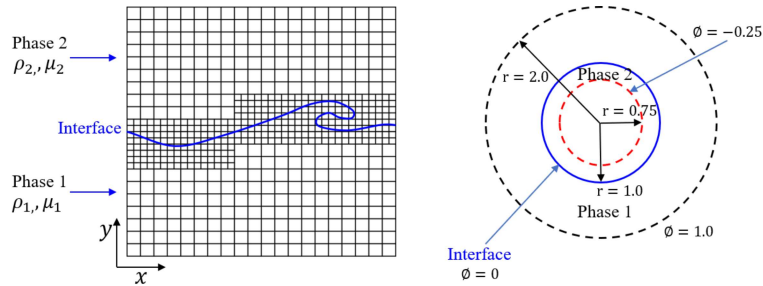

Figure 3.1: Left: two-phase flow on a multi-level Cartesian grid. Right: schematic definition of the LS function.

$$\nabla \cdot \boldsymbol{u} = 0. \tag{3.2}$$

where $\boldsymbol{u}$ is the velocity vector, $p$ is the pressure, $\rho$ is the density, and $\mu$ is the dynamic viscosity. The above equations are normalized by a characteristic velocity $U$, characteristic length $L$, and the density and dynamic viscosity of phase 1, $\rho_1$ and $\mu_1$. In the viscous term $\nabla \cdot 2\mu(\phi)\tilde{\boldsymbol{D}}/Re$, $Re = \rho_1 UL/\mu_1$ is the Reynolds number and $\tilde{\boldsymbol{D}} = (\nabla \boldsymbol{u} + \nabla \boldsymbol{u}^T)/2$ is the strain rate tensor. For the gravitational term $\rho(\phi)\hat{\boldsymbol{g}}/Fr^2$, $\hat{\boldsymbol{g}}$ is the unit vector in the direction of gravity and the Froude number is defined as $Fr = U/\sqrt{gL}$ with $g$ being the gravitational acceleration. In the surface tension term $\kappa(\phi)\delta(\phi)\boldsymbol{n}/We$, $\kappa(\phi)$ is the curvature of the interface, $\delta(\phi)$ is the Dirac function, $\boldsymbol{n}$ is the unit vector of the surface normal, and $We = \rho_1 U^2 L/\sigma$ is the Weber number, with $\sigma$ being the surface tension coefficient.

The two immiscible fluids are tracked by the LS function $\phi$ (Sussman et al., 1999; Sussman and Fatemi, 1999). As illustrated in the right part of Fig. 3.1, $\phi$ is the signed distance from the two-phase interface, with $\phi > 0$ in the phase 1 and $\phi < 0$ in the phase 2. The advection equation of $\phi$ is

$$\frac{\partial \phi}{\partial t} + \nabla \cdot (\boldsymbol{u}\phi) = 0. \tag{3.3}$$

The unit vector $\boldsymbol{n}$ and curvature $\kappa(\phi)$ of the interface are calculated as

$$\boldsymbol{n} = \frac{\nabla \phi}{|\nabla \phi|}, \tag{3.4}$$

$$\kappa(\phi) = \nabla \cdot \boldsymbol{n} = \nabla \cdot \left( \frac{\nabla \phi}{|\nabla \phi|} \right). \tag{3.5}$$

For the numerical treatment, both the Dirac function $\delta(\phi)$ and the Heaviside function $H(\phi)$ are smoothed around the interface as (Sussman et al., 1999)

$$\delta_\epsilon(\phi) = \begin{cases} 0 & |\phi| > \epsilon \\ \frac{1}{2}\left[ \frac{1}{\epsilon} + \frac{1}{\epsilon}\cos\left(\frac{\pi\phi}{\epsilon}\right) \right] & |\phi| \le \epsilon, \end{cases} \tag{3.6}$$

$$H_\epsilon(\phi) = \begin{cases} 0 & \phi < -\epsilon \\ \frac{1}{2}\left[1 + \frac{x}{\epsilon} - \frac{1}{\pi}\sin\left(\frac{\pi\phi}{\epsilon}\right)\right] & |\phi| \leq \epsilon \\ 1 & \phi > \epsilon, \end{cases} \tag{3.7}$$

where $\epsilon$ is the smearing width and is usually set to be twice the grid spacing (Sussman et al., 1999; Sussman and Fatemi, 1999). Finally, the dimensionless density $\rho(\phi)$ and dynamic viscosity $\mu(\phi)$ are given by

$$\rho(\phi) = \lambda + (1-\lambda)H(\phi), \tag{3.8}$$

$$\mu(\phi) = \eta + (1-\eta)H(\phi). \tag{3.9}$$

In the above equations, $\lambda = \rho_2/\rho_1$ and $\eta = \mu_2/\mu_1$ are the normalized density and dynamic viscosity of phase 2.

## 3.2 Time advancement

In this chapter, we use a level-by-level method (Martin and Colella, 2000; Martin et al., 2008) for the time advancement on the multi-level grid. This method updates the solution on each level individually in a certain order and synchronizes the composite solution across different levels. Because the multi-level advancement algorithm is based on the single-level advancement, we first introduce the single-level algorithm and then describe the multi-level algorithm.

### 3.2.1 Single-level advancement

**Time discretization**

For a single level, the momentum equation is advanced by a fractional step method with the approximate projection (Rider, 1995, 1998; Almgren et al., 2000) to enforce the incompressibility condition. The LS advection is updated using the Godunov scheme (Colella, 1990; Bell et al., 1989; Almgren et al., 1998; Sussman et al., 1999).

At the beginning of each time advancement of level $l$, the velocity $\boldsymbol{u}^{n,l}$ and the LS function $\phi^{n,l}$ at time $t^{n,l}$ are given. Owing to the fractional step method we used, the pressure is staggered in time (Almgren et al., 1998; Sussman et al., 1999; Bhalla et al., 2013) and thus the pressure $p^{n-1/2,l}$ at time $t^{n-1/2,l}$ is known. Because the single-level advancement concerns only level $l$, we omit the superscript $^l$ in this section. To obtain the updated velocity $\boldsymbol{u}^{n+1}$, pressure $p^{n+1/2}$, and LS function $\phi^{n+1}$ on level $l$, the solver performs the following steps:

1. Advance the LS function as

$$\phi^{n+1} = \phi^n - \Delta t \left[\nabla \cdot (\boldsymbol{u}\phi)\right]^{n+1/2}. \tag{3.10}$$

The advection term in the above equation is calculated using the Godunov scheme detailed in Appendix A.2.

2. Solve the intermediate velocity $\boldsymbol{u}^*$ semi-implicitly

$$\boldsymbol{u}^{*,n+1} - \frac{\Delta t}{2\rho(\phi^{n+1/2})Re}\nabla \cdot \mu(\phi^{n+1})\nabla\boldsymbol{u}^{*,n+1} = \boldsymbol{u}^n - \Delta t \left[\nabla \cdot (\boldsymbol{u}\boldsymbol{u})\right]^{n+1/2} +$$

$$\frac{\Delta t}{\rho(\phi^{n+1/2})}\left[-\nabla p^{n-1/2} + \frac{1}{2Re}\nabla \cdot \mu(\phi^n)\nabla\boldsymbol{u}^n + \rho(\phi^{n+1/2})\frac{z}{Fr^2} - \frac{1}{We}\kappa(\phi^{n+1/2})\delta(x^{n+1/2})\boldsymbol{n}\right]. \tag{3.11}$$

In Eq. (3.11), the detailed discretization of the advection term $\nabla \cdot (\boldsymbol{u}\boldsymbol{u})$, viscous term $\nabla \cdot (\mu(\phi)\nabla\boldsymbol{u})$, and surface tension term $\kappa(\phi)\delta(x)\boldsymbol{n}/We$ are given in section 3.2.1. The LS function at $t^{n+1/2}$ is calculated by

$$\phi^{n+1/2} = \frac{1}{2}(\phi^n + \phi^{n+1}), \tag{3.12}$$

where $\phi^{n+1}$ is obtained from step 1. The $\rho(\phi^{n+1/2})$, $\mu(\phi^n)$, and $\mu(\phi^{n+1})$ are then obtained from Eqs. (3.8) and (3.9).

3. Apply the projection method to obtain the pressure and a solenoidal velocity field. To conduct the level projection, a temporary variable $\boldsymbol{V}$ is defined as

$$\boldsymbol{V} = \frac{\boldsymbol{u}^{*,n+1}}{\Delta t} + \frac{1}{\rho(\phi^{n+1/2})}\nabla p^{n-1/2}. \tag{3.13}$$

Then the updated pressure $p^{n+1/2}$ is calculated by

$$L^{cc,\text{level}}_{\rho^{n+1/2}} p^{n+1/2} = \nabla \cdot \boldsymbol{V}, \tag{3.14}$$

where $L^{cc,\text{level}}_{\rho^{n+1/2}} p^{n+1/2}$ is a density-weighted approximation to $\nabla \cdot (1/\rho^{n+1/2} \nabla p^{n+1/2})$. Finally, the velocity can be calculated as

$$\boldsymbol{u^{n+1}} = \Delta t \left( \boldsymbol{V} - \frac{1}{\rho^{n+1/2}} \nabla p^{n+1/2} \right). \tag{3.15}$$

As defined in section 2.1.2, $\nabla \cdot$ and $\nabla$ are the cell-centered level divergence operator $D^{cc,\text{level}}$ and level gradient operator $G^{cc,\text{level}}$, respectively. The level gradient operator $G^{cc,\text{level}}$ is not the minus transpose of the level divergence operator $D^{cc,\text{level}}$, i.e., $G^{cc,\text{level}} \neq -(D^{cc,\text{level}})^T$ (Martin and Colella, 2000; Martin et al., 2008; Lal, 1993). As a result, the idempotency of the approximate projection $\boldsymbol{P} = I - G^{cc,\text{level}}(L^{cc,\text{level}})^{-1} D^{cc,\text{level}}$ is not ensured (Almgren et al., 1998), i.e., $\boldsymbol{P}^2 \neq \boldsymbol{P}$. Yet, this nonidempotent approximate projection is stable and appears to be well-behaved in various numerical tests (Almgren et al., 1996; Martin and Colella, 2000; Rider, 1995) and practical applications (Sussman et al., 1999; Martin et al., 2008). Notably, for a uniform single grid with periodic boundary conditions, Lai (Lai, 1993) theoretically proved that this approximate projection method is stable, in that $\|\boldsymbol{P}\| \leq 1$. It should be noted that the approximate projection is applied to the intermediate velocity $\boldsymbol{u^{*,n+1}}$. Compared with the form that projects the increment velocity $\boldsymbol{u^{*,n+1}} - \boldsymbol{u^n}$, e.g. as that used in Almgren et al. (1998), the projection method used here can reduce the accumulation of pressure errors and lead to a more stable algorithm (Rider, 1995; Guy and Fogelson, 2005). We also validate the effectiveness and stability of this approximate projection in section 3.3.1 using a numerical test (Min and Gibou, 2006).

4. Reinitialize the LS function $\phi$ to maintain $\phi$ as a signed distance function of the interface and guarantee the conservation of the mass of the two phases. In this step, a temporary LS function $d(\boldsymbol{x}, \tau)$ is updated iteratively using the following pseudo evolution equation,

$$\frac{\partial d}{\partial \tau} = S(\phi)(1 - |\nabla d|), \tag{3.16}$$

with the initial condition

$$d(\boldsymbol{x}, \tau = 0) = \phi^{n+1}(\boldsymbol{x}), \tag{3.17}$$

where

$$S(\phi) = 2\left(H(\phi) - 1/2\right). \tag{3.18}$$

Here, $\tau$ is the pseudo time for iterations. A second-order essentially non-oscillatory (ENO) scheme is used to discretize the distance function and a second-order Runge–Kutta (RK) method is applied for the pseudo time advancing. To ensure the mass conservation, $d(\boldsymbol{x}, \tau)$ is further corrected by minimizing the differences of the volume of each fluid between $\tau = 0$ and the final iteration (Sussman et al., 1999; Sussman and Fatemi, 1999). Finally, the LS function $\phi$ is re-initialized by the volume corrected $d$.

At last, we give a summary of the single-level advancement algorithm in Algorithm 1 as follows.

---
**Algorithm 1** Single-level advancement algorithm
---
1: Advance the LS function using Eq. (3.10);
2: Solve the intermediate velocity using Eq. (3.11);
3: Apply the projection method to update the pressure and velocity field following Eqs. (3.13–3.15);
4: Re-initialize the LS function on the single level using Eqs. (3.16–3.18).
---

**Discretization of the viscous and surface tension terms**

In this part, all discretized formulas use the level operator (section 2.1.2). For simplicity, only the 2D discretized formulas are given in this section. The 3D formulas can be extended in a straightforward way.

For the discretization of the viscous term, the $x$-component of $\nabla \cdot \mu(\phi)\nabla\boldsymbol{u}$ at point $(i, j)$ is calculated as

$$
\begin{aligned}
(\nabla \cdot \mu(\phi)\nabla u)_{i,j} =& \frac{\mu_{i+1/2,j}(u_{i+1,j} - u_{i,j}) - \mu_{i-1/2,j}(u_{i,j} - u_{i-1,j})}{\Delta x^2} \\
&+ \frac{\mu_{i,j+1/2}(u_{i,j+1} - u_{i,j}) - \mu_{i,j-1/2}(u_{i,j} - u_{i,j-1})}{\Delta y^2},
\end{aligned} \tag{3.19}
$$

where the edge-centered viscosity $\mu_{i+1/2,j}$ and $\mu_{i,j+1/2}$ are defined as

$$\mu_{i+1/2,j} = \frac{1}{2}\left[\mu(\phi)_{i,j} + \mu(\phi)_{i+1,j}\right], \tag{3.20}$$

$$\mu_{i,j+1/2} = \frac{1}{2}\left[\mu(\phi)_{i,j} + \mu(\phi)_{i,j+1}\right]. \tag{3.21}$$

The $y$-component of the viscous term is calculated for the velocity component $v$ in a similar way. For the surface tension term $\kappa(\phi)\delta(x)\boldsymbol{n}/We$, we have

$$(\nabla\phi)_{i,j} = \left(\frac{\phi_{i+1,j} - \phi_{i-1,j}}{2\Delta x}, \frac{\phi_{i,j+1} - \phi_{i,j-1}}{2\Delta y}\right), \tag{3.22}$$

$$\boldsymbol{n}_{i,j} = \frac{(\nabla\phi)_{i,j}}{(|\nabla\phi|)_{i,j}}, \tag{3.23}$$

$$\kappa(\phi)_{i,j} = \frac{n^1_{i+1,j} - n^1_{i-1,j}}{2\Delta x} + \frac{n^2_{i,j+1} - n^2_{i,j-1}}{2\Delta y}, \tag{3.24}$$

where $\boldsymbol{n} := (n^1, n^2)$ and the delta function $\delta(x)$ is calculated from Eq. (3.6).

### 3.2.2 Multi-level advancement

In this section, we describe how we apply the single-level advancement algorithm to the multi-level advancement algorithm. Besides using the subcycling and non-subcycling methods (chapter 2) and applying the synchronization step (chapter 2), we devise a multi-level re-initialization algorithm for the LS function across multiple levels. The multi-level initialization of the flow field is also introduced. At last, a summary of the multi-level advancement algorithm is given. We note that, although some BSAMR (Colella et al., 2009; Zhang et al., 2019) and TBAMR (Burstedde et al., 2011; Min and Gibou, 2006, 2007) frameworks have supported a high refining ratio, in this chapter, we limit the refining ratio to 2 between two consecutive levels for easier implementation.

#### Multi-level re-initialization of the LS function

In the synchronization step, the property of the LS function as a signed distance function is not guaranteed in the multi-level sense. To maintain the regularity of the LS function and improve the mass conservation, a multi-level re-initialization algorithm

---
**Algorithm 2** Multi-level re-initialization of the LS function

---
1: **for** $l = l_{max} - 1, 0, -1$ **do**
2:     **for** $j = 1, N_{iter}$ **do**
3:         $\hat{\phi}^l \leftarrow$ single-level re-initialization of $\phi^l$ on level $l$
4:         $\phi^{l+1} \leftarrow \mathcal{I}_{cons}(\hat{\phi}^l)$
5:         $\hat{\phi}^{l+1} \leftarrow$ single-level re-initialization of $\phi^{l+1}$ on level $l + 1$
6:         $\phi^l \leftarrow$ average $\hat{\phi}^{l+1}$
7:     **end for**
8: **end for**

---

(Algorithm 2) is proposed here. The core part of the algorithm is to synchronize the LS functions on two consecutive levels using the single-level re-initialization and interpolation iteratively, corresponding to the inner loop in Algorithm 2. First, we apply the single-level re-initialization algorithm (section 3.2.1) to the LS function and interpolate the function onto the finer level. Then the single-level re-initialization is carried out on the finer level, after which the LS function is averaged back to the coarser level. From our testing, the LS function can usually be corrected on these two levels after three iterations of the pair re-initialization. The above pair re-initialization process is applied to all levels, from level $l_{max} - 1$ to level 0, as indicated by the outer loop of Algorithm 2. This ensures the LS function as the signed distance function and mitigates the mass loss on all levels, as demonstrated by the test cases. Furthermore, our tests also show that the above re-initialization algorithm is computationally efficient.

We note that our multi-level re-initialization technique is different from the one used in Sussman et al. (1999), which solves the pseudo evolution equation of $\phi$ (Eq. 3.16) from the coarsest level to the finest level using the subcycling method. In this chapter, the multi-level re-initialization step starts from the finest level. The LS function on two consecutive levels are synchronized using the iteration technique, and thus the subcycling method is not needed here.

### Multi-level initialization of flow field

All field values, including the velocity $\boldsymbol{u}$, pressure $p$, and LS function $\phi$, need to be initialized on all levels at the beginning of the simulation. Firstly, the velocity $\boldsymbol{u}$ and LS function $\phi$ on the coarsest level (level 0) are assigned based on the initial conditions.

Ghost cell values for these variables on level 0 are also filled by the physical boundary conditions. Then, the grid on the next level (level 1) is generated based on the refinement criteria. After the refinement, the velocity $\boldsymbol{u}$ and LS function $\phi$ on level 1 are determined based on the initial conditions. This "refining and filling" procedure is repeated until the finest level $l_{max}$ is reached, or until there is no need to refine the grid based on the refinement criteria. The pressure $p$ is initialized as zero on all levels and corrected by the level projection at the first step.

### Summary of multi-level advancement

Algorithm 3 summarizes the unified multi-level advancement algorithm for both the subcycling and non-subcycling methods. After the initialization, we can use either the subcycling or non-subcycling method for time advancement. The synchronization step and the multi-level re-initialization step are then applied when a coarser level catches up with a finer level. Finally, the grid refinement is applied before moving to the next time step. In the multi-level advancement algorithm, the MAC projection, semi-implicit viscous solver (Eq. 3.11), level projection (Eq. 3.14), MAC synchronization (Eq. 2.13), and refluxing (Eq. 2.19) steps use the multigrid (MG) solver on each level. The MLMG solver incorporates the mesh information across multiple levels and is only used with the synchronization projection sub-step.

As a final remark, we emphasize that our multi-level advancement algorithm is a level-by-level advancement method, which is different from the composite advancement method (Bhalla et al., 2013; Nangia et al., 2019a; Sussman et al., 1999; Nangia et al., 2019b) in several aspects. In the level-by-level advancement method, the level variables are used for time advancement. Each level can be advanced individually without considering the finer levels before the synchronization step. Because the time advancements at different levels are decoupled, the constraints of the time step on the coarser levels are alleviated. This is in contrast to the composite advancement method, where the multi-level time advancement is based on the composite variables and only variables in the valid regions are utilized. The MLMG solver is employed to simultaneously update the velocity and pressure in the valid regions of all levels. This distinctly different treatment is the reason why the composite advancement method is not flexible enough to embed both the subcycling and non-subcycling methods in a straightforward way, while the level-by-level method in the present chapter can handle both cycling methods with relative ease.

## 3.3 Results

This section presents several canonical test cases to validate the proposed AMR framework from different aspects. First, we shall clarify some common parameters used by these cases unless stated otherwise. For all of the following cases, $\Delta t_0$ denotes the time step on level 0. We use $\Delta x_0$, $\Delta y_0$, and $\Delta z_0$ to represent the grid spacings in $x$-, $y$-, and $z$-directions, respectively, on level 0. For the multi-level grid, grid spacings on the finer level $l$ satisfy $\Delta x_l = \Delta x_0/2^l$, $\Delta y_l = \Delta y_0/2^l$, and $\Delta z_l = \Delta z_0/2^l$ for all $0 \leq l \leq l_{max}$.

### 3.3.1 Stability of the projection

In this test, we examine the effectiveness and stability of the level projection (Eqs. (3.13) and (3.14)) on the single-level grid and the synchronization projection on the multi-level grid. The projection is performed on the manufactured velocity field $\boldsymbol{u}^* = (u^*, v^*)$ from Min and Gibou (2006), given by

$$u^*(x,y) = \sin(x)\cos(y) + x(\pi - x)y^2 \left(\frac{y}{3} - \frac{\pi}{2}\right), \tag{3.25}$$

$$v^*(x,y) = -\cos(x)\sin(y) + y(\pi - y)x^2 \left(\frac{x}{3} - \frac{\pi}{2}\right). \tag{3.26}$$

This manufactured velocity can be decomposed as $\boldsymbol{u}^* = \boldsymbol{u}_{\text{div}} + \nabla\phi$, where $\boldsymbol{u}_{\text{div}}$ is a divergence-free velocity field and $\phi = -(x^3/3 - \pi x^2/2)(y^3/3 - \pi y^2/2)$. We consider a computational domain $\Omega = [0, \pi]$ with $\boldsymbol{u}^* \cdot \boldsymbol{n} = 0$ on the domain boundary $\partial\Omega$. For the single-level grid, the grid number on level 0 is $64 \times 64$, $128 \times 128$, $256 \times 256$, and $512 \times 512$, respectively. We iteratively apply the level projection to get the approximately divergence-free velocity $\boldsymbol{u}_{\text{appr}}$ on the single level. For the multi-level grid, we statically refine the patches in the upper-right part of the grid. As shown in Fig. 3.2, the grid patches in the rectangular region $(x, y) \in [3\pi/8, 5\pi/8]$ are refined to level 1, and the grid patches in the region $(x, y) \in [\pi/2, 3\pi/4]$ are further refined to level 2. We test four multi-level grids. For each grid, the finest resolutions on level 2 are the same as those on the single-level grid. The synchronization projection is applied to obtain $\boldsymbol{u}_{\text{appr}}$.

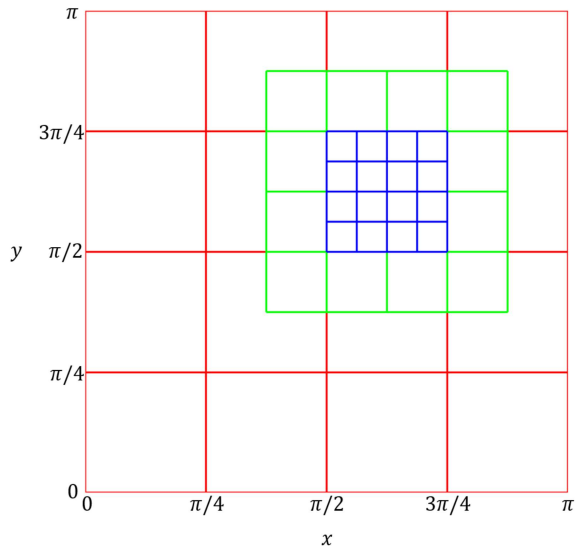

Figure 3.2: Patches on the block-structured adaptive grid for testing the stability of the projections. The red, green, and blue rectangles represent the grid patches on levels 0, 1, and 2, respectively. The equivalent grid resolutions are $64 \times 64$, $128 \times 128$, and $256 \times 256$ on the red, green, and blue patches, respectively.

Figs. 3.3 and 3.4 show the evolution of the $L^{\infty}$ norm $||\boldsymbol{u}_{\text{appr}}^{N} - \boldsymbol{u}_{\text{div}}||_{\infty}$ and the $L^2$ norm $||\boldsymbol{u}_{\text{appr}}^{N}||_{2}$ using the single-level projection and the multi-level synchronization projection, respectively. As shown in Fig. 3.3, the norm errors become almost unchanged after the first several iterations, which proves the stability of the level projection in this test. As the grid number increases, $||\boldsymbol{u}_{\text{appr}}^{N} - \boldsymbol{u}_{\text{div}}||_{\infty}$ decreases (Min and Gibou, 2006). The results of the multi-level synchronization projection in Fig. 3.4 lead to the same conclusion, which indicate that the projection schemes in this chapter maintain the desired stability with the mesh refinement, consistent with the literature (Martin and Colella, 2000; Martin et al., 2008; Brown et al., 2001).

### 3.3.2 Taylor Green Vortex

The Taylor Green Vortex (TGV) is a canonical problem to verify the order of convergence for new algorithms. The theoretical solution of the TGV problem is given

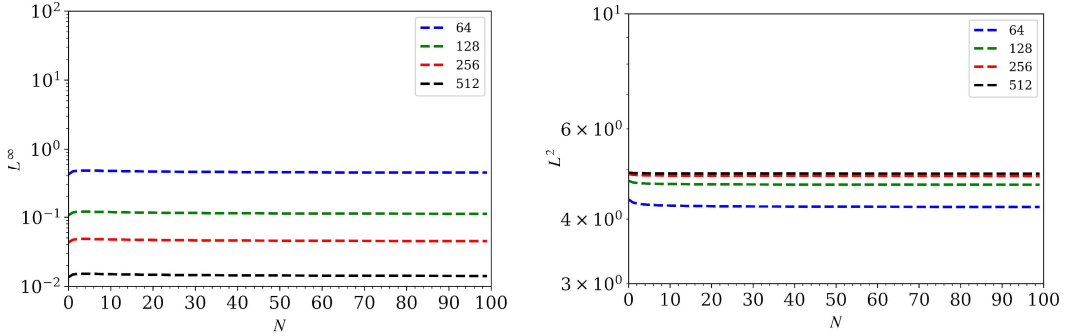

Figure 3.3: Errors of the single-level projection. The $N$ is the iteration number and the resolutions on level 0 are $64^2$, $128^2$, $256^2$, and $512^2$. Left: $||\boldsymbol{u}_{\text{appr}}^N - \boldsymbol{u}_{\text{div}}||_\infty$. Right: $||\boldsymbol{u}_{\text{appr}}^N||_2$.

by

$$u(x, y, t) = -\cos(\pi x)\sin(\pi y)e^{-2\pi^2 \mu t}, \tag{3.27}$$

$$v(x, y, t) = \sin(\pi x)\cos(\pi y)e^{-2\pi^2 \mu t}, \tag{3.28}$$

$$p(x, y, t) = -\frac{\cos(2\pi x) + \sin(2\pi y)}{4}e^{-4\pi^2 \mu t}, \tag{3.29}$$

where $\mu = 0.001$ is the dynamic viscosity.

Both the single-level and multi-level performances of our algorithms are examined here. For tests with a single level, a periodic computational domain with size $1 \times 1$ is employed for all five cases, where the grid number on level 0 is $16 \times 16$, $32 \times 32$, $64 \times 64$, $128 \times 128$, and $256 \times 256$, respectively. The CFL number is kept as a constant 0.5. Numerical results are compared with the theoretical results at $t = 1.0$ and the $L^2$ errors are calculated to obtain the point-wise convergence rate. Fig. 3.5 shows the order of convergence for the single level cases. It is shown that the algorithms achieve the second-order accuracy for the pressure $p$ and a higher-order accuracy (approximately the third order) for the velocity $\boldsymbol{u}$. The higher-order convergence rate for $\boldsymbol{u}$ is also observed in the literature (Guy and Fogelson, 2005; Almgren et al., 1998).

For cases with the multiple levels, the refinement criterion is based on the magnitude of the vorticity, i.e., the grid cells on the coarser levels are refined to the finer levels if $\omega_z > 0.95|\omega_z^{max}|$. The finest level in this problem $l_{max}$ is 2. However, we note that only the static mesh refinement is used here, i.e., the grid cells are only refined at the beginning and kept unchanged throughout the simulation. This is justified because the vortex cores in this TGV problem have no translational motions. For the case with the grid number $32 \times 32$ on level 0, Fig. 3.6 shows the generated grid and it is seen that the grid resolution is higher as the vortex core is approached. Both the subcycling and non-subcycling methods are tested. Similar to the single-level cases, we compare the composite solution with the theoretical results at $t = 1.0$ using the $L^2$ measure of the errors. As shown in Fig. 3.7, the $L^2$ errors of $\boldsymbol{u}$ and $p$ decrease as the grid number increases at a rate of second-order convergence. Moreover, the $L^2$ errors at a given grid spacing for both the subcycling method and the non-subcycling method are comparable, indicating that the two cycling methods produce consistent results. These tests show that our numerical schemes can achieve the desired second order of accuracy on a static multi-level mesh for both the subcycling and non-subcycling methods.

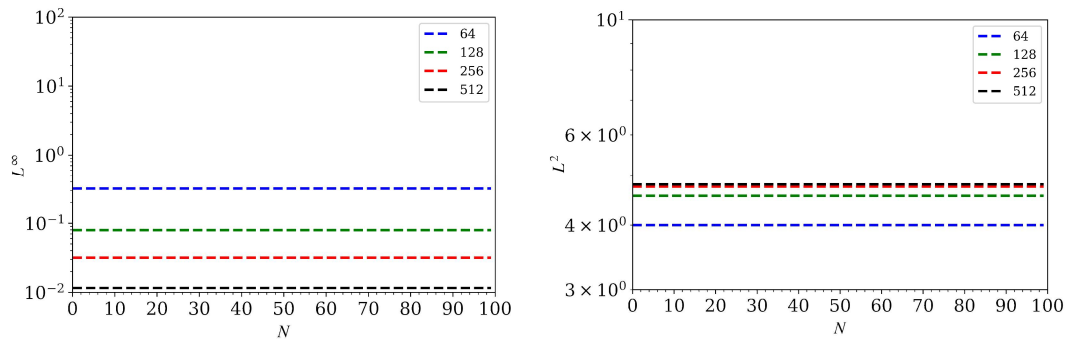

Figure 3.4: Errors of the multi-level synchronization projection. The $N$ is the iteration number and the resolutions on the finest level ($(x, y) \in [\pi/2, 3\pi/4]$) are the same as those of the single-level grid with $64^2$, $128^2$, $256^2$, and $512^2$. Left: $||\boldsymbol{u}_{\mathrm{appr}}^N - \boldsymbol{u}_{\mathrm{div}}||_\infty$. Right: $||\boldsymbol{u}_{\mathrm{appr}}^N||_2$.

### 3.3.3 Four-way vortex merging

The four-way vortex merging problem is used to validate the order of convergence of our algorithms for dynamically refined meshes (Almgren et al., 1998; Popinet, 2003). Here, four vortices are placed in a unit square domain and centered at $(0.5, 0.5)$, $(0.59, 0.5)$, $(0.455, 0.5 + 0.045\sqrt{3})$, and $(0.455, 0.5 - 0.045\sqrt{3})$, respectively. Their vortex strengths are $-150$, $50$, $50$, and $50$, respectively. For each vortex, the vorticity $\omega_z$ decays from the center $(x_i, y_i)$ as $(1 + \tanh(100(0.03 - r_i)))/2$, where $r_i = \sqrt{(x - x_i)^2 + (y - y_i)^2}$. To initialize the velocity field, the vorticity field $\omega_z$ is used as the source term of the Poisson equation for the stream function $\psi$

$$L^{cc,\text{comp}}\psi = \omega_z. \qquad (3.30)$$

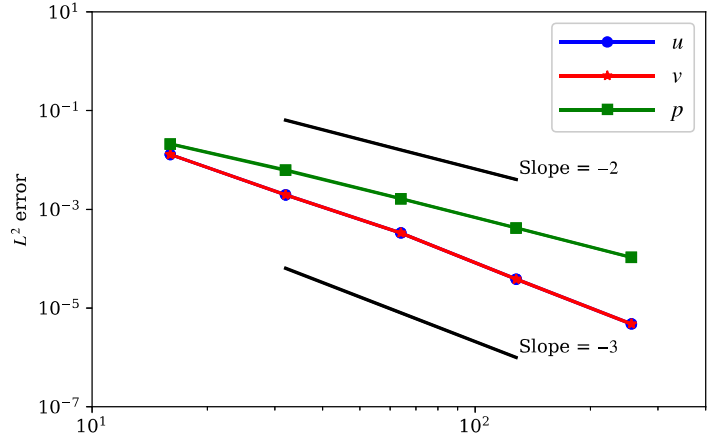

Figure 3.5: Grid convergence of $u$, $v$, and $p$ for the TGV problem on a single-level grid.

The initial velocity field is then calculated as $u(x,y) = \partial\psi/\partial y$ and $v(x,y) = -\partial\psi/\partial y$. The Reynolds number is set to $Re = 1000$. The vorticity criterion is used for the dynamic mesh refinement, i.e., the grid cells on the coarser levels are refined to the finer levels as long as $|\omega_z| > 0.05|\omega_z^{max}|$. The finest level $l_{max}$ is 2 in this problem. As shown in Fig. 3.8, the patches on the finer levels change dynamically to capture the merging vortices. The evolution of the vortices also agrees well with the results in the literature (Almgren et al., 1998; Popinet, 2003).

To obtain the point-wise convergence rate on the multi-level grid, we consider five cases here, of which the grid number on level 0 is $16 \times 16$, $32 \times 32$, $64 \times 64$, $128 \times 128$, and $256 \times 256$, respectively. The CFL number is kept as a constant 0.9. Because there is no exact solution for this problem, we use the result on a $1024 \times 1024$ uniform grid as the reference solution. Numerical results are compared with the reference solution at $t = 0.25$ and the $L^2$ errors are calculated. Fig. 3.9 shows the $L^2$ errors of $\boldsymbol{u}$ and $p$ as a function of the grid number. We can see that our numerical scheme maintains

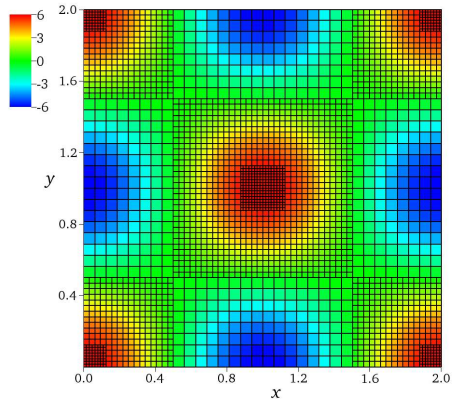

Figure 3.6: Grid hierarchy for the subcycling TGV case with a grid number of $32 \times 32$ on level 0. The contours of vorticity $\omega_z$ at $t = 0$ are also shown.

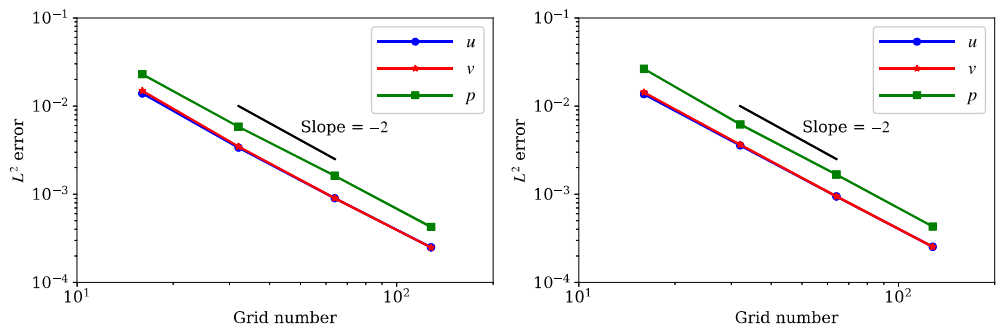

Figure 3.7: Grid convergence of $u$, $v$, and $p$ for the TGV problem on the multiple levels under the static mesh refinement. Left: the subcycling method. Right: the non-subcycling method.

the second-order accuracy in the context of dynamic mesh refinement with two cycling methods.

### 3.3.4 Inviscid shear layer

The inviscid shear layer problem is used to validate the proposed algorithms under the Euler limit within the BSAMR framework. Similar to the setup in Bell et al. (1989), the computational domain is $1 \times 1$ with periodic boundary conditions in both the horizontal and vertical directions. The properties of the inviscid fluid are $\rho(\phi) = \mu(\phi) = 1.0$. The

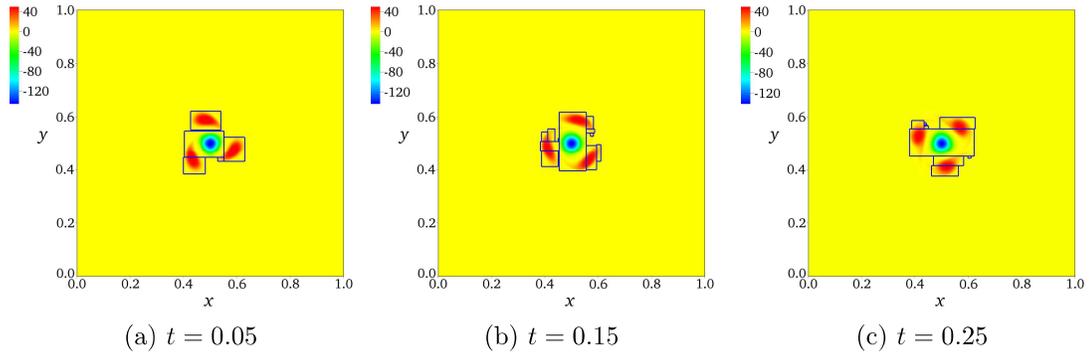

(a) $t = 0.05$      (b) $t = 0.15$      (c) $t = 0.25$

Figure 3.8: Evolution of vorticity $\omega_z$ and grid hierarchy for the non-subcycling four-way vortex merging case. The grid number on level 0 is $64 \times 64$ and the patches are dynamically refined to $l_{max} = 2$. The blue rectangles represent the patches on level 2.

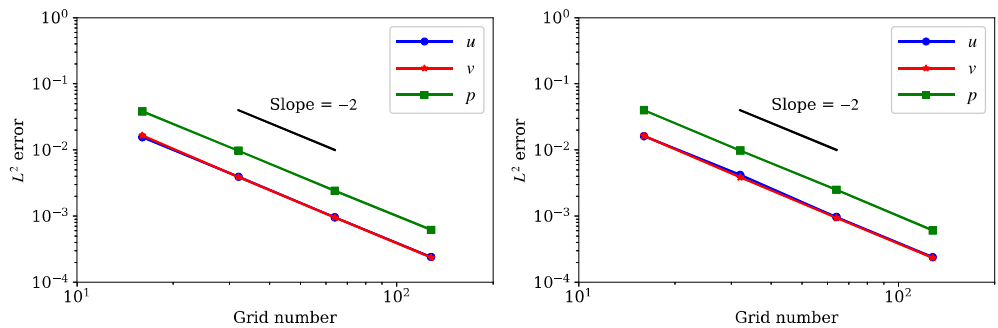

Figure 3.9: Grid convergence of $u$, $v$, and $p$ for the four-way merging vortex problem on the multiple levels with the dynamic mesh refinement. Left: the subcycling method. Right: the non-subcycling method.

Table 3.1: Parameters for cases of the inviscid shear layer problem.

| Case No. | Grid number on level 0 | $l_{max}$ | $\Delta t_0$ | Cycling method |
|:---:|:---:|:---:|:---:|:---:|
| 1 | $128 \times 128$ | 1 | 0.001 | Subcycling |
| 2 | $128 \times 128$ | 1 | 0.001 | Non-subcycling |
| 3 | $128 \times 128$ | 2 | 0.001 | Subcycling |
| 4 | $128 \times 128$ | 2 | 0.001 | Non-subcycling |

initial velocity is given by

$$u(x,y) = \begin{cases} \tanh(\sigma_1(y - 0.25)) & y \le 0.5 \\ \tanh(\sigma_1(0.75 - y)) & y > 0.5, \end{cases} \tag{3.31}$$

$$v(x,y) = \sigma_2 \sin(2\pi x), \tag{3.32}$$

where $\sigma_1 = 30$ and $\sigma_2 = 0.05$. For this problem, we consider four cases, the parameters of which are listed in Table 3.1. These cases have different refinement levels and the vorticity magnitude $|\omega_z| > 0.75|\omega_z^{max}|$ is used as the refinement criterion, same as the four-way merging vortex problem. The comparison between the subcycling and non-subcycling methods are also considered in these cases.

Fig. 3.10 plots the evolution of the vorticity $\omega_z$ and the patches for the three-level subcycling case (case 3). It shows that our simulation captures the very fine vortex structure and has a good agreement with Bell et al. (1989) and Huang et al. (2019).

At the Euler limit, the kinetic energy should remain constant. To validate the property of energy conservation, we evaluate the kinetic energy change $\Delta E(t) = E(t) - E(0)$, where the kinetic energy $E(t) = \int_\Omega [u^2(t) + v^2(t)]/2 dx$. The time series of the relative kinetic energy change, $\Delta E(t)/E(0)$, for each case are plotted in Fig. 3.11. Although $\Delta E(t)/E(0)$ shows small oscillations, which might be caused by the regridding and interpolation operations across multiple levels, the maximum relative kinetic energy variation of all these cases are within 0.12%, comparable to the 0.3% in Huang et al. (2019). Comparing cases 1 and 3 (or cases 2 and 4), one can see that the finer grid cells enabled by the additional level of mesh improve the conservation of the kinetic energy. In summary, our algorithm has a fairly good performance in conserving the kinetic energy for both the subcycling and non-subcycling methods.

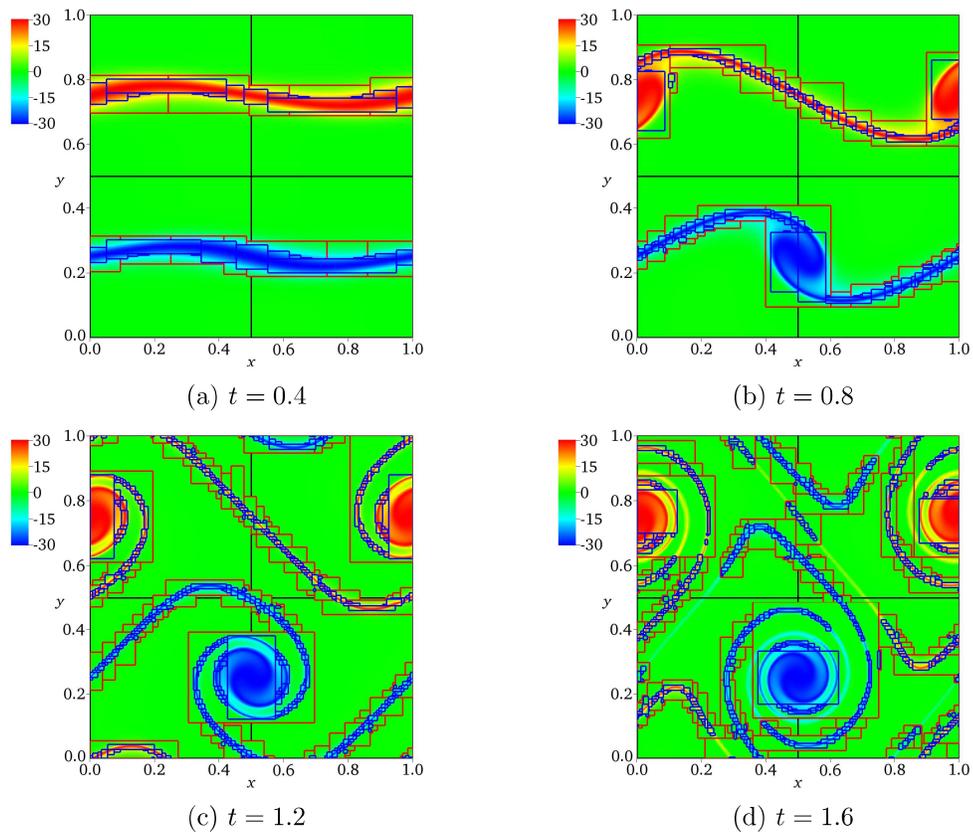

(a) $t = 0.4$

(b) $t = 0.8$

(c) $t = 1.2$

(d) $t = 1.6$

Figure 3.10: Evolution of the vorticity field and grid hierarchy of the inviscid shear layer problem for case 3 (three levels with subcycling). The black, red, and blue rectangles represent the patches on levels 0, 1, and 2, respectively.

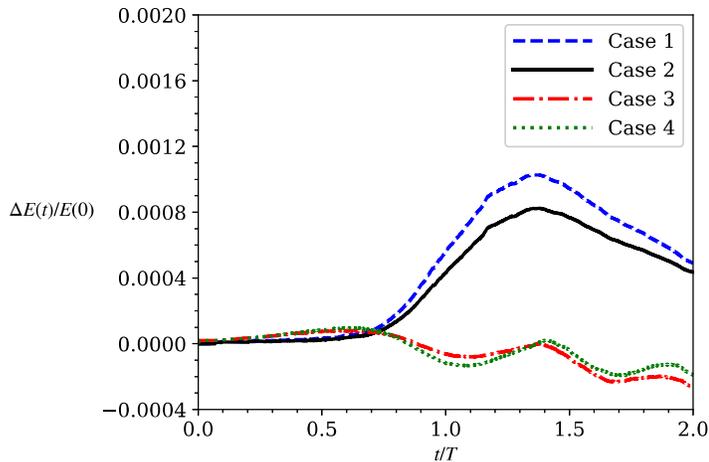

Figure 3.11: Relative kinetic energy error $\Delta E(t)/E(0)$ among the four multi-level cases in the inviscid shear layer problem. Case 1 and case 3 use the subcycling method while case 2 and case 4 use the non-subcycling method. For case 1 and case 2, $l_{max} = 1$. For case 3 and case 4, $l_{max} = 2$.

### 3.3.5 Zalesak's problem

The rotation of a notched disk, i.e., the Zalesak's problem (Zalesak, 1979), is used to validate the advection of the LS function, the single-level re-initialization algorithm, and the multi-level re-initialization algorithm. The computational domain is a $1 \times 1$ periodic rectangle. A notched disk with radius $r = 0.15$ is initially placed at $(0.5, 0.75)$ using the LS function $\phi$, as shown in Figure 3.12. The height and width of the notch are 0.25 and 0.05, respectively. The disk is transported by a prescribed steady velocity field given by

$$u(x,y) = 0.5 - y, \quad v(x,y) = x - 0.5. \tag{3.33}$$

The parameters of the simulation cases are given in Table 3.2. Cases 1 and 2 are single-level simulations and cases 3 and 4 are multi-level simulations. The refinement criterion is the distance to the interface, i.e., the grid cells $(i, j)$ on level $l$ $(0 \leq l < l_{max})$ are refined to the finer level if $|\phi_{i,j}| < 3.0 \max(\Delta x^l, \Delta y^l)$, where $\Delta x^l$ and $\Delta y^l$ are the grid spacings in the $x$ and $y$ directions, respectively. The finest level $l_{max}$ is 2 in this problem. Among the four cases considered here, the re-initialization is only performed in cases 2 and 4 to show the effect of the re-initialization operations.

Table 3.2: Parameters for cases of the Zalesak's problem.

| Case No. | Grid number on level 0 | $l_{max}$ | $\Delta t_0$ | Cycling method | With reinitialization? |
|----------|------------------------|-----------|--------------|----------------|------------------------|
| 1 | $192 \times 192$ | 0 | 0.002 | – | No |
| 2 | $192 \times 192$ | 0 | 0.002 | – | Yes |
| 3 | $48 \times 48$ | 2 | 0.0005 | Subcycling | No |
| 4 | $48 \times 48$ | 2 | 0.0005 | Subcycling | Yes |

The Zalesak disk rotates counterclockwise under the prescribed velocity. Ideally, the shape of the disk should stay the same and return to its initial state after one revolution. Fig. 3.12 shows the shapes of the Zalesak disk, denoted by $\phi = 0$, at the initial moment and after one revolution. For case 1 and case 3, the notched disk deviates from its original shape noticeably. For case 2 and case 4 with the re-initialization process, the disk shape is preserved much better. To quantify the errors in $\phi$, we calculate the relative errors of the disk area, $\delta(t)$, as (Sussman and Puckett, 2000)

$$\delta(t) = \frac{1}{L} \int_\Omega |H(\phi(t)) - H(\phi_e(t))| dx, \qquad (3.34)$$

where $\phi_e(t)$ is the exact LS function and $L$ is the perimeter of the interface. Table 3.3 shows the relative error after one revolution, i.e. $t = 2\pi$, for different cases. The errors in case 2 and case 4 are two orders of magnitude smaller than case 1 and case 3, which shows the efficacy of the single-level and multi-level re-initialization algorithms. Specifically, the errors of case 2 and case 4 have the same order of magnitude as those in (Sussman and Puckett, 2000), in which the coupled level set and volume-of-fluid (CLSVOF) method is used. We also remark that the refluxing issue is not a concern in this problem because the velocity is prescribed, and the non-subcycling results (not shown here) have negligible differences from the subcycling results. Therefore, we conclude that the single-level and the multi-level re-initialization algorithms in this chapter can maintain the LS function as the signed distance function and keep the mass conserved. Our advection schemes for the LS function are also validated with both the subcycling and non-subcycling methods.

Table 3.3: Comparison of the relative difference of the disk area at $t = 2\pi$ after one revolution among the four cases for the Zalesak's problem. Case 1 and case 2 use the single-level grid. Case 3 and case 4 use the multi-level grid with $l_{max} = 2$. Case 2 and 4 have the re-initialization while case 1 and 3 do not.

|            | Case 1 | Case 2  | Case 3 | Case 4  |
|------------|--------|---------|--------|---------|
| $\delta(2\pi)$ | 0.042  | 0.00078 | 0.056  | 0.00081 |

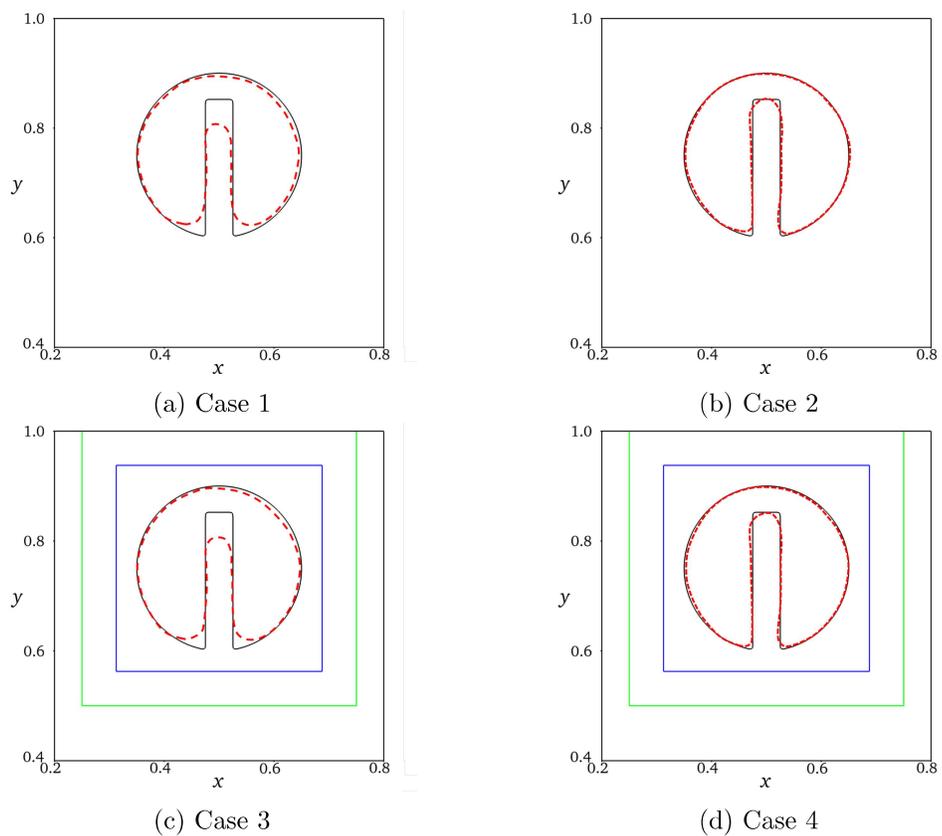

(a) Case 1       (b) Case 2

(c) Case 3       (d) Case 4

Figure 3.12: Comparison of the shapes of the initial Zalesak disk (solid black line) and the disk after one revolution (red dashed line). Case 1 and case 2 use a single-level grid while case 3 and case 4 use the multi-level grid with $l_{max} = 2$. Case 2 and case 4 employ the single-level re-initialization and the multi-level re-initialization, respectively. Case 1 and case 3 do not consider the re-initialization. Only part of the computational domain ($[0.2, 0.8] \times [0.4, 1.0]$) is displayed with LS function for better visualization. The green and blue lines represent patches on level 1 and level 2, respectively.

### 3.3.6 Gravity wave

To confirm that the LS advection scheme works well when coupled with the momentum equations for two-phase flows, a canonical decaying gravity wave case is tested here. The surface profile of a linear deep-water wave is initialized as

$$\eta(x, y) = a_0 \cos(kx - \omega t). \tag{3.35}$$

The velocities are

$$u(x, y) = a_0 \omega e^{ky} \cos(kx - \omega t), \quad v(x, y) = a_0 \omega e^{ky} \sin(kx - \omega t). \tag{3.36}$$

Here, $a_0$ is the initial wave amplitude, $k$ is the wave number, and $\omega = \sqrt{gk}$ is the angular frequency according to the dispersion relationship. In our tests, the wave steepness $a_0 k$ is set to 0.1 such that the linear wave theory is still valid. According to Lamb (Lamb, 1993), the wave decays with time because of viscous dissipation. Its amplitude evolution is

$$a(t) = a_0 e^{-2\nu k^2 t}. \tag{3.37}$$

The Reynolds number is set to $Re = \omega/\nu k^2 = 110$, where $\nu$ is the kinematic viscosity of the water. Other dimensionless number are Froude number $Fr = (\omega k^{-1})/\sqrt{gk^{-1}} = 1.0$, Weber number $We = \rho_w k^{-2} g/\sigma = \infty$, density ratio $\lambda = \rho_2/\rho_1 = 0.0011$, and dynamic viscosity ratio $\eta = \mu_2/\mu_1 = 0.0085$. The computational domain size is $2\pi \times 2\pi$ and the mean water depth is $\pi$. The free slip boundary condition is imposed at the bottom and top of the domain, and the periodic boundary condition is applied at the left and right boundaries. The parameters of the four cases considered in this problem are given in Table 3.4. Cases 1 and 2 are single-level cases with the latter having a higher resolution. Cases 3 and 4 use adaptive grids and their grid spacings on the finest level are the same as case 2. The grid is refined based on the distance to the air–water interface as in the Zalesak's problem (section 3.3.5), therefore the grid resolution near the wave surface in cases 2–4 is the same.

The amplitude evolution of the above four cases is plotted in Fig. 3.13. Results show that case 1, which has a relatively coarse grid resolution, has small deviation from the theoretical result. Meanwhile, the amplitude evolution in the single-level high-resolution case (case 2), the three-level subcycling case (case 3), and the two-level subcycling case (case 4) agree well with the linear wave theory. This result shows that, as the grid resolution increases near the water surface, the simulation results converge to the theoretical solution. Comparing cases 3 and 4 with case 2, we note that the locally refined mesh can yield the same result as the uniform single-level fine mesh. We remark that the small oscillations in the numerical results could be due to using a potential flow solution as the initial condition of a viscous incompressible two-phase flow solver (Lamb, 1993; Xie et al., 2016). In summary, we conclude that our algorithms satisfy the grid convergence and can accurately simulate the two-phase gravity wave flow when multiple levels are considered.

Table 3.4: Parameters for cases of the gravity wave problem.

| Case No. | Grid number on level 0 | $l_{max}$ | $\Delta t_0$ | Cycling method |
|:---:|:---:|:---:|:---:|:---:|
| 1 | $128 \times 128$ | 0 | 0.0005 | – |
| 2 | $256 \times 256$ | 0 | 0.00025 | – |
| 3 | $64 \times 64$ | 2 | 0.001 | Subcycling |
| 4 | $128 \times 128$ | 1 | 0.0005 | Subcycling |

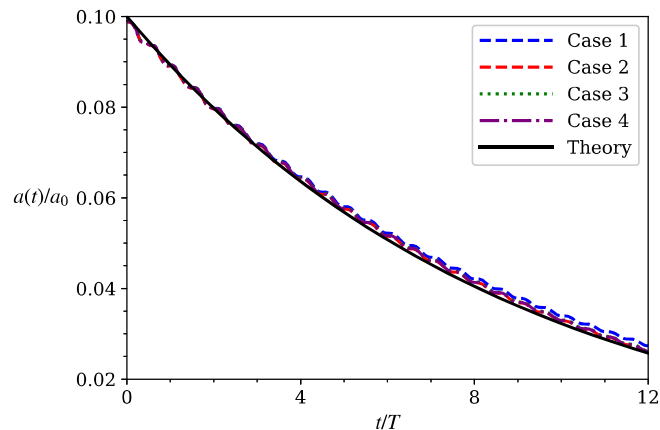

Figure 3.13: Comparison of the evolution of the wave amplitude among the single-level cases (case 1 and case 2), the subcycling three-level case (case 3), the subcycling two-level case (case 4), and the theoretical result for the decaying gravity wave problem. Case 1 has the coarsest resolution while case 2, case 3, and case 4 have the same finest resolution.

Table 3.5: Parameters for cases of the rising bubble problem.

| Case No. | Grid number on level 0 | $l_{max}$ | $\Delta t_0$ | Cycling method |
|----------|------------------------|-----------|--------------|----------------|
| 1 | $512 \times 1024$ | 0 | 0.00025 | – |
| 2 | $128 \times 256$ | 2 | 0.001 | Subcycling |
| 3 | $128 \times 256$ | 2 | 0.00025 | Non-subcycling |

Table 3.6: Comparison of the steady rising velocity ($V_t$) among the cases for the rising bubble problem. Case 1: single-level case; case 2: three-level subcycling case; case 3: three-level non-subcycling case.

|       | Case 1 | Case 2 | Case 3 | Theory |
|-------|--------|--------|--------|--------|
| $V_t$ | 0.971  | 0.966  | 0.965  | 1.0    |

### 3.3.7   Rising bubble

Next, a spherical-cap bubble rising in a liquid is simulated to validate our algorithms for a two-phase flow problem with surface tension. Compared with the gravity wave case, where patches on the finer levels change slowly because of the slow decay of the wave, the grid cells here are refined more dynamically to capture the rising bubble. A large computational domain $[-12, 12] \times [-18, 30]$ is chosen to circumvent wall effects. Free-slip conditions are applied at all boundaries. A spherical bubble of dimensionless radius one is put at $(x, y) = (0, 3)$ surrounded by the stationary fluid as the initial condition. Based on the steady rise velocity $V = 0.215\,\mathrm{m/s}$ and the bubble radius $r = 0.0061\,\mathrm{m}$ in Sussman et al. (1999) and Hnat and Buckmaster (1976), the dimensionless parameters in our simulation are set as $Re = \rho_1 V r / \mu_1 = 9.8$, $Fr = V/\sqrt{gr} = 0.872$, $We = \rho_1 V^2 r / \sigma = 7.6$, $\lambda = \rho_2/\rho_1 = 0.0011$, and $\eta = \mu_2/\mu_1 = 0.0085$. Three cases are simulated, of which the parameters are listed in Table 3.5. For the subcycling case 2 and the non-subcycling case 3, the grid number on level 0 is $128 \times 256$ and then refined to the $l_{max} = 2$, i.e., three levels in total, such that the resolution on the finest level is equivalent to case 1 with grid number $512 \times 1024$. The refinement criterion is based on the distance to the air–water interface, same as the gravity wave problem (section 3.3.6).

Fig. 3.14 shows the movement of the mass centroid of the bubble for different cases. In all of the four cases considered here, the bubble quickly reaches a steady rising velocity. Theoretically, the steady dimensionless rising velocity $V_t$ in this problem is 1.0 (Sussman et al., 1999; Hnat and Buckmaster, 1976). Table 3.6 compares it with the values from the simulations. The relative errors of the four cases are within 4%. Fig. 3.15 shows the time evolution of the bubble shape. Under the combined effects of buoyancy force, viscosity, and surface tension, the bottom of the bubble moves faster than its top at the initial stage, compressing the bubble in the vertical direction and flattening it in the horizontal direction. As shown in Fig. 3.15, at the later stage $t = 6.0$–$8.0$, the bubble rises with a constant speed and its shape remains nearly unchanged. The numerically computed bubble shapes in Fig. 3.15 are in agreement with the experiment of Hnat and Buckmaster (1976) and the simulation of Ryskin and Leal (1984) (results not plotted here). This test proves that our algorithms can correctly capture the dynamics of the two-phase flow when the surface tension effect is involved. The results also indicate that the AMR technique can reproduce the results accurately while using the fewer grid number compared with the single-level fine-grid simulation. Furthermore, the nearly identical results between the subcycling and non-subcycling methods, as shown in Fig. 3.15, validate the consistency of these two cycling methods in our unified BSAMR framework.

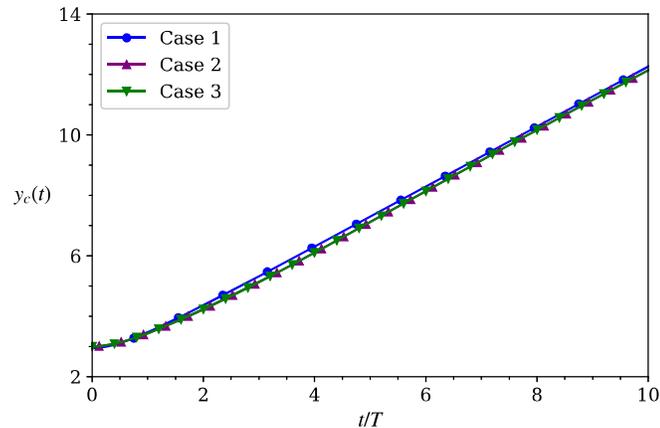

Figure 3.14: Comparison of the time series of the bubble centroid ($y_c(t)$) between the single-level case (case 1), the three-level subcycling case (case 2), and the three-level non-subcycling case (case 3).

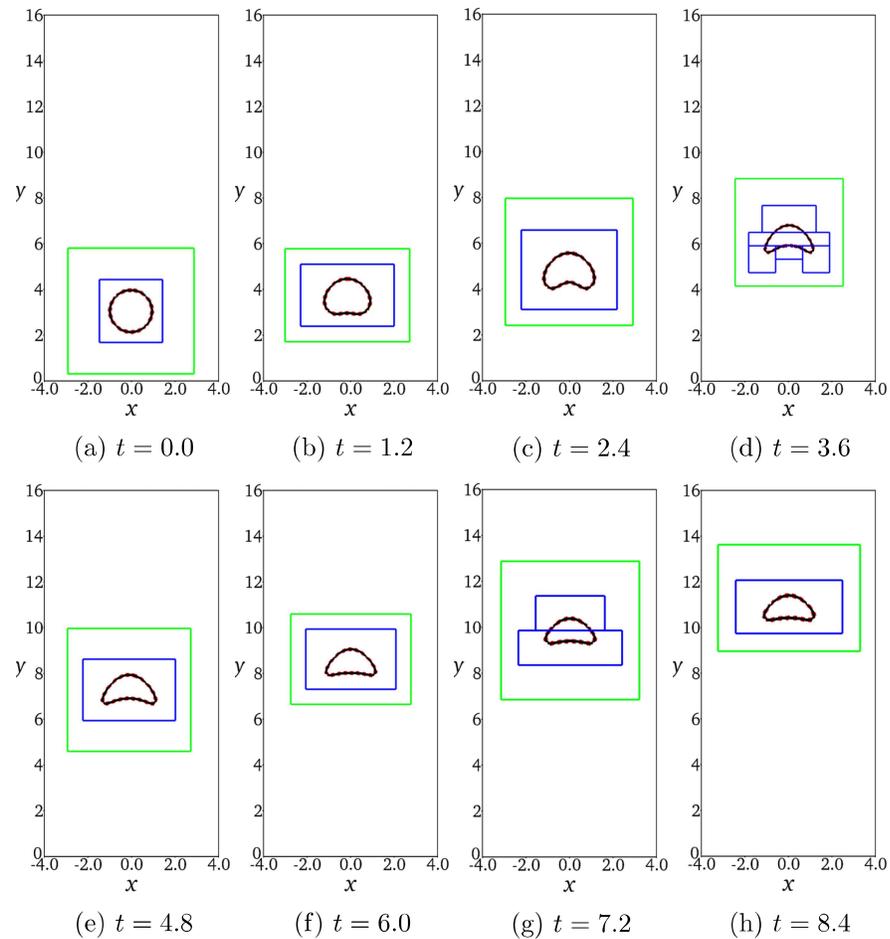

(a) $t = 0.0$     (b) $t = 1.2$     (c) $t = 2.4$     (d) $t = 3.6$

(e) $t = 4.8$     (f) $t = 6.0$     (g) $t = 7.2$     (h) $t = 8.4$

Figure 3.15: Evolution of the shape of the rising bubble for the three-level subcycling case (case 2) and three-level non-subcycling case (case 3). The bubble shape is shown by the isoline of $\phi = 0$. The solid black line and dashed red line correspond to case 2 and case 3, respectively. The green and blue lines represent the patches on level 1 and level 2, respectively.

Table 3.7: Parameters for cases of the Rayleigh–Taylor instability problem.

| Case No. | Grid number on level 0 | $l_{max}$ | $\Delta t_0$ | Cycling method |
|----------|------------------------|-----------|--------------|----------------|
| 1 | $200 \times 800$ | 0 | 0.0005 | – |
| 2 | $100 \times 400$ | 1 | 0.001 | Subcycling |
| 3 | $100 \times 400$ | 1 | 0.0005 | Non-subcycling |
| 4 | $50 \times 200$ | 2 | 0.002 | Subcycling |
| 5 | $50 \times 200$ | 2 | 0.0005 | Non-subcycling |

### 3.3.8 Rayleigh–Taylor instability

The Rayleigh–Taylor (RT) instability problem is simulated here to validate the adaptive two-phase flow algorithms when small vorticity structures are involved. This instability phenomenon occurs for any perturbation to the interface between a lighter fluid ($\rho_2$) at the bottom and a heavier fluid ($\rho_1$) at the top. In the simulation, we follow the same setup as Guermond and Quartapelle (2000). The computational domain is $[0, 1] \times [0, 4]$. The initial interface is given by $y(x) = 2.0 + 0.1 \cos(2\pi x)$. The density ratio is set to $\lambda = \rho_2/\rho_1 = 1/3$ and the Reynolds number is set to be $Re = \rho_1 g^{1/2}/\mu_1 = 3000$. Five cases with different parameters are presented in Table 3.7. The single-level case (case 1) has the same grid number and time step $\Delta t_0$ as in Guermond and Quartapelle (2000) and Ding et al. (2007). For the multi-level cases, the refinement criterion is based on the distance to the air–water interface, same as the gravity problem. From case 2 to case 5, we keep the same resolution on the finest level as case 1 while varying the grid number on level 0. The time step is changed accordingly for the subcycling method or the non-subcycling method.

The evolution of the air–water interface for the three-level non-subcycling case (case 5) is shown in Fig. 3.16. Refined patches on the finer levels are also presented to show the change of the adaptive meshes. We observe a good agreement when comparing the shape of the interface with Ding et al. (2007) (not plotted here). A small perturbation of the interface appears at $t = 0$ and begins to grow due to the gravity effects. The interface then rolls up into the lighter fluid, and a long tail is then formed from $t = 1$ to $t = 1.75$. It is seen that the curling tip of the interface, as well as the secondary vortices of the roll-ups, are fully resolved by the adaptively generated mesh. The left side of Fig. 3.17 compares the transient locations of the falling fluid $y_f(t)$ and rising fluid $y_r(t)$ between the single-level case (case 1) and previous research. A good agreement is obtained. The multi-level cases (cases 2-5) agree well with the single-level fine-grid case (case 1), as shown on the right side of Fig. 3.17. By comparing the transient locations between case 2 and case 3 and between case 4 and case 5, we conclude that the subcycling and non-subcycling methods are consistent with each other, same as the rising bubble problem (section 3.3.7).

The above results show that our algorithms can accurately capture the transient locations of the fluid appearing in the Rayleigh–Taylor instability problem, both for the subcycling method and the non-subcycling method. The refined patches can help increase the resolution of small flow structures. This test further validates the capability of our proposed framework for simulating incompressible two-phase flow problems.

### 3.3.9    3D dam breaking

This section investigates the 3D dam breaking, a dynamic and complex problem which is traditionally computationally expensive. Besides validating the adaptive two-phase flow algorithms for 3D problems, another objective is to compare the computational cost among the single-level, the subcycling, and the non-subcycling cases.

For this problem, a cubic water block with the side length $a = 0.057, 15\,\mathrm{m}$ is put at the left-bottom corner. The computational domain size is $[7a, a, 1.75a]$, as shown in Fig. 3.18. No-slip boundary conditions are imposed on the bottom wall, while all other walls are free-slip boundaries. The $d_f$ and $d_h$ in Fig. 3.18 refer to the dimensional front (point A) and the dimensional height (point B), respectively. The dimensionless front and height are then defined as $\tilde{d}_f = d_f/a$ and $\tilde{d}_h = d_h/a$, respectively. The gravity is in the $-y$ direction. Dimensionless parameters are set as $Re = \rho_1 U a/\mu_1 = 2950$, $Fr = U/\sqrt{ga} = 1.0$, $We = \rho_1 U^2 a/\sigma = 0.54$, $\lambda = \rho_2/\rho_1 = 0.0012$, and $\eta = \mu_2/\mu_1 = 0.016$. Here, $U$ is the characteristic length. Table 3.8 gives the parameters of five simulation cases, where case 1 is the single-level case and all other cases are the multi-level cases. From case 2 to case 5, we refine the grid to $l_{max} = 2$ or 3 using either the subcycling or the non-subcycling method. The refinement criterion is based on the distance to the air–water interface, same as the Rayleigh–Taylor instability problem (section 3.3.8).

Fig. 3.19 compares the dimensionless front $\tilde{d}_f$ and dimensionless height $\tilde{d}_h$ of the single-level case (case 1) with previous experimental results and numerical results. Our

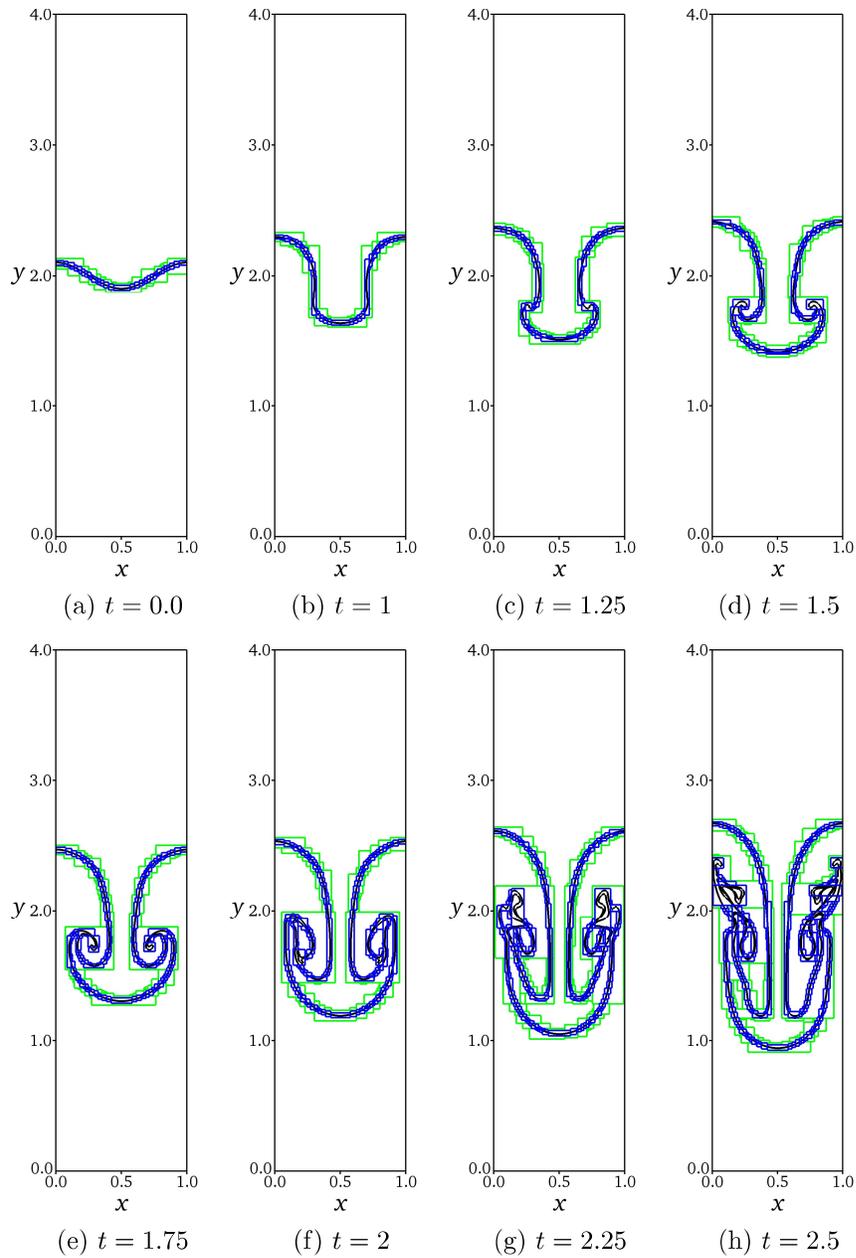

Figure 3.16: Evolution of the air–water interface for the three-level non-subcycling Rayleigh–Taylor instability case (case 5 in Table 3.7). The black line represents the isoline of $\phi = 0$. The green and blue lines represent the patches on level 1 and level 2, respectively.

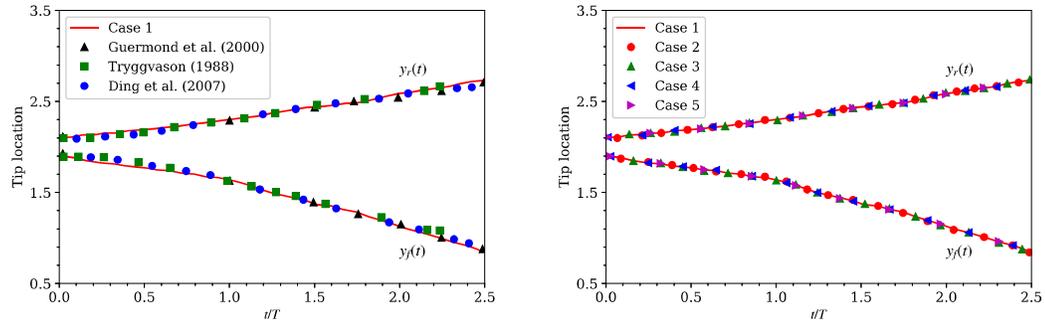

Figure 3.17: Left: comparison of the tip locations of the falling fluid ($y_f(t)$) and the rising fluid ($y_r(t)$) between the single-level case (case 1) and the literature: (▲) **?**; (■) Tryggvason (1988); (●) Ding et al. (2007). Right: comparison of the tip locations of the falling fluid $y_f(t)$ and the rising fluid $y_r(t)$ between the single-level case (case 1), the two-level subcycling case (case 2), the two-level non-subcycling case (case 3), the three-level subcycling case (case 4), and the three-level non-subcycling case (case 5).

.

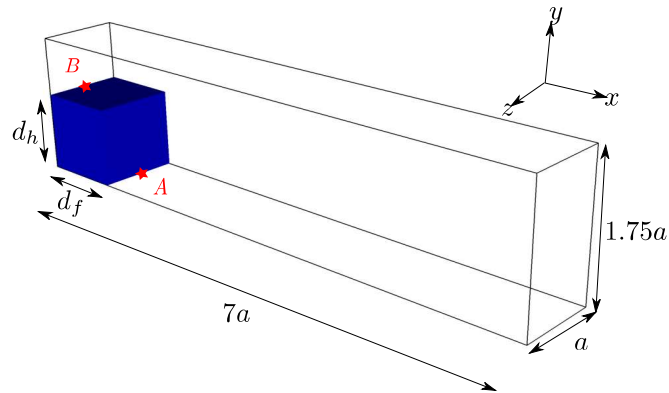

Figure 3.18: Sketch of 3D dam breaking problem. A: dam front position; B: dam height position.

Table 3.8: Parameters for cases of the dam breaking problem.

| Case No. | Grid number on level 0 | $l_{max}$ | $\Delta t_0$ | Cycling method |
|----------|------------------------|-----------|--------------|----------------|
| 1 | $512 \times 128 \times 96$ | 0 | $1.25 \times 10^{-4}$ | – |
| 2 | $256 \times 64 \times 48$ | 1 | $2.50 \times 10^{-4}$ | Subcycling |
| 3 | $256 \times 64 \times 48$ | 1 | $1.25 \times 10^{-4}$ | Non-subcycling |
| 4 | $128 \times 32 \times 24$ | 2 | $5.00 \times 10^{-4}$ | Subcycling |
| 5 | $128 \times 32 \times 24$ | 2 | $1.25 \times 10^{-4}$ | Non-subcycling |

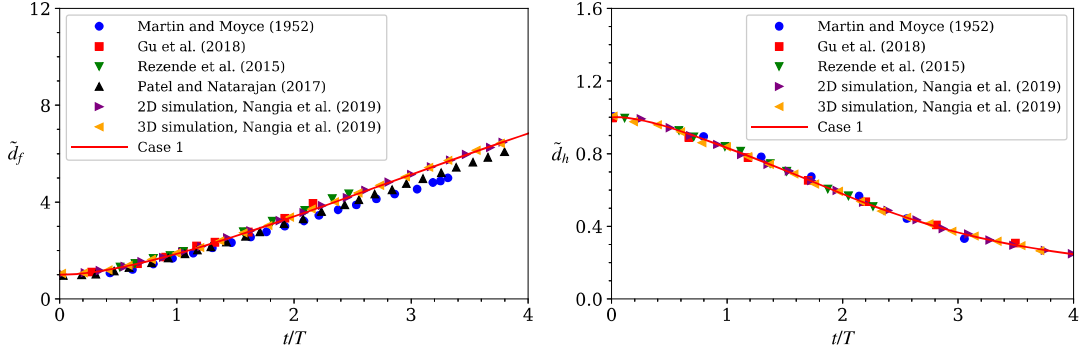

Figure 3.19: Comparison of the evolution of the dimensionless front $\tilde{d}_f$ (left) and the dimensionless height $\tilde{d}_h$ (right) between the single-level case (case 1) and the literature: (●) Martin et al. (1952); (■) Gu et al. (2018); (▼) Rezende et al. (2015); (▲) Patel and Natarajan (2017); (▶) Nangia et al. (2019a); (◀) Nangia et al. (2019a).

.

results agree well with the literature. Fig. 3.20 shows that the time evolution of $\tilde{d}_f$ and $\tilde{d}_h$ for the above five cases, which again indicates the consistency of our numerical algorithms, where the subcycling cases, the non-subcycling cases, and the single-level case can obtain nearly the same accuracy. The evolution of the breaking dam for the three-level subcycling case (case 4) is depicted in Fig. 4.9. Patches are dynamically refined around the interface as time evolves. The shape of the dam is consistent with the results in (Nangia et al., 2019a).

To compare the computational cost of different cases, we profile each case for $t/T = 0 - 0.3$ using 64 CPU cores on the Cray XE6m HPC machine without considering the I/O cost. Table 3.9 shows the total number of grid cells for different cases at $t/T = 0.1$. Compared with the cases with $l_{max} = 2$, the single-level case and the cases with $l_{max} = 1$ have nearly 6.32 times and 1.45 times more cells, respectively, which indicates that the adaptive refinement could reduce the total number of grid cells considerably. We emphasize that the time spent for each case also depends on the time step and the subcycling cases (case 2 and case 4) have less time steps than the non-subcycling cases (case 3 and case 5).

Table 3.10 compares the total wall clock time among different cases. Compared with the single-level case (case 1), it is seen that the two-level subcycling case (case 2) can obtain a 4.8 times speedup and thus save the computational cost. The three-level

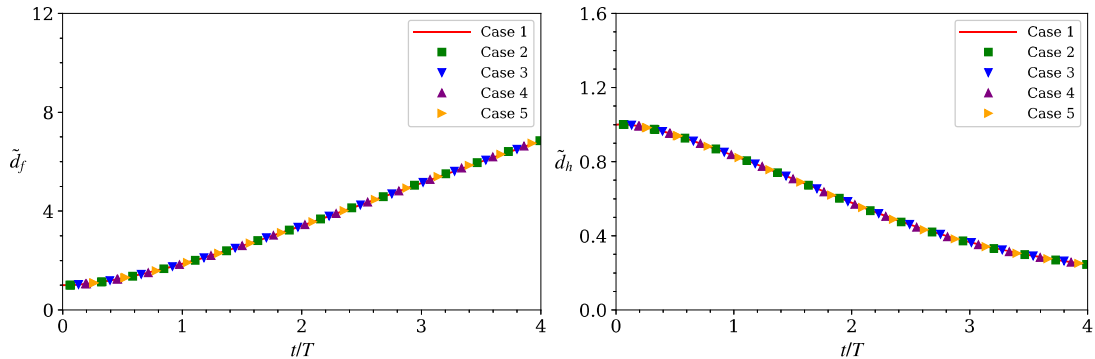

Figure 3.20: Comparison of the evolution of the dimensionless front (left) $\tilde{d}_f$ and the dimensionless height $\tilde{d}_h$ (right) among the cases listed in Table 3.8.

.

Table 3.9: Number of grid cells of the cases in the dam breaking problem at $t/T = 0.1$.

| Case No. | level 0 | level 1 | level 2 | Total cells | Total cells normalized by case 4 |
|---|---|---|---|---|---|
| 1 | 6,291,456 | – | – | 6,291,456 | 6.32 |
| 2 | 786,432 | 657,408 | – | 1,443,840 | 1.45 |
| 3 | 786,432 | 657,408 | – | 1,443,840 | 1.45 |
| 4 | 98,304 | 239,616 | 657,408 | 995,328 | 1 |
| 5 | 98,304 | 239,616 | 657,408 | 995,328 | 1 |

Table 3.10: Comparison of the total wall clock time $T_{total}$ and the speedup among the single-level case (case 1), the two-level subcycling case (case 2), the two-level non-subcycling case (case 3), the three-level subcycling case (case 4), and the three-level non-subcycling case (case 5) for the dam breaking problem.

| | Case 1 | Case 2 | Case 3 | Case 4 | Case 5 |
|---|---|---|---|---|---|
| Total wall time (hrs) | 1.67 | 0.35 | 0.50 | 0.26 | 0.38 |
| Speedup compared with case 1 | 1.0 | 4.8 | 3.3 | 6.4 | 4.4 |

subcycling case (case 4) achieves more speedup (6.4 times) compared with the single-level case. In addition, when comparing the wall clock time of the subcycling cases (case 2 and case 4) with the corresponding non-subcycling cases (case 3 and case 5), we find that subcycling cases lead to a greater reduction in computational cost. The reason is that, compared to the non-subcycling method, the subcycling method can use larger time steps for the coarser levels.

Besides the total wall clock time, the percentages of the time spent on some key parts of the algorithms are also documented, which can help us to identify the most time-consuming parts for optimizing the algorithms in future research. As shown in Fig. 3.22, they include the MAC projection, viscous solver, level projection, and synchronization. Among them, the level projection takes the most time ($> 35\%$), followed by the MAC projection step ($\approx 30\%$). Therefore, optimization of the two projection algorithms is desired. At last, the part denoted as the "Others", including the regridding, the interpolation operations, and the multi-level re-initialization steps, only account for about 5% of the total computation time. This result shows that our multi-level re-initialization algorithm is an economical way to regularize the LS function on the multi-level grid.

## 3.4   Concluding Remarks

In this chapter, we have developed a collocated BSAMR framework for the simulations of incompressible two-phase flows using the inconsistent scheme. The proposed multi-level advancement algorithm based on the level-by-level advancement method uses variables in both the valid and invalid regions and decouples the time advancement for different levels. Because of this decoupling, the time step constraint on the coarser levels is relaxed compared with that on the finest level if the subcycling method is used. On the other hand, the non-subcyling method avoids the time interpolation process across the different levels because data on all levels are located at the same time instant during the simulation.

Compared with the staggered grid and the semi-staggered grid, the collocated grid used here have several benefits. For example, the Godunov scheme, which is robust for flows with a wide range of Reynolds number, can be implemented in a straightforward way when the collocated grid is used, as done in the present chapter. In addition, one set of interpolation schemes and average operations is used for all variables in the context of the collocated grid. Moreover, only the cell-centered MG solver is needed for the velocity and pressure fields.

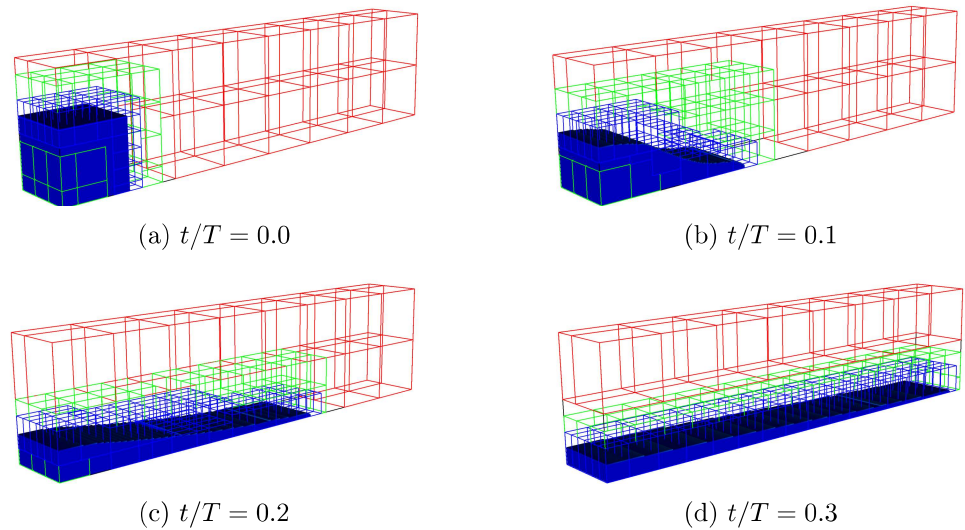

(a) $t/T = 0.0$

(b) $t/T = 0.1$

(c) $t/T = 0.2$

(d) $t/T = 0.3$

Figure 3.21: Profiles of the breaking dam for the three-level subcycling case (case 4) at different time instants. The red, green, and blue lines represent patches on levels 0, 1, 2, respectively.

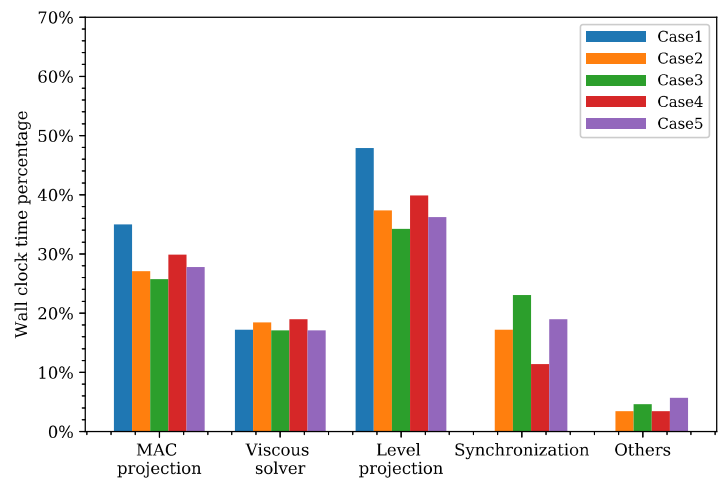

Figure 3.22: Percentages of the wall clock time for different parts in the BSAMR algorithms. The cases considered include the single-level case (case 1), the two-level subcycling case (case 2), the two-level non-subcycling case (case 3), the three-level subcycling case (case 4), and the three-level non-subcycling case (case 5).

We have also developed a novel re-initialization algorithm for the LS function on the multi-level grid to improve the accuracy of the two-phase interface capturing in the BSAMR framework. It employs an iteration technique to synchronize the LS function on two consecutive levels pair by pair. This algorithm leads to a substantial improvement in the mass conservation of the two-phase flow.

The accuracy and robustness of the computational framework are validated with several canonical tests. The results have shown that our numerical schemes obtain the second-order accuracy as designed and conserve the mass, momentum, and energy well. The subcycling and non-subcycling methods produce consistent and accurate results. We have also shown that the multi-level cases can achieve the same level of accuracy with fewer grid cells than the single-level fine-grid cases. In particular, for the 3D dam breaking problem, the multi-level simulation is able to capture the evolution of the dam accurately with substantial speedup compared with the single-level simulation. The synchronization and multi-level re-initialization algorithms developed in this chapter are also shown to be computationally efficient.

---

**Algorithm 3** Multi-level advancement algorithm

---

1: Initialize $\boldsymbol{u}^0$, $\phi^0$, and $p^0$ on level 0
2: $l \leftarrow 0$
3: **while** refinement criteria are satisfied on level $l$ and $l < l_{max}$ **do**
4:     Regrid the patch hierarchy to obtain level $l + 1$
5:     Initialize $\boldsymbol{u}^0$, $\phi^0$, and $p^0$ on level $l + 1$
6:     $l \leftarrow l + 1$
7: **end while**
8: **if** subcycling method is used **then**
9:     $\Delta t^l = 2^{l_{max}-l}\Delta t^{l_{max}}$ for all $0 \le l < l_{max}$
10: **else**
11:     $\Delta t^l = \Delta t^{l_{max}}$ for all $0 \le l < l_{max}$
12: **end if**
13: **for** $n = 1, n_{max}$ **do**         ▷ $n_{max}$ is the number of time steps to be simulated
14:     LEVELCYCLING($0$, $t_n^0$, $t_n^0 + \Delta t^0$, $\Delta t^0$)
15:     Apply the synchronization projection
16:     Perform the multi-level re-initialization of $\phi$     ▷ **Algorithm 2**
17:     Regrid the patch hierarchy and interpolate $\boldsymbol{u}$, $\phi$, and $p$ onto new patches
18: **end for**
19:
20: **procedure** LEVELCYCLING($l$, $t^l$, $t_{max}^l$, $\Delta t^l$)
21:     **while** $t^l < t_{max}^l$ **do**
22:         Perform single-level advancement on level $l$ from $t^l$ to $t^l + \Delta t^l$.     ▷
    **Algorithm 1**
23:         **if** $l < l_{max}$ **then**
24:             LEVELCYCLING($l + 1$, $t^l$, $t^l + \Delta t^l$, $\Delta t^{l+1}$)
25:         **end if**
26:         $t^l \leftarrow t^l + \Delta t^l$
27:     **end while**
28:     **if** $l > 0$ **then**
29:         Average all data from finer levels to the coarser levels
30:     **end if**
31:     **if** $l < l_{max}$ **then**
32:         Perform MAC synchronization and refluxing using Eqs. (2.13–2.22)
33:     **end if**
34: **end procedure**

---

# Chapter 4

# Consistent scheme for two-phase flow

## 4.1 Mathematical formulation

For the consistent scheme, we begin with the conservative form of the incompressible Navier–Stokes equations,

$$\frac{\partial \left( \rho \left( \mathbf{x}, t \right) \mathbf{u} \left( \mathbf{x}, t \right) \right)}{\partial t} + \nabla \cdot \left( \rho(\mathbf{x}, t) \mathbf{u}(\mathbf{x}, t) \mathbf{u}(\mathbf{x}, t) \right) = - \nabla p(\mathbf{x}, t) + \nabla \cdot \left[ \mu(\mathbf{x}, t) \left( \nabla \mathbf{u}(\mathbf{x}, t) + \nabla \mathbf{u}(\mathbf{x}, t)^T \right) \right]$$
$$+ \rho(\mathbf{x}, t) \mathbf{g} + \mathbf{f_s}(\mathbf{x}, t),$$

$$\tag{4.1}$$

$$\nabla \cdot \mathbf{u}(\mathbf{x}, t) = 0, \tag{4.2}$$

where $\mathbf{u}(\mathbf{x}, t)$, $p(\mathbf{x}, t)$, $\rho(\mathbf{x}, t)$, and $\mu(\mathbf{x}, t)$ are the spatially and temporally varying fluid velocity, pressure, density, and dynamic viscosity, respectively, $\mathbf{g}$ is the vector form of the gravitational acceleration, and $\mathbf{f_s}(\mathbf{x}, t)$ is the continuum surface tension force. Note that the continuity constraint Eq. (4.2) is obtained from the incompressibility nature of the fluid $D\rho(\mathbf{x}, t)/Dt = 0$, and mass conservation over the computational domain is expressed as

$$\frac{\partial \rho(\mathbf{x}, t)}{\partial t} + \nabla \cdot \rho(\mathbf{x}, t) \mathbf{u}(\mathbf{x}, t) = \frac{\mathrm{D}\rho(\mathbf{x}, t)}{\mathrm{D}t} + \rho(\mathbf{x}, t) \nabla \cdot \mathbf{u}(\mathbf{x}, t) = 0. \tag{4.3}$$

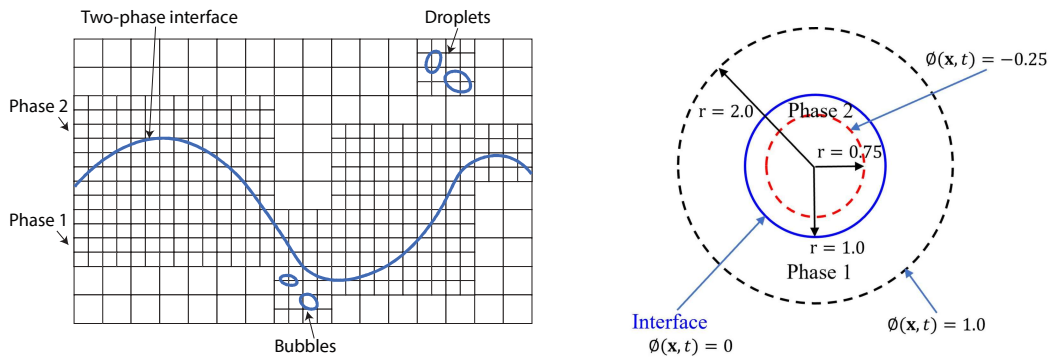

Figure 4.1: Left: two-phase flow on a multilevel Cartesian grid. Right: schematic definition of the LS function.

By combining Eqs. (4.1), (4.2), and (4.3), the incompressible Navier–Stokes equations can also be cast into nonconservative form as

$$\rho(\mathbf{x}, t) \left( \frac{\partial \mathbf{u}(\mathbf{x}, t)}{\partial t} + \nabla \cdot \mathbf{u}(\mathbf{x}, t) \mathbf{u}(\mathbf{x}, t) \right) = -\nabla p(\mathbf{x}, t) + \nabla \cdot \left[ \mu(\mathbf{x}, t) \left( \nabla \mathbf{u}(\mathbf{x}, t) + \nabla \mathbf{u}(\mathbf{x}, t)^T \right) \right]$$
$$+ \rho(\mathbf{x}, t) \mathbf{g} + \mathbf{f}_s(\mathbf{x}, t), \tag{4.4}$$

$$\nabla \cdot \mathbf{u}(\mathbf{x}, t) = 0, \tag{4.5}$$

Although the conservative and nonconservative forms of the Navier–Stokes equations are mathematically equivalent, different numerical treatments lead to different behaviors of the simulation results, as shown in later sections.

In this work, we use the LS function $\phi(\mathbf{x}, t)$ to explicitly capture the evolution of the two-phase interface (Sussman and Smereka, 1997; Sussman and Fatemi, 1999). As illustrated in the left part of Fig. 4.1, the liquid and gas are separated by the two-phase interface. The right part of Fig. 4.1 shows that $\phi(\mathbf{x}, t)$ is a signed distance from the two-phase interface, with $\phi(\mathbf{x}, t) > 0$ in phase 1 and $\phi(\mathbf{x}, t) < 0$ in phase 2. The advection equation of $\phi(\mathbf{x}, t)$ is

$$\frac{\partial \phi(\mathbf{x}, t)}{\partial t} + \nabla \cdot (\mathbf{u}\phi(\mathbf{x}, t)) = 0 \tag{4.6}$$

and the density and dynamic viscosity can be calculated using $\phi(\mathbf{x}, t)$ as

$$\rho(\mathbf{x}, t) = \rho_g \left[ 1 - H\big(\phi(\mathbf{x}, t)\big) \right] + \rho_l H\big(\phi(\mathbf{x}, t)\big), \tag{4.7}$$

$$\mu(\mathbf{x}, t) = \mu_g \left[ 1 - H\big(\phi(\mathbf{x}, t)\big) \right] + \mu_l H\big(\phi(\mathbf{x}, t)\big), \tag{4.8}$$

Here, $H\big(\phi(\mathbf{x}, t)\big)$ is the Heaviside function and can be smoothed around the interface as (Sussman and Smereka, 1997; Sussman et al., 1999)

$$H\big(\phi(\mathbf{x}, t)\big) = \begin{cases} 0, & \phi(\mathbf{x}, t) < -\epsilon, \\ \frac{1}{2} \left[ 1 + \frac{x}{\epsilon} - \frac{1}{\pi} \sin\left(\frac{\pi\phi(\mathbf{x}, t)}{\epsilon}\right) \right], & |\phi(\mathbf{x}, t)| \leq \epsilon, \\ 1, & \phi(\mathbf{x}, t) > \epsilon, \end{cases} \tag{4.9}$$

where $\epsilon$ is the smearing width and is usually set to be one or two grid lengths (Sussman and Smereka, 1997; Nangia et al., 2019a).

## 4.2 Time advancement

In this work, we use a level-by-level method (Martin and Colella, 2000; Martin et al., 2008) for the time advancement of variables on the multilevel grid. As our multilevel advancement algorithm is based on the single-level advancement method, we first introduce the consistent scheme for the single-level advancement and then discuss the multilevel advancement algorithm, which combines the single-level advancement algorithm with two different cycling methods, i.e., the subcycling and non-subcycling methods. A summary of the multilevel advancement algorithm is given at last.

### 4.2.1 Single-level advancement

This section mainly discusses the temporal and spatial discretizations of equations using the consistent scheme for the single-level advancement. Because our main focus is the consistent scheme, the numerical details of the inconsistent scheme (Zeng et al., 2022b) are placed in Appendix A.1.

**Consistent scheme**

For the consistent scheme, the discretizations are applied to the conservative form of the Navier–Stokes equations (Eqs. (4.1) and (4.2)). At time $t^n$, we have the velocity $\mathbf{u}^n$, pressure $p^{n-1/2}$, and LS function $\phi^n$ (Sussman et al., 1999; Nangia et al., 2019a). During the time interval $[t^n, t^{n+1}]$, the time step proceeds as follows.

**Step 1**: The LS function $\phi$, which is used to describe the two-phase interface, is updated by

$$\frac{\phi^{n+1} - \phi^n}{\Delta t} + Q\left(\mathbf{u}_{\text{adv}}^{n+\frac{1}{2}}, \phi^{n+\frac{1}{2}}\right) = 0 \tag{4.10}$$

where $Q\left(\mathbf{u}_{\text{adv}}^{n+\frac{1}{2}}, \phi^{n+\frac{1}{2}}\right)$ is computed using the second-order Godunov scheme (Almgren et al., 1998; Sussman et al., 1999).

**Step 2**: The LS function $\phi^{n+1}$ is then reinitialized to keep its signed distance property and guarantee mass conservation of the two-phase flow. In this step, a temporary LS function $d(\boldsymbol{x}, \tau)$ is updated iteratively using the following pseudoevolution equation:

$$\frac{\partial d}{\partial \tau} = S(\phi)(1 - |\nabla d|), \tag{4.11}$$

with the initial condition

$$d(\tau = 0) = \phi^{n+1}, \tag{4.12}$$

where

$$S(\phi) = 2\left(H(\phi) - 1/2\right). \tag{4.13}$$

Here, $\tau$ is the pseudotime for iterations. A second-order Runge–Kutta method is applied for the pseudo time advancement of $d$ (Sussman et al., 1999; Sussman and Smereka, 1997). To ensure mass conservation, the LS function $\phi^{n+1}$ is corrected by $d$ after pseudotime advancement (Sussman et al., 1999; Sussman and Smereka, 1997; Sussman and Fatemi, 1999). The midpoint value of $\phi$ is then calculated as $\phi^{n+\frac{1}{2}} = (\phi^{n+1} + \phi^n)/2$.

**Step 3**: The viscosity $\mu^{n+1}$ field is reset through the Heaviside function as

$$\mu^{n+1} = \mu_{\text{g}} + (\mu_{\text{l}} - \mu_{\text{g}}) \widetilde{H}\left(\phi^{n+1}\right), \tag{4.14}$$

where $\mu_{\text{l}}$ and $\mu_{\text{g}}$ are the viscosities of the liquid phase and the gas phase, respectively. The smoothed Heaviside function, which has been regularized over $n_{\text{cells}}$ grid cells on either side of the two-phase interface, is defined as

$$\widetilde{H}\left(\phi^{n+1}\right) = \begin{cases} 0, & \phi^{n+1} < -n_{\text{cells}}\Delta x \\ \frac{1}{2}\left(1 + \frac{1}{n_{\text{cells}}\Delta x}\phi^{n+1} + \frac{1}{\pi}\sin\left(\frac{\pi}{n_{\text{cells}}\Delta x}\phi^{n+1}\right)\right), & |\phi^{n+1}| \leq n_{\text{cells}}\Delta x \\ 1, & \text{otherwise} \end{cases}$$

$$(4.15)$$

where the uniform grid spacing $\Delta x = \Delta y$ is assumed (Nangia et al., 2019a) and $n_{cells} = 1$ or 2 are applied for all testing cases. The midpoint value of $\mu$ is then calculated as $\mu^{n+\frac{1}{2}} = (\mu^{n+1} + \mu^n)/2$.

The density field evolves based on Eq. (4.3) using the third-order accurate strong stability preserving Runge–Kutta (SSP-RK3) time integrator (Nangia et al., 2019a,b) as

$$\begin{aligned} \rho^{(1)} &= \rho^n - \Delta t \text{R}\left(\mathbf{u}_{\text{adv}}^n, \rho_{\text{lim}}^n\right), \\ \rho^{(2)} &= \frac{3}{4}\rho^n + \frac{1}{4}\rho^{(1)} - \frac{1}{4}\Delta t \text{R}\left(\mathbf{u}_{\text{adv}}^{(1)}, \rho_{\text{lim}}^{(1)}\right), \\ \rho^{n+1} &= \frac{1}{3}\rho^n + \frac{2}{3}\rho^{(2)} - \frac{2}{3}\Delta t \text{R}\left(\mathbf{u}_{\text{adv}}^{(2)} \cdot \rho_{\text{lim}}^{(2)}\right). \end{aligned}$$

$$(4.16)$$

$\text{R}\left(\mathbf{u}_{\text{adv}}, \tilde{\rho}_{\text{lim}}\right) \approx \left[\left(\nabla \cdot \left(\mathbf{u}_{\text{adv}}\tilde{\rho}_{\text{lim}}\right)\right)_{i,j} \cdot \left(\nabla \cdot \left(\mathbf{u}_{\text{adv}}\tilde{\rho}_{\text{lim}}\right)\right)_{i,j}\right]$ is an explicit CUI approximation to the cell-centered advection term of the density. The details of the CUI scheme are given in Section 4.2.1. Here, the subscript 'lim' indicates the limited face-centered variables, and these variables are interpolated from the corresponding cell-centered variables using the CUI scheme (Section 4.2.1). In the variables $\mathbf{u}_{\text{adv}}^n$, $\mathbf{u}_{\text{adv}}^1$, and $\mathbf{u}_{\text{adv}}^2$, the subscript 'adv' represents the face-centered advection velocities. To obtain these velocities, we define two auxiliary variables $\mathbf{u}^{(1)}$ and $\mathbf{u}^{(2)}$ as

$$\mathbf{u}^{(1)} = 2\mathbf{u}^n - \mathbf{u}^{n-1}, \tag{4.17}$$

$$\mathbf{u}^{(2)} = \frac{3}{2}\mathbf{u}^n - \frac{1}{2}\mathbf{u}^{n-1}. \tag{4.18}$$

Note that $\mathbf{u}^{n-1}$ and $\mathbf{u}^n$ are the cell-centered velocity in the previous time step and the current time step, respectively. As shown in Nangia et al. (2019a,b), it is crucial to use these interpolated and extrapolated velocities to maintain the time accuracy of the consistent scheme while calculating the advection velocities. The cell-centered velocities, including $\mathbf{u}^n$, $\mathbf{u}^{(1)}$, and $\mathbf{u}^{(2)}$, are then averaged onto the face of the cell-centered control volume to obtain $\mathbf{u}_{fc}^{n}$, $\mathbf{u}_{fc}^{(1)}$, and $\mathbf{u}_{fc}^{(2)}$. Finally, the marker and cell (MAC) projection is applied to these face-centered velocities to obtain the divergence-free advection velocities $\mathbf{u}_{adv}^{n}$, $\mathbf{u}_{adv}^{(1)}$, and $\mathbf{u}_{adv}^{(2)}$ (Almgren et al., 1998; Sussman et al., 1999). The midpoint value of the density is calculated by $\rho^{n+\frac{1}{2}} = (\rho^{n+1} + \rho^n)/2$.

**Step 4**. The intermediate velocity $\tilde{\mathbf{u}}^{*,n+1}$ is solved semi-implicitly as

$$\frac{\rho^{n+1}\tilde{\mathbf{u}}^{n+1,*} - \rho^n\mathbf{u}^n}{\Delta t} + \mathbf{C}^{n+\frac{1}{2}} = -\nabla p^{n-\frac{1}{2}} + \frac{1}{2}(\nabla \cdot (\mu^{n+\frac{1}{2}}\nabla\mathbf{u}^n) + \nabla \cdot (\mu^{n+\frac{1}{2}}\nabla\tilde{\mathbf{u}}^{n+1,*})) + \rho^{n+\frac{1}{2}}\mathbf{g} + \mathbf{f}_s^{n+\frac{1}{2}},$$

$$(4.19)$$

where the approximation to the convective derivative is given by

$$\mathbf{C}\left(\mathbf{u}_{adv}^{(2)}, \boldsymbol{\rho}_{lim}^{(2)}\mathbf{u}_{lim}^{(2)}\right) \approx \left[\left(\nabla \cdot \left(\mathbf{u}_{adv}^{(2)}\rho_{lim}^{(2)}u_{lim}^{(2)}\right)\right)_{i,j}, \left(\nabla \cdot \left(\mathbf{u}_{adv}^{(2)}\rho_{lim}^{(2)}v_{lim}^{(2)}\right)\right)_{i,j}\right] \qquad (4.20)$$

using the CUI scheme (Section 4.2.1). Here, we use the same advection velocity $\mathbf{u}_{adv}$ and the limited density $\rho_{lim}^{(2)}$ as in Eq. (4.16). This is the key requirement to ensure consistent mass and momentum transport (Nangia et al., 2019b). Finally, we note that although the density field $\rho^{n+1}$ is not updated using the LS function in **Step 3**, it needs to be reset by the LS function $\phi^{n+1}$ using Eq. (A.2) after **Step 4**. This postprocessing step helps synchronize the density field and LS field to avoid significant distortions in the interface for high-density-ratio flows (Nangia et al., 2019a).

**Step 5**: After obtaining the intermediate velocity, a level projection is applied to obtain the updated velocity $\tilde{\mathbf{u}}^{n+1}$ and pressure $p^{n+1/2}$ fields. An auxiliary variable $\boldsymbol{V}$ is first calculated by

$$\boldsymbol{V} = \frac{\tilde{\mathbf{u}}^{*,n+1}}{\Delta t} + \frac{1}{\rho^{n+1/2}}\boldsymbol{\nabla}p^{n-\frac{1}{2}}. \qquad (4.21)$$

Then, $\boldsymbol{V}$ is projected on the divergence-free velocity field to obtain the updated pressure $p^{n+1/2}$ via

$$L_{\rho^{n+1/2}}^{cc,l}p^{n+1/2} = \boldsymbol{\nabla} \cdot \boldsymbol{V}, \qquad (4.22)$$

where $L_{\rho^{n+1/2}}^{cc,l}p^{n+1/2}$ is the density-weighted approximation to $\boldsymbol{\nabla} \cdot (1/\rho^{n+1/2}\boldsymbol{\nabla}p^{n+1/2})$ on level $l$. Finally, the divergence-free velocity $\tilde{\mathbf{u}}^{n+1}$ on level $l$ is obtained as

$$\widetilde{\mathbf{u}}^{n+1} = \Delta t \left( \boldsymbol{V} - \frac{1}{\rho^{n+1/2}} \boldsymbol{\nabla} p^{n+1/2} \right). \tag{4.23}$$

We remark that the level projection is applied to the intermediate velocity $\boldsymbol{u^{*,n+1}}$. Compared with the form that projects the increment velocity $\boldsymbol{u^{*,n+1}} - \boldsymbol{u^n}$, e.g., as used in Almgren et al. (1998), the projection method used here can reduce the accumulation of pressure errors and lead to a more stable algorithm (Rider, 1995; Guy and Fogelson, 2005; Zeng et al., 2022b).

**Discretization of the convective term and surface tension term**

To obtain the convective terms in Eqs. (4.16) and (4.20), a third-order accurate Koren's limited CUI scheme is applied, which was first proposed by Roe and Baines (1982) and further studied by Waterson and Deconinck (2007) and Patel and Natarajan (2015) in the multiphase flow simulation. For simplicity, only the 2D discretized formulas of the CUI scheme are given in this section. The 3D formulas can be extended in a straightforward way.

A cell-centered variable $\psi_{\text{lim}}$, which can be $\rho_{\text{lim}}$ in Eqs. (4.16) or $\rho_{\text{lim}}\mathbf{u}_{\text{lim}}$ in Eqs. (4.20), is advected by $\mathbf{u}_{\text{adv}}$ using

$$\nabla \cdot (\mathbf{u}_{\text{adv}} \psi_{\text{lim}})_{i,j} = \frac{u_e \psi_e - u_w \psi_w}{\Delta x} + \frac{v_n \psi_n - v_s \psi_s}{\Delta y}, \tag{4.24}$$

where $u_e$, $u_w$, $v_n$, and $v_s$ are edge-centered divergence-free velocities located on the east, west, north, and south sides of cell $(i,j)$. With these associated velocities, the CUI scheme is utilized to obtain $\psi_w$, $\psi_e$, $\psi_n$, and $\psi_s$ on the edges of the control volume, which is marked by the green dashed line in Fig. 4.2. For a given face of the cell centered control volume $f \in \{e, w, n, s\}$, the upwind $\psi_C$, the far upwind $\psi_U$ and the downwind $\psi_D$ are labeled according to the direction of the edge-centered advection velocity. The upwinded approximation of $\psi_f$ is then calculated as (Nangia et al., 2019a)

$$\widetilde{\psi}_f = \begin{cases} 3\widetilde{\psi}_C, & 0 < \widetilde{\psi}_C \leq \frac{2}{13} \\ \frac{5}{6}\widetilde{\psi}_C + \frac{1}{3}, & \frac{2}{13} < \widetilde{\psi}_C \leq \frac{4}{5} \\ 1, & \frac{4}{5} < \widetilde{\psi}_C \leq 1 \\ \widetilde{\psi}_C, & \text{otherwise} \end{cases} \tag{4.25}$$

where the normalized value is given by Waterson and Deconinck (2007)

$$\widetilde{\psi} = \frac{\psi - \psi_U}{\psi_D - \psi_U}. \tag{4.26}$$

After obtaining $\psi_w$, $\psi_e$, $\psi_n$, and $\psi_s$ on the edges of the control volume $f$ in Eq. (4.25), the convective terms can be calculated using Eq. (4.24).

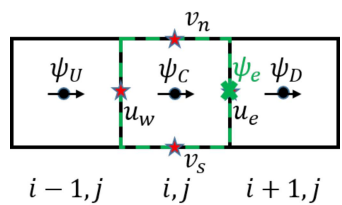

(a) CUI stencil for $\psi_e$ ($u_e \geq 0$)

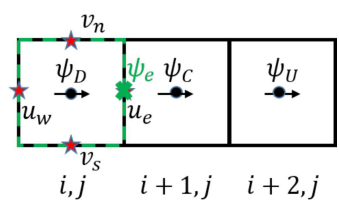

(b) CUI stencil for $\psi_e$ ($u_e < 0$)

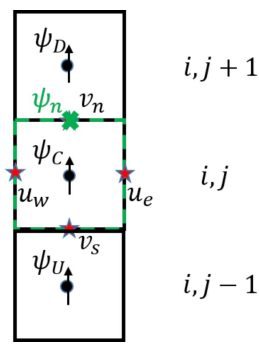

(c) CUI stencil for $\psi_n$ ($v_n \geq 0$)

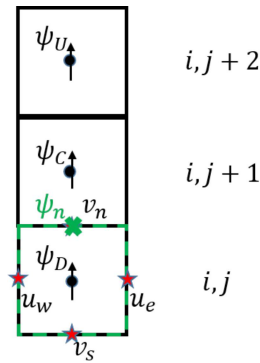

(d) CUI stencil for $\psi_n$ ($v_n < 0$)

Figure 4.2: In each diagram, the green dashed line represents the control volume over which the convective term $\nabla \cdot (\mathbf{u}_{\text{adv}} \psi_{\text{lim}})_{i,j}$ is computed. The upwind, centered, and downwind variables are labeled as $\psi_U$, $\psi_C$, and $\psi_D$, respectively. (a) Required CUI stencil to compute $\psi_e$ when $u_e \geq 0$. (b) Required CUI stencil to compute $\psi_e$ when $u_e < 0$. (c) Required CUI stencil to compute $\psi_n$ when $v_n \geq 0$. (d) Required CUI stencil to compute $\psi_n$ when $v_n < 0$.

Following the continuum surface tension model in Brackbill et al. (1992), the cell-centered surface tension force $\mathbf{f}_s$ in Eqs. (A.4) and (4.19) is defined as

$$\mathbf{f}_s = \sigma \kappa \nabla H = -\sigma \nabla \cdot \left( \frac{\nabla \phi}{|\nabla \phi|} \right) \nabla H, \tag{4.27}$$

where $\sigma$ is the surface tension coefficient and $H$ is the smoothed Heaviside function defined in Eq. (A.3). This discretization yields a good balance between the surface tension force and the pressure gradient, as indicated by (Nangia et al., 2019a; Brackbill et al., 1992).

## 4.2.2 Multilevel advancement

Algorithm 4 summarizes the unified multilevel advancement algorithm for both the subcycling and non-subcycling methods. We first initialize the velocity $\boldsymbol{u}$, pressure $p$, and LS function $\phi$ on all levels at the beginning of the simulation based on some refinement criteria. After the initialization, we can use either the subcycling or the non-subcycling method for time advancement. We can also choose either the inconsistent or the consistent scheme for spatial discretization. The synchronization operations are then applied when a coarser level catches up with a finer level. Finally, the grid is either refined or derefined before moving to the next step.

To summarize, the single-level consistent scheme is extended to the multilevel grid so that the multilevel advancement algorithm can accurately simulate the high-density-ratio and high-Reynolds-number two-phase flows using AMR. We emphasize that our multilevel advancement algorithm is a level-by-level advancement method, which differs from the composite advancement method (Bhalla et al., 2013; Nangia et al., 2019a; Sussman et al., 1999; Nangia et al., 2019b). The level variables are used for time advancement in the level-by-level advancement method. Before the synchronization step, each level can be advanced individually without taking into account the finer levels. The time step restrictions on the coarser levels are alleviated because the time advancements at various levels are decoupled. This is in contrast to the composite advancement technique, in which the multilevel multigrid (MLMG) solver is used to update the velocity and pressure in the valid areas of all levels at the same time. Because of this different treatment, the composite advancement method is not flexible enough to easily integrate both the subcycling and the non-subcycling methods, while the level-by-level method used in this study can easily manage both.

## 4.3 Results

This section presents several multiphase flow problems to test the convergence, consistency, and stability of the proposed schemes within the BSAMR framework from various aspects. The consistent and inconsistent schemes are compared to show the importance and necessity of using the consistent scheme in the high-density-ratio and high-Reynolds-number two-phase flow problems. In these problems, either the subcycling or non-subcycling methods are employed for time advancement when the multilevel grid is involved. We first define some common parameters here. For each problem, $\Delta t_0$ denotes the time step on level 0. We use $\Delta x_0$, $\Delta y_0$, and $\Delta z_0$ to represent the grid spacings in the $x$-, $y$-, and $z$-directions, respectively, on level 0. For the multilevel grid, the grid spacings on level $l$ satisfy $\Delta x_l = \Delta x_0/2^l$, $\Delta y_l = \Delta y_0/2^l$, and $\Delta z_l = \Delta z_0/2^l$ for all $0 \leq l \leq l_{max}$. For the time step size, the finest level $\Delta t^{l_{max}}$ is first determined by restrictions of the CFL condition, gravity, viscosity, and surface tension (Zeng et al., 2022b; Sussman et al., 1999). The time step sizes on the coarser levels, depending on whether the subcycling or the non-subcycling method is used, are then calculated based on Algorithm 4.

### 4.3.1 2D reversed single vortex problem

In this section, we test the order of convergence for the consistent scheme in a 2D reversed single vortex problem (Chiu and Lin, 2011; Mirjalili et al., 2020) on the multilevel grid. In this problem, a 2D circular drop with radius $R = 0.15$ is placed at $(0.5, 0.75)$ in a unit computational domain $[0, 1] \times [0, 1]$. The velocity field is given by the stream function,

$$\Psi(x, y, t) = \frac{1}{\pi} \sin^2(\pi x) \sin^2(\pi y) \cos\left(\frac{\pi t}{T}\right), \tag{4.28}$$

in which the rotational period is $T = 4.0$. The velocities in the $x$ and $y$ directions are defined as $u(x, y, t) = \partial \Psi / \partial y$ and $v(x, y, t) = -\partial \Psi / \partial x$, respectively. The periodic boundary condition is applied in both $x$ and $y$ directions.

To obtain the pointwise convergence rate on the multilevel grid, we consider four cases here, of which the grid number $N_x \times N_y$ on level 0 is $16 \times 16$, $32 \times 32$, $64 \times 64$, and $128 \times 128$. The finest level $l_{max}$ is 2, and the refinement criterion is based on the distance to the drop interface, i.e., grid cells $(i, j)$ on level $l$ ($0 \leq l < l_{max}$) are refined to the finer level if $|\phi_{i,j}| < 8.0 \max(\Delta x_l, \Delta y_l)$, which ensures that the surface of the drop is always placed on the finest level. The upper part of Fig. 4.3 shows the state of the drop at $t = T/2$ and $t = T$ for two different resolutions using the subcycling method. As we increase the resolution, the interface at the tail of the stretching vortex becomes sharper. At $t = T$, the interfacial profile of the drop at the high resolution converges to its initial profile, which represents the exact solution. On the other hand, the circular shape of the drop is distorted for the simulation with low resolution. The lower part of Fig. 4.3 shows the grid hierarchy and the shape of the drop using the non-subcycling method. The subcycling and non-subcycling methods produce consistent and accurate results in this simulation. The error of the results is defined as (Chiu and Lin, 2011; Mirjalili et al., 2020)

$$E_{\text{shape}} = \sum_{i=1}^{N_x} \sum_{j=1}^{N_y} |\phi(i, j, t = T) - \phi(i, j, t = 0)| \Delta x \Delta y, \tag{4.29}$$

where $\phi(i, j, t = 0)$ is the exact reference solution. The corresponding rates of convergence are given in Fig. 4.4. The errors decrease with an approximately second-order convergence rate for both the subcycling and the non-subcycling methods. These errors are also comparable with the values reported by (Mirjalili et al., 2020).

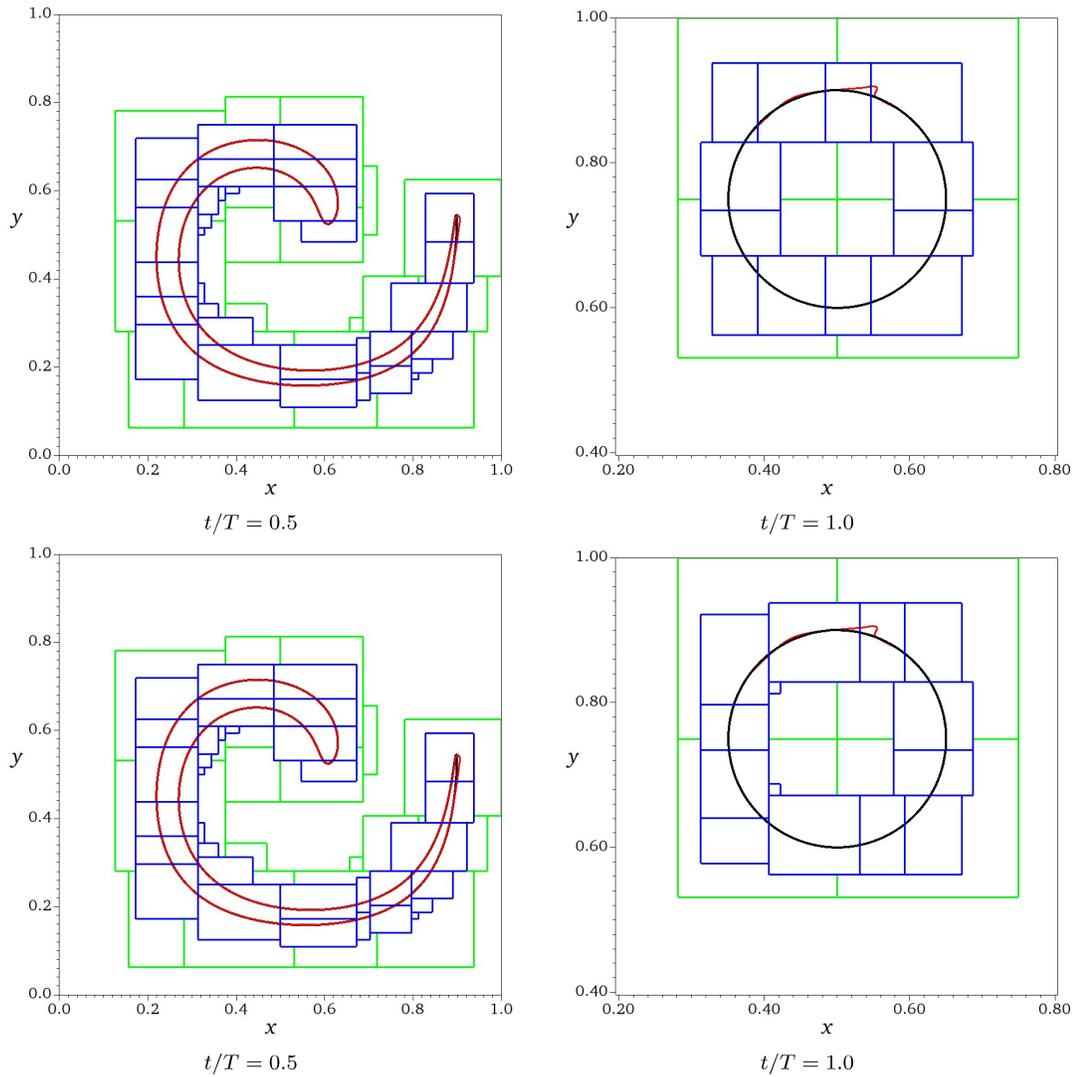

Figure 4.3: Profiles of the drop interface in the reverse single vortex problem using the subcycling method (upper) and non-subcycling method (lower) at different time instants. The red and black lines represent the drop interface with the coarsest grid numbers $32 \times 32$ and $64 \times 64$, respectively. The green and blue lines represent patches on levels 1 and 2, respectively.

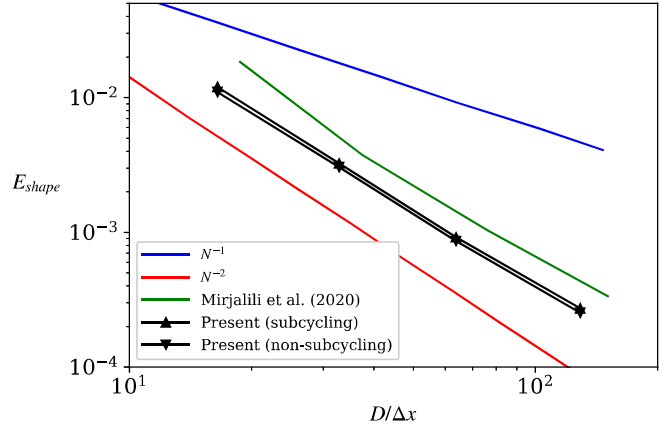

Figure 4.4: Spatial–temporal convergence for the 2D reversed single-vortex problem on a multilevel grid.

### 4.3.2   2D convected droplet with a high density ratio

The convected droplet is a canonical problem to demonstrate the importance of consistent mass and momentum transport in flows with a high density ratio. We first consider a single-level 2D test by employing both the consistent and the inconsistent schemes. A dense droplet is placed in a periodic computational domain with size $[0, 1] \times [0, 1]$. The initial center position of the droplet is $(X_0, Y_0) = (0.25, 0.5)$, and the diameter of the droplet is $D = 0.4$. The density ratio between the inner and outer parts of the droplet is $\rho_i/\rho_o = 10^6$, where $\rho_o = 1.0$. The viscosity, gravity, and surface tension force are all set to zero. The vertical velocity is initialized as zero, and the horizontal velocity is $u_{i,j} = 1 - \widetilde{H}_{i,j}^f$, where $\widetilde{H}_{i,j}^f$ is a smoothed Heaviside function defined as

$$\widetilde{H}_{i,j}^f = \begin{cases} 1, & \phi_{i,j} < -\epsilon, \\ \frac{1}{2}\left(1 + \frac{1}{\epsilon}\phi_{i,j} + \frac{1}{\pi}\sin\left(\frac{\pi}{\epsilon}\phi_{i,j}\right)\right), & |\phi_{i,j}| \le \epsilon, \\ 0, & \text{otherwise.} \end{cases} \quad (4.30)$$

Here, $\phi(\mathbf{x}, 0) = D/2 - \sqrt{(x - X_0)^2 + (y - Y_0)^2}$ is the initial LS function, and $\epsilon = \min(\Delta x_0, \Delta y_0)$ is the smearing width (Nangia et al., 2019a). After initialization, the level projection is applied to generate a divergence-free initial velocity field (Yang et al., 2021b). This problem has been investigated by Bussmann et al. (2002), Desjardins and Moureau (2010), Ghods and Herrmann (2013), Patel and Natarajan (2015) and Nangia et al. (2019a). In the following 2D cases, the grid size is discretized as $N_x \times N_y = 128 \times 128$. Each case is run until $t = 0.5$ with a constant time step size $\Delta t = 1/(31.25 N_x)$ (Nangia et al., 2019a).

Because of the high density contrast between the droplet and the outer fluid, it is expected that the droplet moves at a constant velocity and maintains its circular shape. As shown in the lower part of Fig. 4.5, the simulations are stable when using the consistent mass and momentum transport scheme. For the cases using the inconsistent scheme, as plotted in the upper part of Fig. 4.5, the surface distortion of the original droplet is clearly seen, and small spurious droplets are then generated when the simulation becomes unstable.

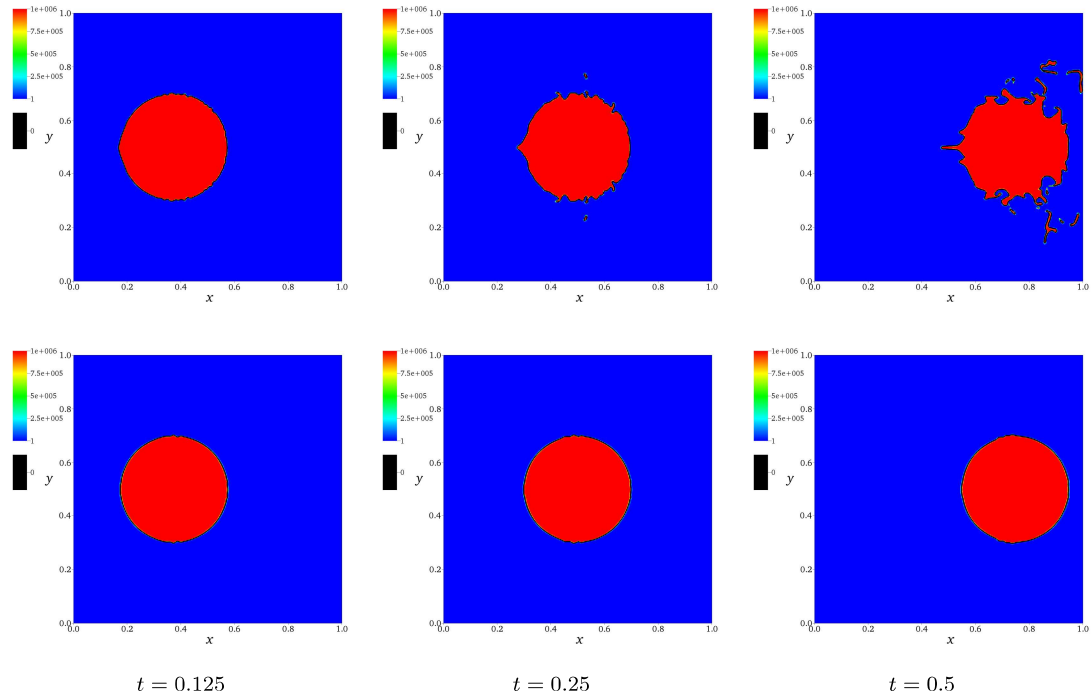

$t = 0.125$                    $t = 0.25$                    $t = 0.5$

Figure 4.5: Geometry and density evolution of the 2D convected droplet problem. The upper and lower parts show the results of the inconsistent and consistent schemes, respectively.

### 4.3.3   2D dam breaking

This section investigates a dynamic 2D dam-breaking problem to further validate the robustness and necessity of the consistent scheme when the multilevel grid is involved. As shown in Fig. 4.6, a square block with side length $a = 0.057\,\mathrm{m}$ is placed at the left-bottom corner, and the computational domain size is $7a \times 1.75a$. A no-slip boundary condition is imposed on the bottom wall, while all other walls are considered free-slip boundaries (Rezende et al., 2015; Patel and Natarajan, 2017). The dimensionless front is defined as $\tilde{d}_f = d_f/a$, in which $d_f$ refers to the dimensional distance between the front position (point A in Fig. 4.6) and the origin. Other dimensionless parameters are set as $Re = \rho_l U a/\mu_l = 2950$, $Fr = U/\sqrt{ga} = 1.0$, and $We = \rho_l U^2 a/\sigma = 0.54$. Here, $U$ is the characteristic length. The density ratio and viscosity ratio are set as $\rho_g/\rho_l = 0.0012$ and $\mu_g/\mu_l = 0.016$, respectively (Nangia et al., 2019a). Table 4.1 gives the parameters of four simulation cases. The refinement criterion is based on the distance to the air–water interface, the same as the 2D reversed vortex problem in Section 4.3.1.

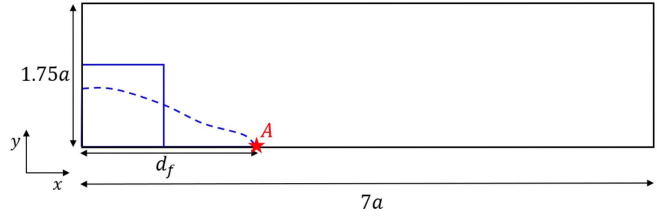

Figure 4.6: Sketch of the 2D dam-breaking problem. Solid line: dam interface at $t = 0$. Dashed line: dam interface at $t > 0$. $A$: dam front position at $t > 0$.

Table 4.1: Parameters for cases of the dam-breaking problem.

| Case No. | Grid number on level 0 | $l_{max}$ | $\Delta t^0$ | Scheme | Cycling method |
|----------|------------------------|-----------|--------------|--------|----------------|
| 1 | $512 \times 128$ | 0 | $1.25 \times 10^{-4}$ | Consistent | – |
| 2 | $512 \times 128$ | 0 | $1.25 \times 10^{-4}$ | Inconsistent | – |
| 3 | $256 \times 64$ | 1 | $2.50 \times 10^{-4}$ | Consistent | Subcycling |
| 4 | $256 \times 64$ | 1 | $2.50 \times 10^{-4}$ | Inconsistent | Subcycling |

Fig. 4.7 compares the dimensionless front $\tilde{d}_f$ of the single-level cases (Cases 1 and 2) and the two-level cases (Cases 3 and 4) with previous experimental and numerical results (Gu et al., 2018; Rezende et al., 2015; Nangia et al., 2019a). The results of the cases that use the consistent scheme (Cases 1 and 3) agree well with the literature. For cases using the inconsistent scheme (Cases 2 and 4), the magnitude of the dimensionless front $\tilde{d}_f$ is underestimated. As shown in Fig. 4.7, $\tilde{d}_f$ decreases and then increases at approximately $t/T = 2.4$. This phenomenon is also obtained in (Patel and Natarajan, 2017) because the liquid part of the dam front becomes unstable and breaks into small droplets in the inconsistent scheme cases.

To further compare the differences of the results between the consistent scheme and the inconsistent scheme, the time evolution of the normalized velocity amplitude and normalized fluid mass are plotted in Fig. 4.8. Here, the normalized velocity amplitude is defined as the maximum value of $||\mathbf{u}(t)||/||\mathbf{u}(0)||$, where $||\mathbf{u}|| = \sqrt{u^2 + v^2}$ is the $L^2$ norm of the velocity vector. The normalized fluid mass is defined as $||m(t)||/||m(0)||$, where $m = \int \rho \, dV$ is the total fluid mass. As shown in the left part of Fig. 4.8, the normalized velocity amplitude agrees well between Cases 3 and 4 when $t/T < 1.0$.

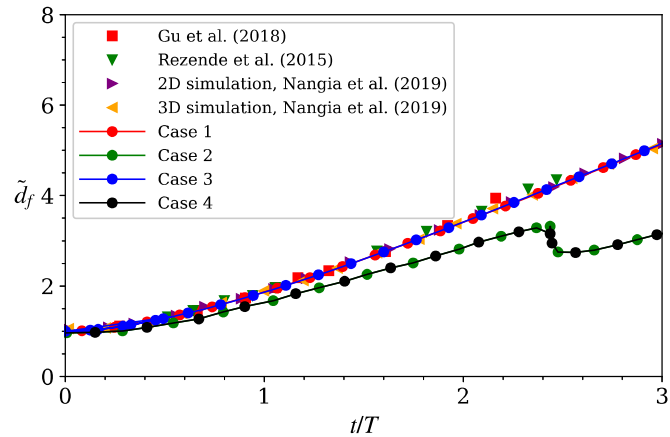

Figure 4.7: Comparison of the evolution of the dimensionless front $\tilde{d}_f$ among the single-level cases (Cases 1 and 2), the two-level subcycling cases (Cases 3 and 4), and the literature: (■) Gu et al. (2018); (▼) Rezende et al. (2015); (▶, ◀) Nangia et al. (2019a). Cases 1 and 3 use the consistent scheme, while Cases 2 and 4 use the inconsistent scheme.

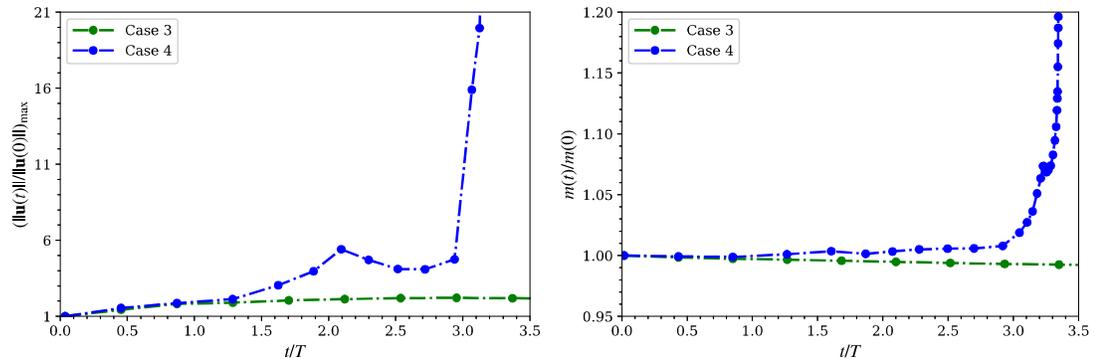

Figure 4.8: Comparison of the time evolution of the normalized velocity amplitude (left) and the normalized fluid mass (right) between the case using the consistent scheme (Case 3) and the case using the inconsistent scheme (Case 4) in the 2D dam-breaking problem.

Over time, the normalized velocity in the case with the inconsistent scheme becomes unreasonably large, and the simulation quickly becomes unstable. Compared with the consistent scheme, the instability induced by the inconsistent scheme also affects the conservation of the fluid mass. Although the same LS reinitialization method is used for both Cases 3 and 4, the right part of Fig. 4.8 shows that the mass loss of Case 3 is less than 2%, while the simulation of Case 4 quickly becomes unstable.

Next, the evolution of the breaking dam for the two-level consistent scheme case (Case 3) and the two-level inconsistent scheme case (Case 4) is depicted in Fig. 4.9, in which patches are dynamically refined around the interface as time evolves. The left part of Fig. 4.9 shows that the results of the consistent case remain stable and produce a physically reasonable solution. The shape of the dam also compares favorably to the experimental results (Martin et al., 1952) and the numerical results (Nangia et al., 2019a) (not plotted). On the other hand, in the right part of Fig. 4.9, unphysical deformations appear because of the dissimilar transport of mass and momentum caused by the inconsistent scheme. We again emphasize that consistent mass and moment transport are important for stable simulations of this dam breaking problem with a large liquid–gas density ratio.

### 4.3.4   2D droplet splashing on a thin liquid film

In this section, we aim to validate the consistency of the results between the subcycling and non-subcycling methods when a multilevel grid is involved. We consider the problem of droplet splashing on a thin liquid film, which has applications in inkjet printing (Singh et al., 2010) and spray cooling (Kim, 2007). The computational domain is $8D \times 2D$ with a no-slip boundary condition at all sides. The droplet is initially placed at $(X_0, Y_0) = (4D, 0.75D)$ with a diameter of $D = 1.0$ and an initial downward velocity of $U = 1.0$. The thin film fills one-tenth of the computational domain with an initial height of $0.2D$ (Nangia et al., 2019a). The density ratio and the viscosity ratio between the liquid (or the thin film) and the surrounding gas are $\rho_l/\rho_g = 815$ and $\mu_l/\mu_g = 55$, respectively. Owing to the high impact velocity, the gravitational force does not play an important role and can be neglected (Nangia et al., 2019a; Patel and Natarajan, 2017). The Reynolds number is $Re = \rho_g U D/\mu_g = 66$, and the Weber number is $We = \rho_g U^2 D/\mu_g = 0.126$.

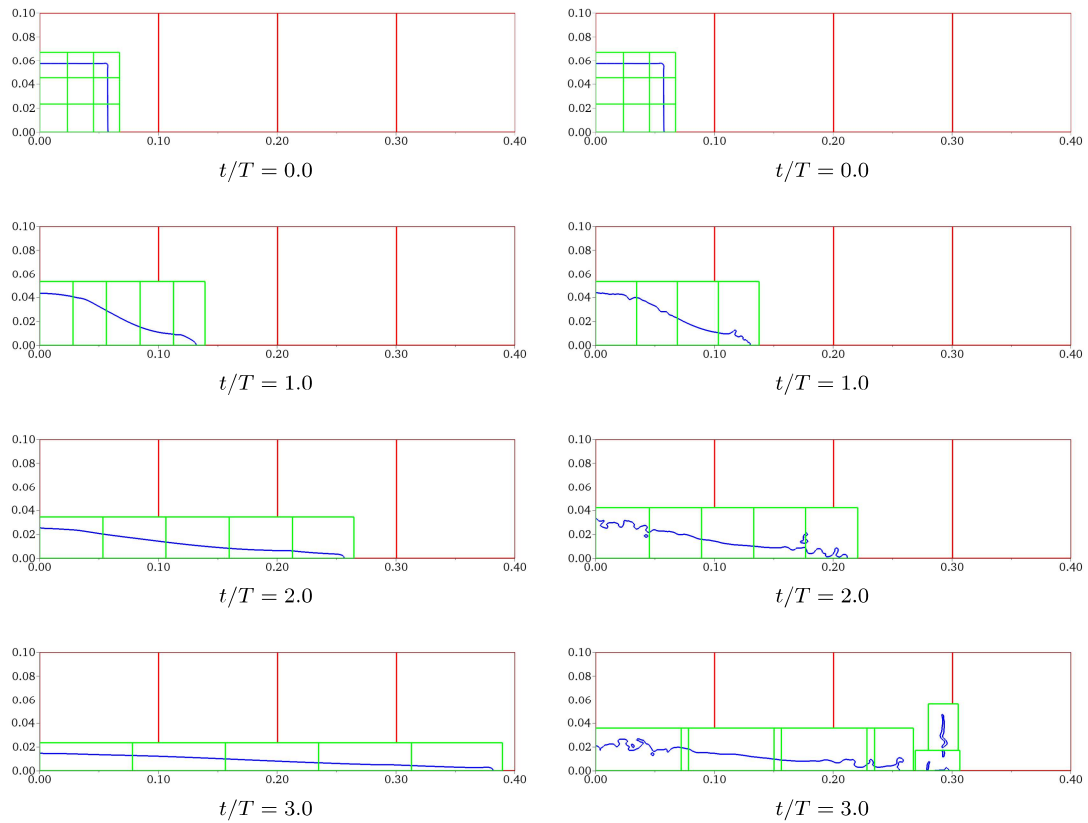

Figure 4.9: Profiles of water surface in the breaking dam for the two-level subcycling cases using the consistent scheme (left) and inconsistent scheme (right) at different time instants. The red and green lines represent patches on levels 0 and 1, respectively.

Two multilevel cases are considered in Table 4.2. For these multilevel cases, the finest level $l_{max}$ is 3, and the refinement criterion is based on the distance to the gas–liquid interface; i.e., grid cells $(i, j, k)$ on level $l$ ($0 \leq l < l_{max}$) are refined to the finer level if $|\phi_{i,j,k}| < 8.0 \max(\Delta x_l, \Delta y_l, \Delta z_l)$, which ensures that the thin film and the splashing droplet are always placed on the finest level. Simulations of these two cases are carried out until $t/T = 1.5$.

The left part of Fig. 4.10 shows the jet base location $x_J$, which is the midpoint of the two neck points of the splashing sheet. The right part shows the temporal evolution of the dimensionless spread factor $x_J/D$, which follows a power law defined by $x_J/D \propto (Ut/D)^{1/2}$ (Coppola et al., 2011; Josserand and Zaleski, 2003; Howison et al., 2005), where $U$ is the characteristic velocity. Our simulation results agree with Nangia et al. (2019a). The slope is consistent with the power law, and the results of the subcycling case (Case 1) and non-subcycling case (Case 2) are similar. Fig. 4.11 plots the time evolution of the splashing at different time instances for both the subcycling and the non-subcycling methods. Although the grid patches for different cycling methods are different, the shapes of the thin film and the splashing droplet are almost identical between these two cycling methods. We note that this problem was also simulated in (Li et al., 2013), where a lattice Boltzmann method was used, and in (Patel and Natarajan, 2017), where a VOF method was applied. Although we are using a different method, our simulations agree favorably with those results.

Table 4.2: Parameters for cases of a 2D droplet splashing on a thin film.

| Case No. | Grid number on level 0 | $l_{max}$ | $\Delta t^0$ | Scheme | Cycling method |
|---|---|---|---|---|---|
| 1 | $256 \times 64$ | 3 | $1.2 \times 10^{-3}$ | Consistent | Subcycling |
| 2 | $256 \times 64$ | 3 | $1.5 \times 10^{-4}$ | Consistent | Non-subcycling |

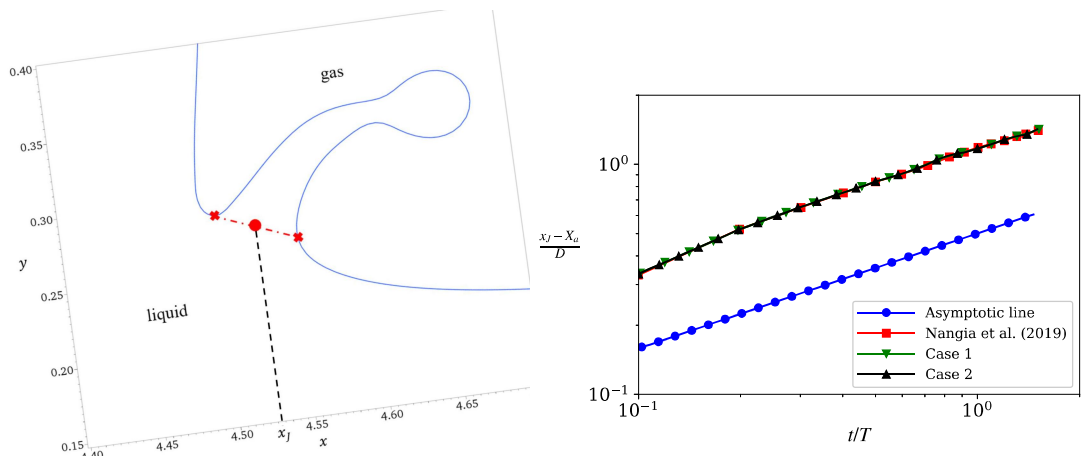

Figure 4.10: Left: sketch of the jet base location $x_J$; right: comparison of the temporal evolution of the dimensionless jet base location among the four-level subcycling case (Case 1), four-level non-subcycling case (Case 2), Nangia et al. (2019a), and the asymptotic results from theoretical analysis (Coppola et al., 2011; Josserand and Zaleski, 2003; Howison et al., 2005).

In summary, the proposed framework produces consistent results between the subcycling and non-subcycling methods when multiple levels of grids are involved in the simulation. The surface tension and droplet splashing dynamics can be accurately captured using the consistent scheme.

### 4.3.5   2D breaking wave

Simulations of breaking waves are of great interest in ocean science and engineering research. To further compare the robustness and stability between the consistent scheme and inconsistent scheme in simulations with very high Reynolds numbers, a 2D breaking wave test is presented in this section. The numerical sketch of the initial wave geometry and the computational domain is given by Fig. 4.12. The surface profile of a deep-water wave is initialized as

$$\eta(x,y) = a\left(\cos(kx) + \frac{1}{2}\varepsilon\cos(2kx) + \frac{3}{8}\varepsilon^2\cos(3kx)\right), \qquad (4.31)$$

where $a$ is the wave amplitude, $k = 2\pi/\lambda$ is the wavenumber, $\lambda$ is the wavelength, and $\varepsilon = ak$ is the initial wave steepness. The velocities are

$$u(x,y) = \Omega a\exp(ky)\cos(kx), \quad v(x,y) = \Omega a\exp(ky)\sin(kx), \qquad (4.32)$$

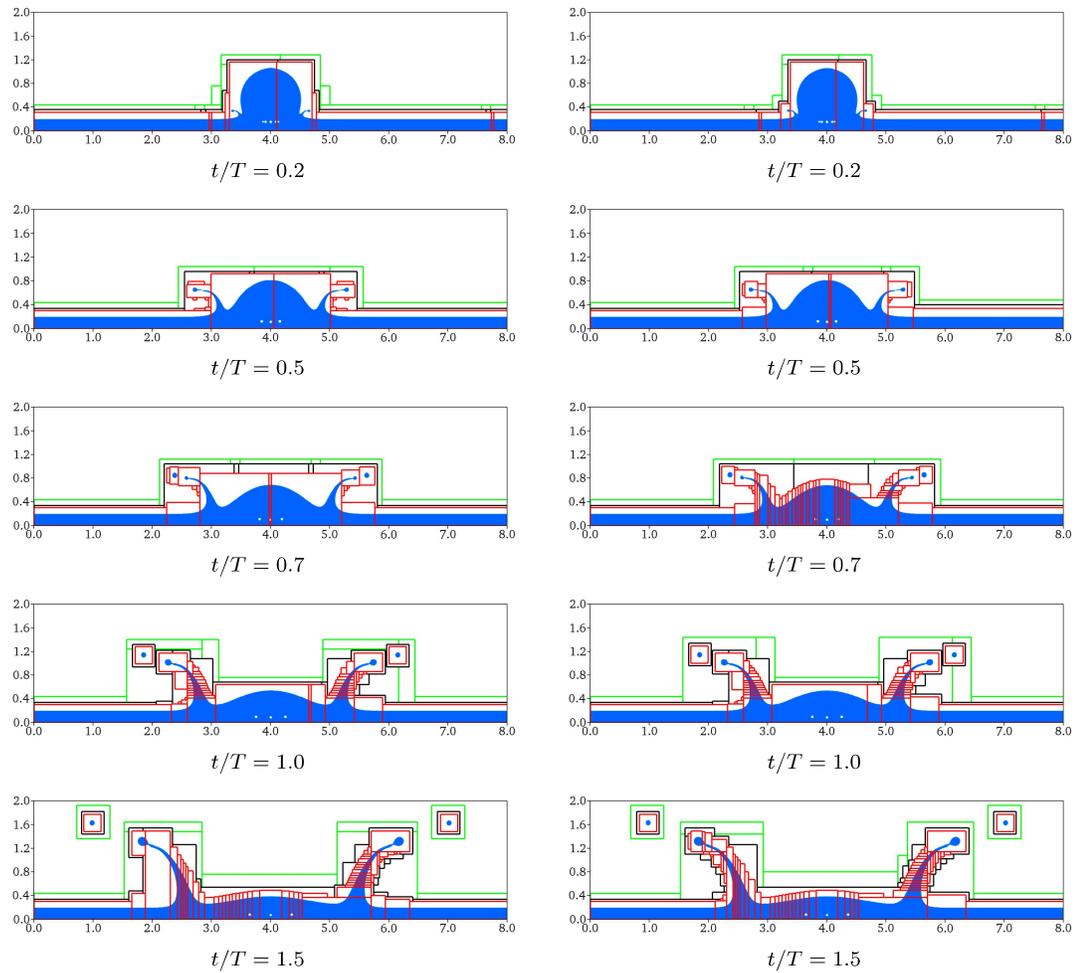

Figure 4.11: Grid hierarchy and evolution of droplet splashing on a thin liquid film at five different time instances. Left: subcycling method; right: non-subcycling method. Green patches: level 1; black patches: level 2; red patches: level 3.

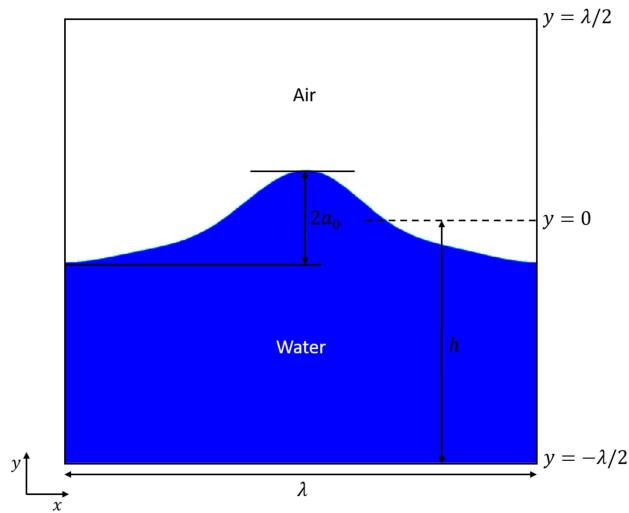

Figure 4.12: Sketch of the initial wave geometry and the computational domain for the 2D breaking wave problem.

Table 4.3: Parameters of cases for the 2D breaking wave problem.

| Case no. | Grid number on level 0 | $l_{max}$ | $\Delta t^0$ | Scheme | Cycling method |
|----------|------------------------|-----------|--------------|--------|----------------|
| 1 | $128 \times 128$ | 2 | $8.0 \times 10^{-5}$ | Inconsistent | Subcycling |
| 2 | $128 \times 128$ | 2 | $2.0 \times 10^{-5}$ | Inconsistent | Non-subcycling |
| 3 | $128 \times 128$ | 2 | $8.0 \times 10^{-5}$ | Consistent | Subcycling |
| 4 | $128 \times 128$ | 2 | $2.0 \times 10^{-5}$ | Consistent | Non-subcycling |

where $\Omega = \sqrt{gk(1 + \varepsilon^2)}$ (Iafrati, 2009). The free slip boundary condition is imposed in the $y$ direction, and the periodic boundary condition is applied in the $x$ direction. The mean water depth is $h = \lambda/2$, as shown in Fig. 4.12. The parameters of the four cases are listed in Table 4.3. In all these cases, we follow the parameters in (Iafrati, 2009; Yang et al., 2021b), in which $\lambda = 0.27$ m, $a = 0.0236$ m, and $\varepsilon = 0.55$. Let $\lambda$ be the characteristic length scale and $T = \sqrt{\lambda/g}$ denote the characteristic time scale. The Reynolds number is $Re = \rho_w \lambda^{3/2} g^{1/2}/\mu_w = 4.0 \times 10^6$. Other dimensionless numbers are the Froude number $Fr = \sqrt{\lambda/g}/T = 1.0$, Weber number $We = \rho_w \lambda^2 g/\sigma = 99.6$, density ratio $\rho_a/\rho_w = 0.0012$, and dynamic viscosity ratio $\mu_a/\mu_w = 0.04$.

We note that all cases in Table 4.3 have the same finest resolution on the multilevel grid, in which Cases 1 and 2 use the inconsistent scheme and Cases 3 and 4 use the consistent scheme. In addition, Cases 1 and 3 use the subcycling method, while Cases 2 and 4 use the non-subcycling method. The grid is refined based on the distance to the air–water interface in the same way as in the dam-breaking problem (Section 4.3.3).

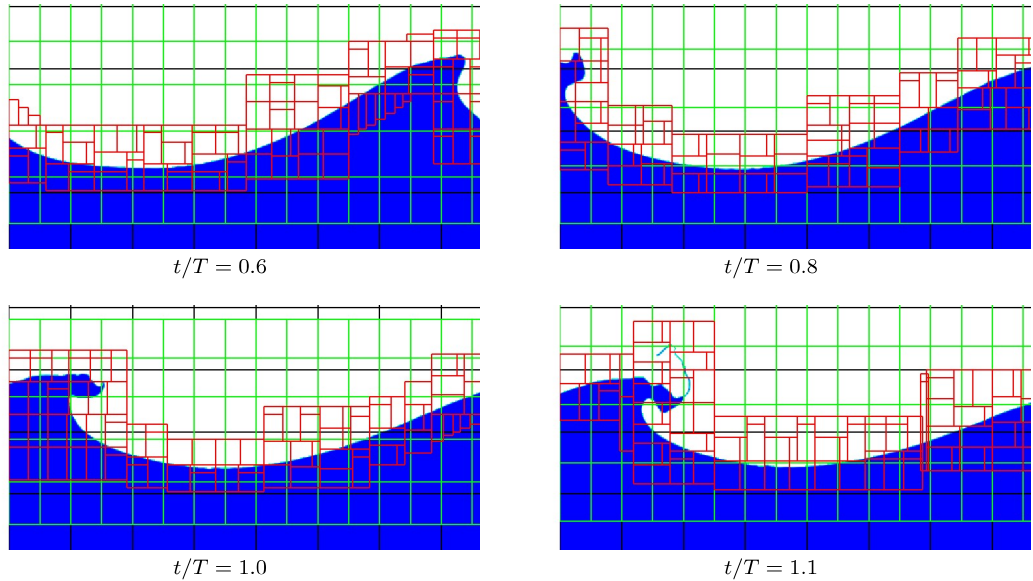

Figure 4.13: Evolution of a 2D plunging breaker obtained from the three-level non-subcycling case (Case 2) using the inconsistent scheme. The black, green, and red lines represent patches on levels 0, 1, and 2, respectively.

Fig. 4.13 shows the time evolution of the wave geometry obtained from the three-level nonsubcycled case (Case 2), in which the inconsistent scheme is used. Over time, small distortions appear at the wave crest due to numerical instability. The wave surface becomes rough, and a spurious long thin sheet is generated and then moves upward at approximately $t/T = 1.1$. The sheet further becomes thinner, and the simulation quickly halts because a very large spurious velocity appears at the sheet location. Similar to the dam-breaking problem in Section 4.3.3, the left part of Fig. 4.14 compares the time evolution of the normalized velocity amplitude among the above four cases. For Cases 1 and 2, the normalized velocity amplitude suddenly becomes very large at approximately $t/T = 1.1$, which corresponds to the last panel in Fig. 4.13. For Cases 3 and 4, the time evolution of the normalized velocity gradually increases and is physically reasonable, which demonstrates the stability of the consistent scheme in simulating high-Reynolds-number two-phase flow problems.

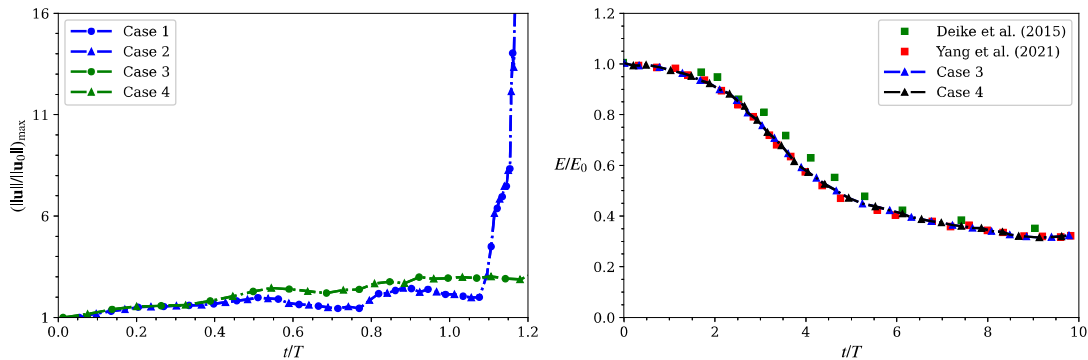

Figure 4.14: Left: comparison of the time evolution of the velocity amplitude among Cases 1–4 in the 2D breaking wave problem. Right: comparison of the time evolution of the total mechanical energy among Case 3, Case 4, and previous results (Deike et al. (2015); Yang et al. (2021b)) in the 2D breaking wave problem. Cases 1 and 2 use the inconsistent scheme, while Cases 3 and 4 use the consistent scheme. The subcycling method is applied to Cases 1 and 3, while the non-subcycling method is applied to Cases 2 and 4.

To show the accuracy of the consistent scheme, the right part of Fig. 4.14 compares the total mechanical energy $E = E_k + E_p$ among Case 3, Case 4, and previous results in Deike et al. (2015); Yang et al. (2021b) using different codes. Here, $E_k = \frac{1}{2} \int \rho_w u_i u_i dV$ is the kinetic energy, and $E_p = \int (\rho_w g y) dV - \frac{1}{2} \rho_w g V_w h$ is the potential energy. The present results for the total mechanical energy agree with the previous results of Deike et al. (2015) and Yang et al. (2021b). This again demonstrates the consistency between the non-subcycling method and the subcycling methods. Fig. 4.15 plots the variation in the mean velocity $u_{avg}$ along the water depth at $t/T = 10$ and $t/T = 15$. Here, the mean velocity $u_{avg}$ is the average in the horizontal direction. The maximum averaging velocity appears near the wave surface, gradually decreases as the water depth increases, and approaches zero at the water bottom. The present work agrees with Iafrati (2009) and Yang et al. (2021b).

Fig. 4.16 shows the successive contours of the wave geometry in the three-level subcycling case (Case 3), which uses the consistent scheme. At the beginning of the breaking process, the wave profile steepens, followed by the breaking event. The jet plunges onto the free surface, entrains a large cavity, and generates a splashing sheet at around $t/T = 1.8$. Then, new jet impacts are generated, and different sizes of bubbles are entrained. Next, the entrained bubbles collapse and fragment into small bubbles, which slowly rise back to the surface because of buoyancy. At approximately $t/T = 10.0$, most bubbles and air pockets disappear, and the water surface becomes flat. We note that the breaking process agrees qualitatively with the previous results in Chen et al. (1999); Iafrati (2009), which shows that the consistent scheme in the present adaptive framework is stable and robust for the simulation of breaking waves at a high Reynolds number.

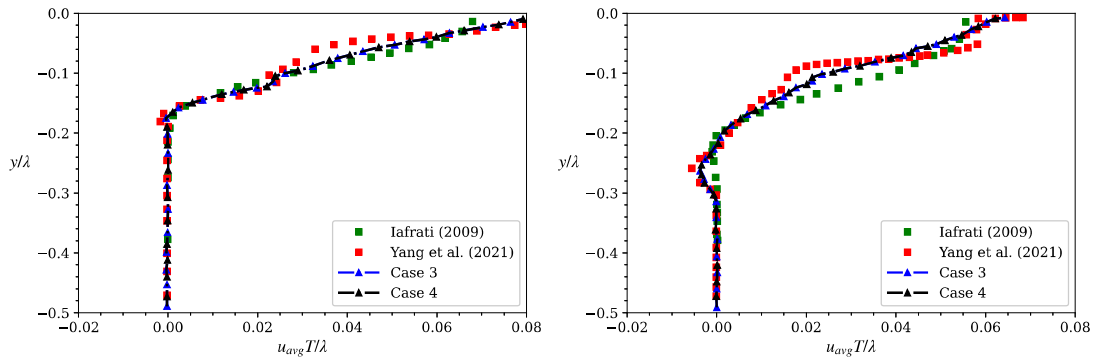

Figure 4.15: Comparison of the mean velocity $u_{avg}$ in the horizontal direction at $t/T = 10$ (left) and $t/T = 15$ (right) among the three-level subcycling case (Case 3), the three-level non-subcycling case (Case 4), and previous results (Iafrati (2009); Yang et al. (2021b)).

### 4.3.6 3D breaking wave

This section investigates a 3D Stokes breaking wave, which is a dynamic and complex problem that is considered computationally expensive. In addition to validating the consistent scheme for 3D problems, another objective of this test is to compare the computational cost among the single-level and multilevel subcycling and non-subcycling cases. The computational domain is $L_x \times L_y \times L_z = \lambda \times \lambda/2 \times \lambda$, where $\lambda = 0.25\,\text{m}$ is the wavelength. The boundary conditions in the $x$ and $z$ directions are periodic, and the free slip boundary condition is applied in the $y$ direction. The initial wave geometry, Weber number, and Reynolds number are the same as those in the 2D breaking wave simulation (Section 4.3.5). Three cases using the consistent scheme are considered, as listed in Table 4.4. For the cases on the multilevel grid, grid cells are refined near the air–water interface.

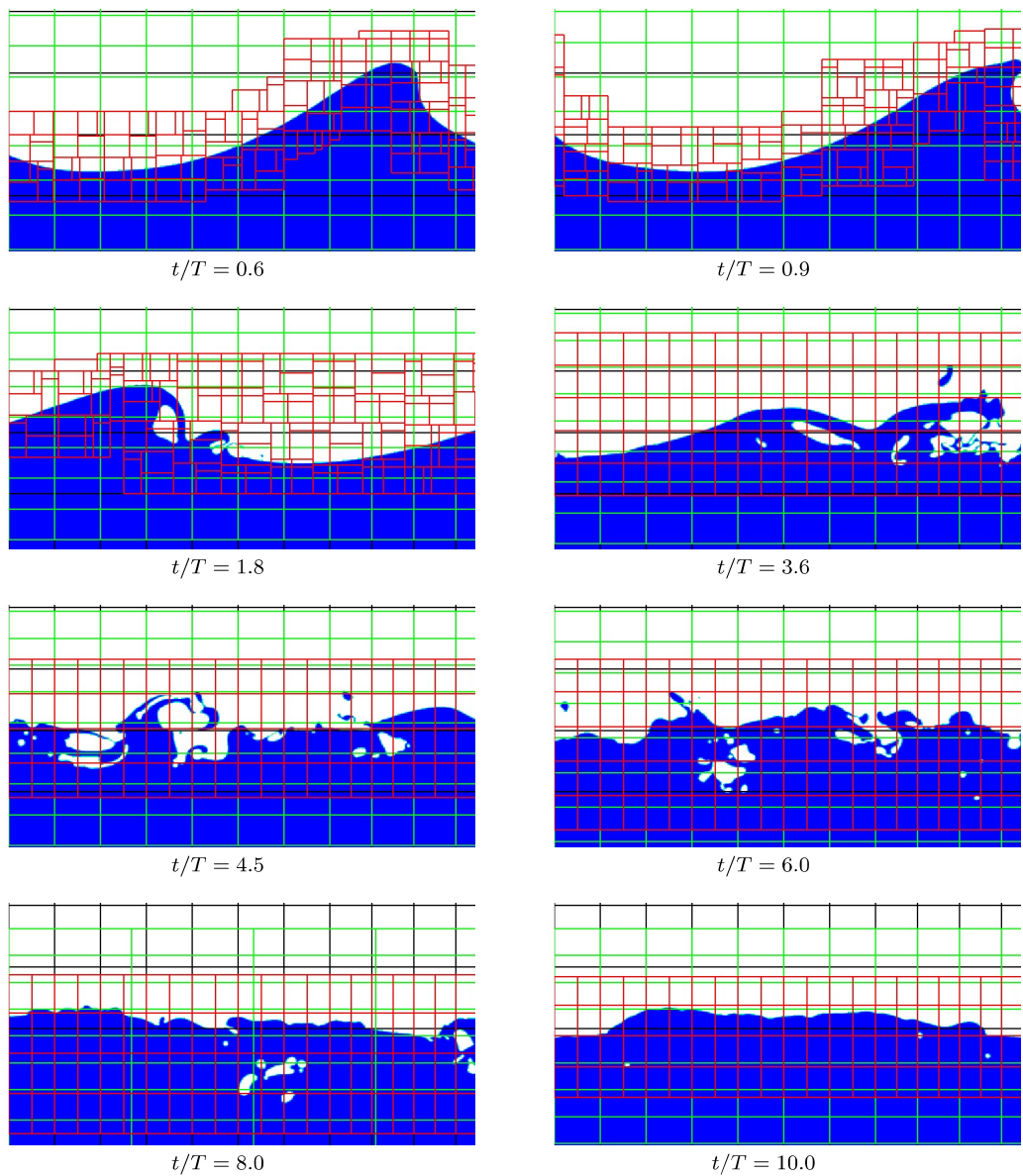

Figure 4.16: Evolution of a 2D plunging breaker obtained from the three-level subcycling case (Case 3) using the consistent scheme. The black, green, and red lines represent patches on levels 0, 1, and 2, respectively.

Table 4.4: Parameters of cases for the 3D breaking wave problem.

| Case No. | Grid number on level 0 | $l_{max}$ | $\Delta t^0$ | Cycling method |
|----------|------------------------|-----------|--------------|----------------|
| 1 | $512 \times 256 \times 512$ | 0 | $1.6 \times 10^{-5}$ | – |
| 2 | $128 \times 64 \times 128$ | 2 | $6.4 \times 10^{-5}$ | Subcycling |
| 3 | $128 \times 64 \times 128$ | 2 | $1.6 \times 10^{-5}$ | Non-subcycling |

Fig. 4.17 shows the evolution of the air–water interface obtained from the three-level subcycling case (Case 2). Grid patches are dynamically refined around the interface as time evolves. At the initial stage of the simulation, the 3D wave geometry is similar to that in the 2D simulation. Then, the plunging jets form at the wave crest and strike the front wave face. At approximately $t/T = 2.0$, air bubbles are entrained by the plunging jets, and 3D structures are observed. Next, upward splashes are generated, and more air is entrained at approximately $t/T = 3.0$. Water droplets and bubbles are also generated ($t/T = 4.0$) by the plunging breaker. From $t/T = 8.0$, the bubbles burst out of the surface, the droplets fall into the water, and the wave surface gradually becomes smooth. The breaking process obtained in our 3D simulation is consistent with the results in Yang et al. (2021b). The left part of Fig. 4.18 compares the total mechanical energy in the above three cases with results in the literature. The time series of the total mechanical energy obtained from the consistent scheme agrees with the results of Wang et al. (2016) and Yang et al. (2021b).

To compare the computational cost for different cases, we profile each case for $t/T = 0 - 0.5$ using 256 CPUs on a Cray XC40/50 (Onyx) system at the U.S. Army Engineer Research and Development Center, excluding the I/O costs. Table 4.5 shows the total number of grid cells for different cases at $t/T = 1.0$. Compared with the adaptive cases with $l_{\max} = 2$ (Cases 2 and 3), the single-level case (Case 1) has nearly five times more cells; i.e., the adaptive refinement considerably reduces the total number of grid cells.

The right part of Fig. 4.18 compares the wall clock time between the single-level case and the multilevel cases for the time range $t/T = 0 - 0.5$. Compared with the single-level case (Case 1), the three-level subcycling case (Case 2) achieves more than a $10\times$ speedup in terms of wall clock time, which significantly reduces the computational cost of the 3D simulation. By comparing the nonsubcycled case (Case 3) with the subcycled case (Case 2), we find that the subcycled case further lowers the computational cost by a factor of 1.8. The reason is that, compared to the non-subcycling method, the subcycling method uses a larger time step size for the coarser levels. In addition to the total wall clock time, the wall clock time spent on some key parts of the algorithm is also documented, including advection, viscous solver, level projection, and synchronization. Among them, the level projection takes the most time, followed by the advection step and the viscous solver step. Therefore, optimization of these three parts is desired in future work. The part denoted as the "Others", including the regridding, the interpolation operations, and the reinitialization step, accounts for only approximately 2% of the total computation time.

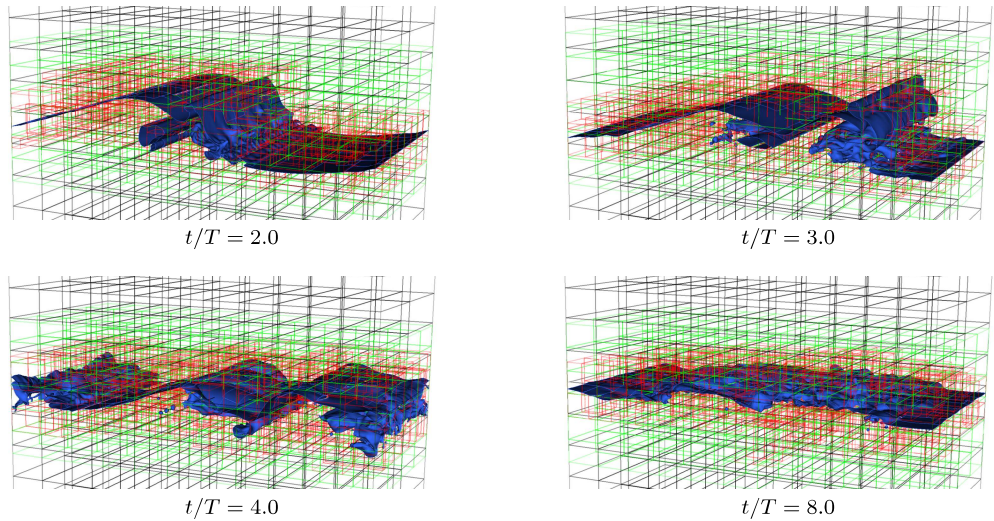

Figure 4.17: Evolution of a 3D plunging breaker obtained from the three-level subcycling case (Case 2) using the consistent scheme. The black, green, and red lines represent patches on levels 0, 1, and 2, respectively.

Table 4.5: Number of grid cells of the cases in the 3D breaking wave problem at $t/T = 0.5$.

| Case No. | level 0 | level 1 | level 2 | Total cells | Total cells normalized by Case 3 |
|----------|---------|---------|---------|-------------|----------------------------------|
| 1 | 67,108,864 | – | – | 67,108,864 | 4.92 |
| 2 | 1,048,576 | 4,194,304 | 8,388,608 | 13,631,488 | 1 |
| 3 | 1,048,576 | 4,194,304 | 8,388,608 | 13,631,488 | 1 |

Finally, we emphasize that only the cases using the consistent scheme are presented here. For 3D breaking wave cases with a relatively low Reynolds number and a low density ratio, we simulated them using both the inconsistent scheme and the consistent scheme and then compared their computational cost. We found that the computational time in the consistent scheme is approximately the same as that in the inconsistent scheme, which indicates that the consistent scheme has negligible computational overhead compared with the inconsistent scheme.

## 4.4   Concluding Remarks

In this chapter, we have developed a consistent adaptive framework for the simulations of incompressible two-phase flows with high density ratios and high Reynolds numbers. It is found that a consistent discretization of convective terms in mass and momentum equations is important to ensure the stability of the simulations for high-density-ratio flows. This is achieved by employing the bounded and monotonic CUI scheme for the discretization of the convective fluxes. Although the inconsistent scheme, in which a Godunov scheme is applied in place of the CUI scheme for momentum transport, could work well in the rising bubble and Rayleigh–Taylor instability cases (Zeng et al., 2022b; Patel and Natarajan, 2015), it does not perform satisfactorily at high density ratios, as evidenced in the convected droplet, dam breaking, and droplet splashing cases, and at high Reynolds numbers, as shown in the breaking wave cases.

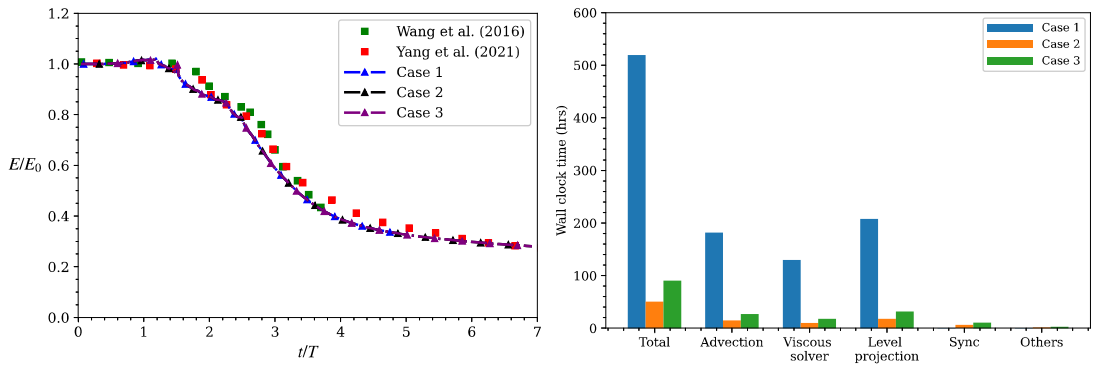

Figure 4.18: Left: comparison of the time evolution of the total energy among the single-level case (Case 1), three-level subcycling case (Case 2), three-level non-subcycling case (Case 3), and previous results (Wang et al. (2016); Yang et al. (2021b)) for the 3D breaking wave problem. Right: comparison of the wall clock time of key advancing steps among the single-level case (Case 1), three-level subcycling case (Case 2), and three-level non-subcycling case (Case 3) for the 3D breaking wave problem.

Different from the previous work (Nangia et al., 2019a; Patel and Natarajan, 2017), a purely collocated grid is employed in the present framework, in which all variables are defined at the center of the grid cells. This design eases the implementation of the CUI scheme because the divergence-free advective velocity $\mathbf{u}_{adv}$ for the cell-centered case can be directly obtained after the MAC projection. Moreover, only one set of interpolation and the averaging scheme is needed when the multilevel grid is involved.

The consistent scheme provides a numerically stable and reasonably accurate solution to realistic multiphase flows, such as breaking waves with a high Reynolds number. It is found that an unphysical spurious thin sheet is generated at the wave crest when the inconsistent scheme is used. When AMR is applied to locally resolve the complex flow physics near the wave surface, the multilevel cases can achieve the same level of accuracy with fewer total grid cells compared with the single-level fine-grid cases. In particular, for the 3D breaking wave problem, the multilevel simulation can capture the evolution of the total mechanical energy accurately with substantial speedup compared with the single-level simulation.

---

**Algorithm 4** Multilevel advancement algorithm

---

1: Initialize $\boldsymbol{u}^0$, $\phi^0$, and $p^0$ on level 0
2: $l \leftarrow 0$
3: **while** refinement criteria are satisfied on level $l$ and $l < l_{max}$ **do**
4:     Regrid the patch hierarchy to obtain level $l + 1$
5:     Initialize $\boldsymbol{u}^0$, $\phi^0$, and $p^0$ on level $l + 1$
6:     $l \leftarrow l + 1$
7: **end while**
8: Determine the time step $\Delta t^{l_{max}}$ on the finest level
9: **if** subcycling method is used **then**
10:     $\Delta t^l = 2^{l_{max}-l} \Delta t^{l_{max}}$ for all $0 \le l < l_{max}$
11: **else**
12:     $\Delta t^l = \Delta t^{l_{max}}$ for all $0 \le l < l_{max}$
13: **end if**
14: **for** $n = 1, n_{max}$ **do**                    ▷ $n_{max}$ is the number of time steps to be simulated
15:     LEVELCYCLING$(0, t_n^0, t_n^0 + \Delta t^0, \Delta t^0)$
16:     Regrid the patch hierarchy, and interpolate $\boldsymbol{u}$, $\phi$, and $p$ onto new patches
17: **end for**
18:
19: **procedure** LEVELCYCLING$(l, t^l, t_{max}^l, \Delta t^l)$
20:     **while** $t^l < t_{max}^l$ **do**
21:         **if** consistent scheme is used **then**
22:             Perform single-level advancement on level $l$ from $t^l$ to $t^l + \Delta t^l$ using the consistent scheme
23:         **else**
24:             Perform single-level advancement on level $l$ from $t^l$ to $t^l + \Delta t^l$ using the inconsistent scheme in chapter 3
25:         **end if**
26:         **if** $l < l_{max}$ **then**
27:             LEVELCYCLING$(l + 1, t^l, t^l + \Delta t^l, \Delta t^{l+1})$
28:         **end if**
29:         $t^l \leftarrow t^l + \Delta t^l$
30:     **end while**
31:     Apply the synchronization step
32: **end procedure**

---

# Chapter 5

# DLM method for FSI problems

## 5.1  Mathematical formulation

In this chapter, we use the distributed Lagrange multiplier (DLM) method to simulate the fluid structure interaction (FSI) problems.

We begin with the governing equations for a fluid-structure system occupying a multilevel Cartesian grid $\Omega \subset \mathbb{R}^d$, where $d = 2, 3$ denotes the spatial dimension. The left part of Fig. 5.1 shows a schematic of two solid bodies on a two-dimensional multilevel Cartesian grid. The momentum and material incompressibility equations are described using a fixed Eulerian coordinate system $\mathbf{x} = (x_1, \dots, x_d) \in \Omega$. The immersed body is described using a Lagrangian coordinate system, where $\mathbf{s} = (s_1, \dots s_d) \in \Omega_c$ denotes the fixed material coordinate system attached to the structure and $\Omega_c \subset \mathbb{R}^d$ is the Lagrangian curvilinear coordinate domain. The position of the structure at time $t$ is $\mathbf{X}(\mathbf{s}, t)$; it occupies a volumetric region $V_b(t) \subset \Omega$. The equations of motion of the coupled fluid-structure system are

$$\rho(\mathbf{x}, t) \left( \frac{\partial \mathbf{u}}{\partial t}(\mathbf{x}, t) + \boldsymbol{\nabla} \cdot (\mathbf{u}(\mathbf{x}, t) \mathbf{u}(\mathbf{x}, t)) \right) = - \boldsymbol{\nabla} p(\mathbf{x}, t) + \boldsymbol{\nabla} \cdot \left[ \mu(\mathbf{x}, t) \left( \boldsymbol{\nabla} \mathbf{u}(\mathbf{x}, t) + \boldsymbol{\nabla} \mathbf{u}(\mathbf{x}, t)^T \right) \right]$$
$$+ \wp(\mathbf{x}, t) \mathbf{g} + \mathbf{f}_c(\mathbf{x}, t), \tag{5.1}$$

$$\boldsymbol{\nabla} \cdot \mathbf{u}(\mathbf{x}, t) = 0, \tag{5.2}$$

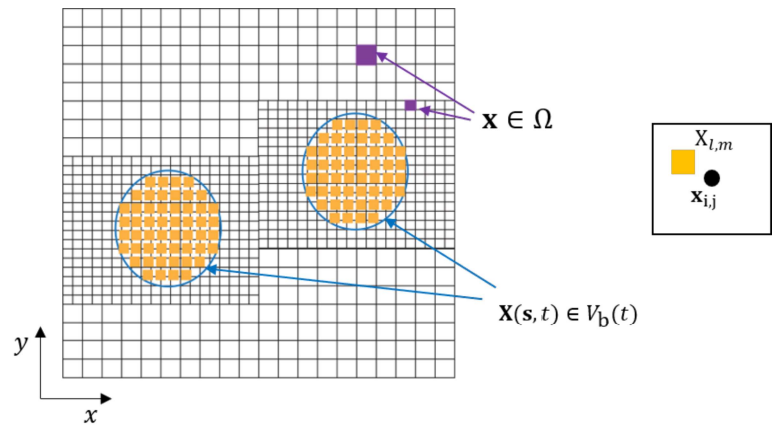

Figure 5.1: Left: two solid bodies (confined by blue lines) on a multilevel Cartesian grid. The Eulerian grid cells (■, purple) discretize the $\Omega$ region, and the Lagrangian markers (■, orange) discretize the $V_{\mathrm{b}}(t)$ region. Right: schematic representation of a single Cartesian grid cell. The Eulerian quantities are defined at the cell center (●, black); the Lagrangian quantities are defined on the marker points (■, orange), which are free to move on the Eulerian grid.

$$\mathbf{f}_{\mathrm{c}}(\mathbf{x}, t) = \int_{V_{\mathrm{b}}(t)} \mathbf{F}_{\mathrm{c}}(\mathbf{s}, t) \delta(\mathbf{x} - \mathbf{X}(\mathbf{s}, t)) \mathrm{d}\mathbf{s}, \tag{5.3}$$

$$\frac{\partial \mathbf{X}}{\partial t}(\mathbf{s}, t) = \mathbf{U}(\mathbf{s}, t), \tag{5.4}$$

$$\mathbf{U}(\mathbf{s}, t) = \int_{V_{\mathrm{b}}(t)} \mathbf{u}(\mathbf{x}, t) \delta(\mathbf{x} - \mathbf{X}(\mathbf{s}, t)) \mathrm{d}\mathbf{x}. \tag{5.5}$$

Here, $\mathbf{u}(\mathbf{x}, t)$ is the Eulerian velocity of the coupled fluid–structure system, $p(\mathbf{x}, t)$ is the pressure, $\rho(\mathbf{x}, t)$ is the Eulerian density field (fluid density in the fluid region $\Omega_{\mathrm{f}} = \Omega \setminus \Omega_{\mathrm{c}}$ and solid density in the solid region $\Omega_{\mathrm{c}}$), $\mu(\mathbf{x}, t)$ is the dynamic viscosity of the fluid–structure system, and $\wp(\mathbf{x}, t)$ is the modified density, which will be detailed later. Because $\rho(\mathbf{x}, t)$, $\wp(\mathbf{x}, t)$, and $\mu(\mathbf{x}, t)$ are allowed to change spatially and temporally in this chapter, the solid structure can be heavier (or lighter) and more viscous (or less viscous) than the surrounding fluid. The gravitational acceleration is written as $\mathbf{g} = (g_1, \ldots, g_d)$. In Eq. (5.1), $\mathbf{f}_{\mathrm{c}}(\mathbf{x}, t)$ represents the Eulerian force density, which accounts for the presence of the solid in the domain. $\delta(\mathbf{x}) = \Pi_{i=1}^{d} \delta(x_i)$ represents the $d$-dimensional Dirac delta function, which is employed to facilitate the information exchange between the Eulerian quantity and Lagrangian quantity. Specifically, Eq. (5.3) converts the Lagrangian force density $\mathbf{F}_c(\mathbf{s}, t)$ to an equivalent Eulerian force density $\mathbf{f}_c(\mathbf{x}, t)$, in an operation that is referred to as *force spreading*. Eq. (5.5) maps the Eulerian velocity $\mathbf{u}(\mathbf{x}, t)$ to the Lagrangian marker velocity $\mathbf{U}(\mathbf{s}, t)$, in an operation that is referred to as *velocity interpolation*. Constrained by the no-slip boundary condition at the fluid-solid interface, the velocity of a Lagrangian marker $\mathbf{U}(\mathbf{s}, t)$ follows the local fluid velocity. For notational convenience, we denote the force spreading operation in Eq. (5.3) as

$$\mathbf{f}_c = \boldsymbol{\mathcal{S}}[\mathbf{X}]\mathbf{F}, \tag{5.6}$$

where $\boldsymbol{\mathcal{S}}[\mathbf{X}]$ is the force spreading operator. Similarly, the velocity interpolation operation in Eq. (5.5) is written in shorthand notation as

$$\mathbf{U} = \boldsymbol{\mathcal{J}}[\mathbf{X}]\mathbf{u}, \tag{5.7}$$

where $\boldsymbol{\mathcal{J}}[\mathbf{X}]$ is the velocity interpolation operator. As shown in Peskin (2002); Nangia et al. (2019b), the Lagrangian-Eulerian coupling operators conserve energy as long as $\boldsymbol{\mathcal{S}}$ and $\boldsymbol{\mathcal{J}}$ are adjoint.

## 5.2 Spatial discretization on a multilevel adaptive grid

This section describes the spatial discretization of Eulerian and Lagrangian quantities and their coupling on a multilevel adaptive grid. The discretization of two-dimensional spatial operators is presented in this section. The discrete version of the three-dimensional spatial operators can be defined analogously.

### 5.2.1 Eulerian discretization

All Eulerian variables, including the fluid velocity $\mathbf{u}$, pressure $p$, and LS functions $\phi$ and $\psi$, are defined at cell centers following the collocated grid variable arrangement setup. The discretization of these Eulerian variables are in chapter 2.

### 5.2.2  Lagrangian discretization

In the IB approach, the Lagrangian markers that define the solid structure are free to move on the background Eulerian grid. Following the convention of Nangia et al. (Nangia et al., 2019b), the Lagrangian markers are indexed by $(q, m)$ with mesh spacings $(\Delta s_1, \Delta s_2)$ in the two curvilinear directions. Because the present work only considers rigid body kinematic constraints, explicit connectivity information among the marker points is not required. Moreover, the Lagrangian markers are placed on the finest grid level $l_{\max}$, and therefore, the relevant Lagrangian quantities, including the position $(\mathbf{X})_{q,m}$, velocity $(\mathbf{U})_{q,m}$, and force $(\mathbf{F})_{q,m}$ of the marker points, are defined only on the finest grid level.

### 5.2.3  Lagrangian-Eulerian coupling

The Eulerian and Lagrangian quantities are transformed through two coupling operators described in Section 5.1: the force spreading operator $\boldsymbol{\mathcal{S}}[\mathbf{X}]$ and the velocity interpolation operator $\boldsymbol{\mathcal{J}}[\mathbf{X}]$. In this work, a canonical $d$-dimensional delta function with the tensor product form $\delta_h(\mathbf{x}) = \prod_{i=1}^{d} \delta_h(x_i)$ is applied to approximate the coupling operators. In each dimension, $\delta_h(x_i)$ is defined as $\delta_h(x_i) = \frac{1}{h}\varphi_4\left(\frac{x_i}{h}\right)$, where $\varphi_4(r)$ is the four-point IB kernel of Peskin (1972, 2002) given by

$$\varphi_4(r) = \begin{cases} \frac{1}{8}(3 - 2|r| + \sqrt{1 + 4|r| - 4r^2}), & 0 \leqslant |r| < 1, \\ \frac{1}{8}(5 - 2|r| - \sqrt{-7 + 12|r| - 4r^2}), & 1 \leqslant |r| < 2, \\ 0, & 2 \leqslant |r|. \end{cases} \tag{5.8}$$

The solid structure is completely enclosed by the finest level $l_{\max}$, and the Lagrangian markers within the solid region $\Omega_{\mathrm{b}}$ are identified by the marker set $\mathcal{M}_{\mathrm{b}}$. Given a Lagrangian force density $\mathbf{F} = (F_1, F_2)$ defined on $\mathcal{M}_{\mathrm{b}}$, the corresponding Eulerian force density $\mathbf{f}^{l_{\max}} = \left(f_1^{l_{\max}}, f_2^{l_{\max}}\right)$ is obtained through the force spreading operation (Eq. (5.6)) as

$$\begin{aligned} (f_1)_{i,j}^{l_{\max}} &= \sum_{(q,m) \in \mathcal{M}_b} (F_1)_{q,m}\, \delta_{h^{l_{\max}}}\left(\mathbf{x}_{i,j}^{l_{\max}} - \mathbf{X}_{q,m}\right) \Delta s_1 \Delta s_2, \\ (f_2)_{i,j}^{l_{\max}} &= \sum_{(q,m) \in \mathcal{M}_b} (F_2)_{q,m}\, \delta_{h^{l_{\max}}}\left(\mathbf{x}_{i,j}^{l_{\max}} - \mathbf{X}_{q,m}\right) \Delta s_1 \Delta s_2. \end{aligned} \tag{5.9}$$

Likewise, with $\mathbf{u}^{l_{\max}} = \left(u_1^{l_{\max}}, u_2^{l_{\max}}\right)$ being the Eulerian velocity on the finest level $l_{\max}$ and $\mathbf{U} = (U_1, U_2)$ being the velocity of the Lagrangian markers, the velocity interpolation operation (Eq. (5.7)) can be written as

$$
\begin{aligned}
(U_1)_{q,m} &= \sum_{(i,j)} (u_1)_{i,j}^{l_{\max}} \, \delta_{h^{l_{\max}}} \left(\mathbf{x}_{i,j}^{l_{\max}} - \mathbf{X}_{q,m}\right) \Delta x^{l_{\max}} \Delta y^{l_{\max}}, \\
(U_2)_{q,m} &= \sum_{(i,j)} (u_2)_{i,j}^{l_{\max}} \, \delta_{h^{l_{\max}}} \left(\mathbf{x}_{i,j}^{l_{\max}} - \mathbf{X}_{q,m}\right) \Delta x^{l_{\max}} \Delta y^{l_{\max}}.
\end{aligned}
\tag{5.10}
$$

## 5.3    Time advancement

In this section, we use a level-by-level method (Martin and Colella, 2000; Martin et al., 2008) for the time advancement of variables on a multilevel grid. As our multilevel advancement algorithm (Section 5.3.4) is based on the single-level advancement method, we introduce a single-level advancement algorithm (Section 5.3.1) in which different types of kinematic constraints (Section 5.3.2) are considered and the hydrodynamic force and torque acting on the immersed object are calculated (Section 5.3.3) if needed. The multilevel advancement algorithm, which combines the single-level advancement algorithm with the subcycling and non-subcycling methods, is discussed in Section 5.3.4. To match the data across multiple levels, various synchronization operations are performed every time a finer level catches up with a coarser level; the details of the synchronization operations are discussed in Section 5.3.4.

### 5.3.1    Single-level advancement

Our numerical method uses a time-splitting approach, in which we first solve the momentum equation Eq. (5.1) using the approximate projection method (Rider, 1995, 1998; Almgren et al., 2000) to enforce the incompressibility condition Eq. (5.2) before correcting the Eulerian velocity to enforce the rigid constraint in the solid domain. At time $t^n$, we are given the collocated Eulerian velocity $\mathbf{u}^n$ and time-staggered pressure $p^{n-1/2}$ (Almgren et al., 1998; Sussman et al., 1999; Bhalla et al., 2013). For the rigid body, the mass center position $\mathbf{X}_r^n$, velocity $\mathbf{U}_r^n$, and angular velocity $\mathbf{W}_r^n$ are also known at time $t^n$. Our objective is to obtain the updated fluid velocity $\mathbf{u}^{n+1}$, pressure $p^{n+1/2}$, and updated rigid body quantities $\mathbf{X}_r^{n+1}$, $\mathbf{U}_r^{n+1}$, and $\mathbf{W}_r^{n+1}$ at the next time level $t^{n+1}$. The details of the single-level time advancement are given below.

1. The LS function $\psi^n$ is calculated based on $\mathbf{X}_r^n$ and the geometry of the solid body, and the LS function $\phi^n$ are calculated based on the location of the gas-liquid interface, both at time $t^n$. To capture the movement of the fluid-solid and gas-liquid interfaces, $\psi^n$ and $\phi^n$ are advanced to $\psi^{n+1}$ and $\phi^{n+1}$ using

$$\psi^{n+1} = \psi^n - \Delta t \left[\boldsymbol{\nabla} \cdot (\mathbf{u}\psi)\right]^{n+1/2}, \tag{5.11}$$

$$\phi^{n+1} = \phi^n - \Delta t \left[\boldsymbol{\nabla} \cdot (\mathbf{u}\phi)\right]^{n+1/2}, \tag{5.12}$$

where the advection terms $\left[\boldsymbol{\nabla} \cdot (\mathbf{u}\psi)\right]^{n+1/2}$ and $\left[\boldsymbol{\nabla} \cdot (\mathbf{u}\phi)\right]^{n+1/2}$ are computed using the Godunov scheme (Colella, 1990; Bell et al., 1989; Almgren et al., 1998; Sussman et al., 1999), as described in Appendix A.2. The midpoint values of $\psi$ and $\phi$ are calculated as $\psi^{n+\frac{1}{2}} = (\psi^{n+1} + \psi^n)/2$ and $\phi^{n+\frac{1}{2}} = (\phi^{n+1} + \phi^n)/2$, respectively.

2. To set the midpoint density field $\rho^{n+\frac{1}{2}}$ appearing on the left-hand side of the momentum equation (Eq. (5.1)), two Heaviside functions are defined as

$$\widetilde{H}^{\text{flow}}\left(\phi^{n+\frac{1}{2}}\right) = \begin{cases} 0, & \phi^{n+\frac{1}{2}} < -n_{\text{cells}}\Delta x \\ \frac{1}{2}\left(1 + \frac{1}{n_{\text{cells}}\Delta x}\phi^{n+\frac{1}{2}} + \frac{1}{\pi}\sin\left(\frac{\pi}{n_{\text{cells}}\Delta x}\phi^{n+\frac{1}{2}}\right)\right), & \left|\phi^{n+\frac{1}{2}}\right| \leq n_{\text{cells}}\Delta x \\ 1, & \text{otherwise}, \end{cases} \tag{5.13}$$

$$\widetilde{H}^{\text{body}}\left(\psi^{n+\frac{1}{2}}\right) = \begin{cases} 0, & \psi^{n+\frac{1}{2}} < -n_{\text{cells}}\Delta x \\ \frac{1}{2}\left(1 + \frac{1}{n_{\text{cells}}\Delta x}\psi^{n+\frac{1}{2}} + \frac{1}{\pi}\sin\left(\frac{\pi}{n_{\text{cells}}\Delta x}\psi^{n+\frac{1}{2}}\right)\right), & \left|\psi^{n+\frac{1}{2}}\right| \leq n_{\text{cells}}\Delta x \\ 1, & \text{otherwise}, \end{cases} \tag{5.14}$$

where $n_{\text{cells}}$ is the smearing width of the fluid-solid or gas-liquid interface assuming a uniform grid spacing in all directions, i.e., $\Delta x = \Delta y$. The density $\rho^{n+\frac{1}{2}}$ is then set via a two-step process (Nangia et al., 2019b; Bhalla et al., 2020) as

$$\widetilde{\rho}^{n+\frac{1}{2}} = \rho_{\text{g}} + (\rho_{\text{l}} - \rho_{\text{g}})\,\widetilde{H}^{\text{flow}}\left(\phi^{n+\frac{1}{2}}\right), \tag{5.15}$$

$$\rho^{n+\frac{1}{2}} = \rho_{\text{s}} + \left(\widetilde{\rho}^{n+\frac{1}{2}} - \rho_{\text{s}}\right)\widetilde{H}^{\text{body}}\left(\psi^{n+\frac{1}{2}}\right), \tag{5.16}$$

where $\rho_{\mathrm{l}}$, $\rho_{\mathrm{g}}$, and $\rho_{\mathrm{s}}$ are the liquid density, gas density, and solid density, respectively. Depending on the type of kinematic constraint (Section 5.3.2), the modified density $\wp$ appearing in the gravitational term of the momentum equation is set to

$$\wp = \begin{cases} \rho_{\mathrm{l}}, & \text{for prescribed motion of the solid} \\ \rho_{\mathrm{s}}, & \text{for free motion and prescribed shape of the solid} \end{cases} \tag{5.17}$$

so that the inertia and buoyancy effects due to the weight of the solid are properly included in the FSI simulation without producing spurious gravitational currents (Nangia et al., 2019b; Bhalla et al., 2020).

3. Neither $\psi$ nor $\phi$ are guaranteed to retain the signed distance property after the advections in Eqs. (5.11) and (5.12), even if they are initialized as the signed distance function at the beginning of the simulation. For the structures with relatively simple geometries that are considered in this work, the solid LS function $\psi$ can be directly computed using the centroid information of the body. For immersed bodies with complex geometries, the constructive solid geometry (CGS) and/or R-functions (Shapiro, 2007) can be employed to determine the analytical expressions for $\psi$. The fluid LS function $\phi$ is reinitialized by computing the steady-state solution to the Hamilton-Jacobi equation

$$\frac{\partial d}{\partial \tau} = S(\phi)(1 - |\nabla d|), \tag{5.18}$$

where

$$S(\phi) = 2\left(H(\phi) - 1/2\right). \tag{5.19}$$

The initial condition of Eq. (5.18) is

$$d(\boldsymbol{x}, \tau = 0) = \phi^{n+1}(\boldsymbol{x}), \tag{5.20}$$

where $\tau$ is the pseudotime for iterations. A classical second-order Runge-Kutta (RK) method is applied for the pseudotime advancement of $d$ (Sussman et al., 1999; Sussman and Smereka, 1997), which helps minimize the volume change of each fluid phase and ensure mass conservation (Sussman et al., 1999; Sussman and Smereka, 1997; Sussman and Fatemi, 1999). The flow LS function $\phi$ is updated by $d$ after this pseudotime advancement.

4. The intermediate velocity $\widetilde{\mathbf{u}}^{*,n+1}$ is solved semi-implicitly without considering the rigidity constraint as

$$\rho^{n+\frac{1}{2}}\left(\frac{\widetilde{\mathbf{u}}^{*,n+1} - \mathbf{u}^n}{\Delta t} + \boldsymbol{\nabla} \cdot (\mathbf{uu})^{n+\frac{1}{2}}\right) = -\boldsymbol{\nabla}p^{n-\frac{1}{2}} + \frac{1}{2}\left(\boldsymbol{\nabla} \cdot \mu(\psi^{n+1}, \phi^{n+1})\boldsymbol{\nabla}\widetilde{\mathbf{u}}^{*,n+1} + \boldsymbol{\nabla} \cdot \mu(\psi^n, \phi^n)\boldsymbol{\nabla}\mathbf{u}^n\right)$$
$$+ \wp^{n+\frac{1}{2}}\mathbf{g}, \tag{5.21}$$

where the convective term $\boldsymbol{\nabla} \cdot (\mathbf{uu})^{n+\frac{1}{2}}$ is calculated using the Godunov scheme (Appendix A.2). In this work, the viscosity $\mu(\psi^{n+1}, \phi^{n+1})$ or $\mu(\psi^n, \phi^n)$ does not depend on the solid LS function $\psi$ and is set to the surrounding fluid viscosity, i.e., $\mu^{n+1} = \mu_{\mathrm{f}}(\phi^{n+1})$, following Nangia et al. (2019b).

5. After obtaining the intermediate velocity, a level projection is applied to obtain the updated velocity $\widetilde{\mathbf{u}}^{n+1}$ and pressure $p^{n+\frac{1}{2}}$ fields. An auxiliary variable $\boldsymbol{V}$ is first calculated by

$$\boldsymbol{V} = \frac{\widetilde{\mathbf{u}}^{*,n+1}}{\Delta t} + \frac{1}{\rho^{n+1/2}}\boldsymbol{\nabla}p^{n-\frac{1}{2}}. \tag{5.22}$$

Next, $\boldsymbol{V}$ is projected on the divergence-free velocity field to obtain the updated pressure $p^{n+1/2}$ via

$$L^{cc,l}_{\rho^{n+1/2}}p^{n+1/2} = \boldsymbol{\nabla} \cdot \boldsymbol{V}, \tag{5.23}$$

where $L^{cc,l}_{\rho^{n+1/2}}p^{n+1/2}$ is the density-weighted approximation to $\boldsymbol{\nabla} \cdot (1/\rho^{n+1/2}\boldsymbol{\nabla}p^{n+1/2})$ on level $l$. The divergence-free velocity $\widetilde{\mathbf{u}}^{n+1}$ on level $l$ is then obtained as

$$\widetilde{\mathbf{u}}^{n+1} = \Delta t\left(\boldsymbol{V} - \frac{1}{\rho^{n+1/2}}\boldsymbol{\nabla}p^{n+1/2}\right). \tag{5.24}$$

We note that the intermediate velocity on each level is projected without the pressure gradient term, as the pressure gradient term is subtracted in Eq. (5.22). This step reduces the accumulation of pressure errors and produces a more stable algorithm (Rider, 1995; Guy and Fogelson, 2005). Moreover, the approximate projection approach effectively removes the issue of the pressure checker-boarding problem that appears on collocated grids (Martin and Colella, 2000; Almgren et al., 2000; Zeng et al., 2022b;

Zeng and Shen, 2019).

6. The updated velocity $\widetilde{\mathbf{u}}^{n+1}$ satisfies the incompressibility condition but needs to be corrected to satisfy the constraints of motions of the rigid body within the solid region $V_{\mathrm{b}}(t)$. To achieve this, the Lagrangian velocity $(\mathbf{U}_{\mathrm{b}})_{q,m}^{n+1}$ and the marker points position $\mathbf{X}_{q,m}^{n+1}$ need to be approximated. Because these approximations may vary based on the kinematic constraints, for now, we assume that they are known variables; the detailed steps for calculating $(\mathbf{U}_{\mathrm{b}})_{q,m}^{n+1}$ and $\mathbf{X}_{q,m}^{n+1}$ are presented in Section 5.3.2.

To proceed, the difference between the Lagrangian velocity $(\mathbf{U}_{\mathrm{b}})_{q,m}^{n+1}$ and the background Eulerian velocity $\widetilde{\mathbf{u}}^{n+1}$ is calculated via the velocity interpolation operation as

$$(\Delta \mathbf{U}_{\mathrm{c}})_{q,m}^{n+1} = \begin{cases} (\mathbf{U}_{\mathrm{b}})_{q,m}^{n+1} - \left( \boldsymbol{\mathcal{J}} \left[ \mathbf{X}_{q,m}^{n+1} \right] \widetilde{\mathbf{u}}^{n+1} \right)_{q,m}, & \text{for } (q,m) \in V_{\mathrm{b}}(t), \\ \mathbf{0}. & \text{for } (q,m) \notin V_{\mathrm{b}}(t). \end{cases} \quad (5.25)$$

This velocity difference is then employed to approximate the Lagrangian and Eulerian constraint forces as

$$\mathbf{F}_{q,m}^{n+1} = \frac{\rho_{\mathrm{s}}}{\Delta t} (\Delta \mathbf{U}_{\mathrm{c}})_{q,m}^{n+1}, \quad (5.26)$$

$$\mathbf{f}_{\mathrm{c}}^{n+1} = \boldsymbol{\mathcal{S}} \left[ \mathbf{X}_{q,m}^{n+1} \right] \mathbf{F}_{q,m}^{n+1}. \quad (5.27)$$

The Eulerian velocity field is corrected by the Eulerian constraint force as

$$\mathbf{u}^{n+1} = \widetilde{\mathbf{u}}^{n+1} + \frac{\Delta t}{\rho_{\mathrm{s}}} \mathbf{f}_{\mathrm{c}}^{n+1}, \quad (5.28)$$

and the solid effects are properly included in the fluid-solid system (Nangia et al., 2019b).

## 5.3.2  Types of kinematic constraints

The approximation to the constrained Lagrangian velocity $(\mathbf{U}_{\mathrm{b}})_{q,m}^{n+1}$ and position $\mathbf{X}_{q,m}^{n+1}$ depends on the type of kinematic constraint in the FSI. In this section, we consider three types of kinematic constraints: prescribed motion of the structure, free motion of the structure, and prescribed shape of the structure.

**Prescribed motion**

If the motion of the structure is prescribed, then the velocity and position of the body are known *a priori* and not influenced by the surrounding fluid. Thus, the centroid position $\mathbf{X}_r^n$, centroid velocity $\mathbf{U}_r^{n+1}$, and angular velocity $\mathbf{W}_r^{n+1}$ of the body are given. The constrained Lagrangian velocity $(\mathbf{U}_b)_{q,m}^{n+1}$ of the markers is calculated as

$$(\mathbf{U}_b)_{q,m}^{n+1} = \mathbf{U}_r^{n+1} + \mathbf{W}_r^{n+1} \times \mathbf{R}_{q,m}^n, \tag{5.29}$$

where $\mathbf{R}_{q,m}^n = \mathbf{X}_{q,m}^n - \mathbf{X}_r^n$. Since the marker position $\mathbf{X}_{q,m}^n$ is already known from the calculation at the previous time step, the new position $\mathbf{X}_{q,m}^{n+1}$ of the marker points is updated using the midpoint scheme as

$$\mathbf{X}_{q,m}^{n+1} = \mathbf{X}_{q,m}^n + \frac{\Delta t}{2} \left( (\mathbf{U}_b)_{q,m}^n + (\mathbf{U}_b)_{q,m}^{n+1} \right). \tag{5.30}$$

**Free motion**

In contrast to the prescribed kinematics case, the motion of a freely moving body is influenced by the surrounding fluid. To account for this two-way coupled interaction, the linear and angular momentum of the fluid occupying the solid domain is redistributed as rigid body momentum, which provides an estimate of the centroid velocities $\mathbf{U}_r^{n+1}$ and $\mathbf{W}_r^{n+1}$ at the new time level. This momentum projection step is carried out using the principle of conservation of linear and angular momentum in the solid domain and is written as

$$\mathrm{M}_b \mathbf{U}_r^{n+1} = \sum_{\mathbf{X}_{q,m} \in V_b} \rho_s \left( \boldsymbol{\mathcal{J}}_h \left[ \mathbf{X}^{n+\frac{1}{2}} \right] \widetilde{\mathbf{u}}^{n+1} \right)_{q,m} \Delta s_1 \Delta s_2 + \mathbf{F}_{\mathrm{ext}} \, \Delta t, \tag{5.31}$$

$$\mathbf{I}_b \mathbf{W}_r^{n+1} = \sum_{\mathbf{X}_{q,m} \in V_b} \rho_s \mathbf{R}_{q,m}^{n+\frac{1}{2}} \times \left( \boldsymbol{\mathcal{J}}_h \left[ \mathbf{X}^{n+\frac{1}{2}} \right] \widetilde{\mathbf{u}}^{n+1} \right)_{q,m} \Delta s_1 \Delta s_2, \tag{5.32}$$

where

$$\mathrm{M}_b = \sum_{\mathbf{X}_{q,m} \in V_b} \rho_s \Delta s_1 \Delta s_2 \tag{5.33}$$

is the mass of the solid body and

$$\mathbf{I}_{\mathrm{b}} = \sum_{\mathbf{X}_{q,m} \in V_{\mathrm{b}}} \rho_{\mathrm{s}} \left( \mathbf{R}_{q,m}^{n+\frac{1}{2}} \cdot \mathbf{R}_{q,m}^{n+\frac{1}{2}} \, \mathbf{I} - \mathbf{R}_{q,m}^{n+\frac{1}{2}} \otimes \mathbf{R}_{q,m}^{n+\frac{1}{2}} \right) \tag{5.34}$$

is the moment of inertia tensor, with $\mathbf{I}$ being the $d$-dimensional identity tensor. In Eq. (5.31), $\mathbf{F}_{\mathrm{ext}}$ is utilized to account for the effects of external (nonhydrodynamic) forces, such as those provided by springs and dampers that may be attached to a solid body. The centroid position $\mathbf{X}_r^{n+1}$ is then updated using the midpoint scheme as

$$\mathbf{X}_r^{n+1} = \mathbf{X}_r^n + \frac{\Delta t}{2}(\mathbf{U}_r^{n+1} + \mathbf{U}_r^n). \tag{5.35}$$

The constrained Lagrangian velocity $(\mathbf{U}_{\mathrm{b}})_{q,m}^{n+1}$ and position $\mathbf{X}_{q,m}^{n+1}$ of the marker points are updated using Eqs. (5.29) and (5.30), respectively, in a manner similar to the prescribed motion case.

**Prescribed shape**

In some cases of freely moving bodies, particularly those encountered in aquatic locomotion, the shape of the body changes with time. The deformation of the body can be prescribed as shape mapping $\boldsymbol{\chi} = \boldsymbol{\chi}(\mathbf{s}, t)$. The deformation velocity of the body $\mathbf{U}_{\mathrm{k}}$ can be obtained as $\mathbf{U}_{\mathrm{k}} = \partial \boldsymbol{\chi}(\mathbf{s}, t)/\partial t$. In general, the net linear and angular momentum of the deformation velocity is nonzero. A direct use of such deformation velocity in the FSI algorithm would manifest as a spurious external force or torque acting on the freely swimming body. Therefore, the additional momentum due to the prescribed deformation velocity must be removed from the equations of motion to conserve the system momentum. To project the deformation kinematics onto a space of net zero momentum, the linear and angular momentum redistribution Eqs. (5.31) and (5.32) are modified to remove the extraneous momentum as

$$\mathrm{M}_{\mathrm{b}} \mathbf{U}_{\mathrm{r}}^{n+1} = \sum_{\mathbf{X}_{q,m} \in V_{\mathrm{b}}} \rho_{\mathrm{s}} \left( (\boldsymbol{\mathcal{J}}_h \left[ \mathbf{X}^{n+\frac{1}{2}} \right] \widetilde{\mathbf{u}}^{n+1})_{q,m} - (\mathbf{U}_{\mathrm{k}})_{q,m}^{n+1} \right) \Delta s_1 \Delta s_2, \tag{5.36}$$

$$\mathbf{I}_{\mathrm{b}}\mathbf{W}_{\mathrm{r}}^{n+1} = \sum_{\mathbf{X}_{q,m} \in V_{\mathrm{b}}} \rho_{\mathrm{s}}\mathbf{R}_{q,m}^{n+\frac{1}{2}} \times \left( (\boldsymbol{\mathcal{J}}_{h}\left[\mathbf{X}^{n+\frac{1}{2}}\right]\widetilde{\mathbf{u}}^{n+1})_{q,m} - (\mathbf{U}_{\mathrm{k}})_{q,m}^{n+1} \right) \Delta s_{1}\Delta s_{2}. \quad (5.37)$$

In these equations, it is assumed that the deformation velocity $(\mathbf{U}_{\mathrm{k}})_{q,m}^{n+1}$ has been obtained from the prescribed body shape $\boldsymbol{\chi}(\mathbf{s}, t)$. In addition, the deformation velocity $(\mathbf{U}_{\mathrm{k}})_{q,m}^{n+1}$ needs to be considered while updating the constrained Lagrangian velocity $(\mathbf{U}_{\mathrm{b}})_{q,m}^{n+1}$ as

$$(\mathbf{U}_{\mathrm{b}})_{q,m}^{n+1} = \mathbf{U}_{\mathrm{r}}^{n+1} + \mathbf{W}_{\mathrm{r}}^{n+1} \times \mathbf{R}_{q,m}^{n} + (\mathbf{U}_{\mathrm{k}})_{q,m}^{n+1}. \quad (5.38)$$

The position of the marker points $\mathbf{X}_{q,m}^{n+1}$ is updated using Eq. (5.30).

### 5.3.3 Hydrodynamic force and torque calculation

In the DLM/IB method, the fluid-structure interaction is handled implicitly. Therefore, there is no need to iterate between fluid domain integrators and solid domain integrators to maintain stability, which is a strict requirement for some sharp-interface immersed boundary methods (Yang and Balaras, 2006; Borazjani et al., 2008; Calderer et al., 2014). Moreover, there is no need to explicitly evaluate the hydrodynamic force and torque acting on the body to determine its motion. These features make the DLM/IB scheme more efficient than some sharp-interface immersed boundary methods that require velocity interpolation and pressure reconstruction around the solid surface to calculate the required hydrodynamic forces and torques. However, if needed, the hydrodynamic force and torque acting on the immersed object in the DLM/IB method can still be calculated as a postprocessing step using the following equations (Nangia et al., 2019b,a):

$$\boldsymbol{\mathcal{F}}^{n+1} = \sum_{\mathbf{X}_{q,m} \in V_{\mathrm{b}}} \rho_{\mathrm{s}} \left[ \frac{(\mathbf{U}_{\mathrm{b}})_{q,m}^{n+1} - (\mathbf{U}_{\mathrm{b}})_{q,m}^{n}}{\Delta t} - \frac{\Delta \mathbf{U}_{q,m}^{n+1}}{\Delta t} \right] \Delta s_{1}\Delta s_{2}, \quad (5.39)$$

$$\boldsymbol{\mathcal{M}}^{n+1} = \sum_{\mathbf{X}_{q,m} \in V_{\mathrm{b}}} \rho_{\mathrm{s}}\mathbf{R}_{q,m}^{n+1} \times \left[ \frac{(\mathbf{U}_{\mathrm{b}})_{q,m}^{n+1} - (\mathbf{U}_{\mathrm{b}})_{q,m}^{n}}{\Delta t} - \frac{\Delta \mathbf{U}_{q,m}^{n+1}}{\Delta t} \right] \Delta s_{1}\Delta s_{2}. \quad (5.40)$$

### 5.3.4 Multilevel advancement

This section describes the extension of the single-level advancement algorithm presented in Section 5.3.1 to the multilevel advancement algorithm using subcycling and non-subcycling methods (Section 5.3.4). A force averaging scheme is proposed to average the Eulerian forces from the finest level to the coarser levels, which conserves the momentum of the system at a discrete level. A synchronization step is then applied to match the variables on multiple levels, which provides a better representation of the composite solution on multiple levels. Next, a multilevel initialization of the fluid and solid fields is introduced. Finally, a summary of the multilevel advancement algorithm is provided.

**Subcycling and non-subcycling methods**

To advance variables on a multilevel grid, we consider two types of cycling methods, namely, the subcycling method and non-subcycling method. In the subcycling method, variables on different levels are advanced with different time step sizes. The main advantage of the subcycling method is that when the Courant–Friedrichs–Lewy (CFL) number is kept the same on different grid levels, a larger grid spacing on a coarser level allows for a larger time step size on this level. For example, if the refinement ratio between two neighboring levels is two and the $L^\infty$-norm of velocity on both levels is approximately the same, then the time step size on the coarser level $\Delta t^l$ can be twice as large as that on the finer level $\Delta t^{l+1}$. In contrast, in the non-subcycling method, variables on different levels advance with the same time step size that is dictated by the finest level $l_{\max}$. In this case, variables on all levels are always at the same time instant.

Fig. 5.2 schematically shows how the subcycling and non-subcycling methods are used to advance the variables on a multilevel grid with $l_{\max} = 2$. As shown in the sketch, only 7 substeps are needed to advance all levels from $t^n$ to $t^n + \Delta t^0$ using the subcycling method. In contrast, 12 substeps are needed for the non-subcycling method. Within each substep of the two cycling methods, the single-level advancement algorithm described in Section 5.3.1 is employed to time march the solution. Although the non-subcycling method allocates more steps than the subcycling method, the latter requires an additional time interpolation of variables due to the mismatch of time step sizes among different levels. For example, to fill the ghost cell values at the coarse-fine boundary of a finer level, spatiotemporal interpolation of variables on the coarser level is required in the subcycling method. In contrast, in the non-subcycling method, spatial interpolation only suffices. In Section 5.4.9, the time savings for the subcycling and non-subcycling methods are carefully quantified in the simulations of the three-dimensional swimming eel.

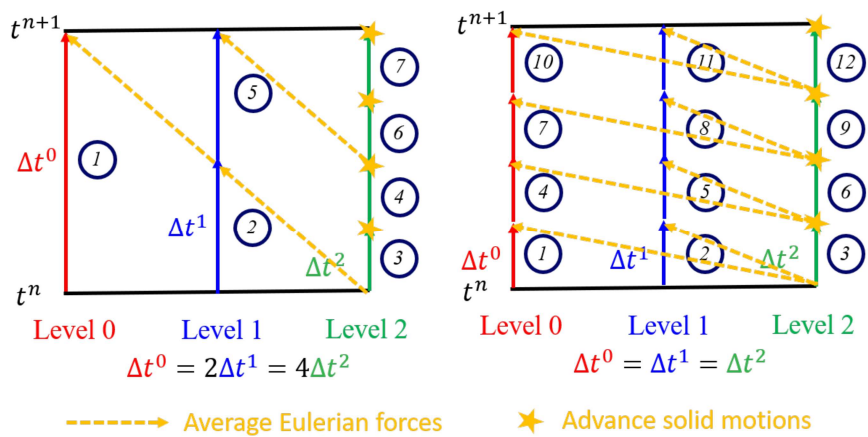

Figure 5.2: Schematic of the substeps in the level-by-level advancement method for a three-level grid ($l_{\max} = 2$). Left: the subcycling method. Right: the non-subcycling method. The substeps are represented by circled numbers. The yellow dashed lines represent the averaging of Eulerian forces from the finest level to the coarser levels, and the yellow stars represent the time advancement of the immersed body on the finest level.

In the context of FSI problems, there are two points of consideration when using the level-by-level time advancement technique. First, the Lagrangian markers are distributed only on the finest level. Thus, the Lagrangian quantities are updated only after the time advancement of the flow variables on the finest level, as indicated by the yellow stars in Fig. 5.2. Consequently, the solution on the finest level represents the final solid motion. Second, the Eulerian IB forces on a coarser level need to be properly considered. If the IB forces are not included on a coarser level, then the flow field on this level cannot "feel" the existence of the solid structure. Consequently, the updated flow velocity on a coarser level that is used to provide Dirichlet boundary conditions for the finer level will be incorrect, which will further lead to incorrect solutions on the finer level. To resolve this problem, we propose a force averaging algorithm to average the latest Eulerian forces on the finest level onto coarser levels in a sequential manner, as depicted in Fig. 5.2. Specifically, $\tilde{\mathbf{f}}_c^{t+\Delta t^l, l}$ denotes the Eulerian force that needs to be approximated at time $t + \Delta t^l$ on the coarser level $l$ for all $0 \leq l < l_{\max}$. Because the Eulerian force $\hat{\mathbf{f}}_c^{t, l_{\max}}$ on the finest level $l_{\max}$ is known at time $t$, we obtain the following Algorithm 5:

---

**Algorithm 5** Force averaging algorithm

---

1: **if** subcycling method is used **then**
2:      $\Delta t^l = 2^{l_{\max}-l} \Delta t^{l_{\max}}$ for all $0 \leq l < l_{\max}$
3: **else**
4:      $\Delta t^l = \Delta t^{l_{\max}}$ for all $0 \leq l < l_{\max}$
5: **end if**
6: **for** $l = l_{\max} - 1, 0, -1$ **do**
7:      **if** $l = l_{\max} - 1$ **then**
8:          $\tilde{\mathbf{f}}_c^{t+\Delta t^l, l} \leftarrow$ average $\tilde{\mathbf{f}}_c^{t, l+1}$
9:      **else**
10:          $\tilde{\mathbf{f}}_c^{t+\Delta t^l, l} \leftarrow$ average $\tilde{\mathbf{f}}_c^{t+\Delta t^{l+1}, l+1}$
11:      **end if**
12: **end for**
13: Replace $\mathbf{f}_c^{n+1}$ in Eq. (5.28) with $\tilde{\mathbf{f}}_c^{t+\Delta t^l, l}$ for updating the velocity

---

In Algorithm 5, four-point averaging operators and eight-point averaging operators (Martin and Colella, 2000; Martin et al., 2008) are employed in two spatial dimensions and three spatial dimensions, respectively.

**Synchronization**

The synchronization operation is utilized to make the solution data consistent across multiple levels and to obtain the composite solution from the level data. This operation is applied in both cycling methods (Almgren et al., 1998; Martin and Colella, 2000; Martin et al., 2008). There are three substeps involved in the synchronization operation:

Substep 1. Averaging

The fluid velocity $\mathbf{u}$, pressure $p$, and LS functions $\psi$ and $\phi$ on coarser levels are replaced by the corresponding data on the finer levels after the averaging substep. The variable values on the finer levels are considered to be better resolved than those on the coarser levels. Because a collocated grid framework is applied in this work, the same averaging operator can be employed for all of the aforementioned variables. We note that the averaging operation used here is different from the averaging step of the force averaging algorithm. Here, the averaging operator aims to synchronize all flow variables on multiple levels; it is applied to variables at the same time instance whenever a finer level catches up with a coarser level. On the other hand, the averaging operator in the force averaging algorithm is applied only to the Eulerian IB force variable at different time instants to obtain an approximate value of the IB force on a coarser level.

Substep 2. MAC synchronization and refluxing

At the CF boundary, the advection velocity $\mathbf{u}^{\mathrm{adv,l}}$ on the coarser level $l$ is generally not equal to the edge average of the advection velocity $\mathbf{u}^{\mathrm{adv,l+1}}$ on the next finer level $l+1$. This difference can create an imbalance of the momentum and scalar fluxes at the CF boundary. As a result, the free stream is not preserved while advancing the variables level by level. To remedy this problem, MAC synchronization and refluxing algorithms are carefully designed to maintain the conservation of momentum and scalar in the entire domain.

Substep 3. Composite grid projection

Because the level projection is applied on a level-by-level basis, it does not guarantee that the fluid velocity is divergence-free across all levels (Almgren et al., 1998; Martin et al., 2008). As a last step of the synchronization operation, a composite grid projection is applied to enforce the divergence-free condition on the velocity field across the entire hierarchy (Almgren et al., 1998; Martin and Colella, 2000). Using the composite grid operators defined in Section 5.2.1, a multilevel multigrid (MLMG) solver (Minion, 1996) is employed to project the fluid velocity on a divergence-free space as

$$L^{cc,\mathrm{comp}}_{\rho^{n+1}}\Theta = \frac{1}{\Delta t^{\mathrm{sync}}}D^{cc,\mathrm{comp}}\mathbf{u}^{n+1}, \tag{5.41}$$

$$\mathbf{u}^{n+1} := \mathbf{u}^{n+1} - \Delta t^{\mathrm{sync}}\boldsymbol{G}^{cc,\mathrm{comp}}\Theta, \tag{5.42}$$

where $\Delta t^{\text{sync}}$ is the time step of level 0, i.e., $\Delta t^{\text{sync}} = \Delta t^0$, and $L_{\rho^{n+1}}^{cc,\text{comp}}\Theta$ is the density-weighted approximation to the term $\boldsymbol{\nabla} \cdot (1/\rho^{n+1}\boldsymbol{\nabla}\Theta)$. We note that $\mathbf{u}^{n+1}$ becomes divergence-free in a multilevel sense after the composite grid projection substep.

**Multilevel initialization**

All field values need to be initialized on all levels at the beginning of the simulation. For example, the fluid velocity $\mathbf{u}$ on the coarsest level (level 0) is assigned based on the initial condition. The solid LS function $\psi$ is computed analytically based on the position of the center of mass of the body $\mathbf{X}_r^0$ and its geometry. The flow LS function $\phi$ is initialized based on the position of the gas-liquid interface when multiphase flow effects are included in the FSI simulation. The grid cells on the next level (level 1) are generated based on refinement and nesting criteria. After the level refinement, the velocity $\mathbf{u}$ and LS functions $\psi$ and $\phi$ values on level 1 are assigned based on the initial conditions. This "refining and filling" procedure is repeated until the finest level $l_{\max}$ is reached or until there is no need to refine the grid based on the refinement criteria.

Next, the Lagrangian markers are initialized on the finest level by placing one Lagrangian marker per Eulerian grid cell. The physical position $\mathbf{X}_{q,m}^0$, velocity $\mathbf{U}_{q,m}^0$ and radius vector $\mathbf{R}_{q,m}^0$ of the Lagrangian markers are determined from the known initial condition of the solid body. The pressure $p$ is initialized to zero on all levels.

**Summary of the multilevel advancement algorithm**

Algorithm 6 summarizes the unified multilevel advancement algorithm for both the sub-cycling methods and non-subcycling methods. After the variable initialization, we can use either the subcycling method or the non-subcycling method for time advancement. The force averaging algorithm is employed before each time step to approximate the Eulerian IB forces on all coarser levels. The synchronization step is applied whenever a finer level catches up with a coarser level. Grid refinement is applied before moving to the next time step.

**Algorithm 6** Multilevel advancement algorithm

1: Initialize $\mathbf{X}_r^0$, $\mathbf{U}_r^0$, $\mathbf{W}_r^0$, $\mathbf{u}^0$, $\psi^0$, $\phi^0$, and $p^0$ on level 0

2: $l \leftarrow 0$

3: **while** refinement criteria are satisfied on level $l$ and $l < l_{\max}$ **do**

4:     Regrid the patch hierarchy to obtain level $l + 1$

5:     Initialize $\mathbf{X}_r^0$, $\mathbf{U}_r^0$, $\mathbf{W}_r^0$, $\mathbf{u}^0$, $\psi^0$, $\phi^0$, and $p^0$ on level $l + 1$

6:     $l \leftarrow l + 1$

7: **end while**

8: Initialize $\mathbf{X}_{q,m}^0$, $\mathbf{U}_{q,m}^0$ and $\mathbf{R}_{q,m}^0$ for all Lagrangian markers on level $l_{\max}$

9: **if** subcycling method is used **then**

10:     $\Delta t^l = 2^{l_{\max} - l} \Delta t^{l_{\max}}$ for all $0 \le l < l_{\max}$

11: **else**

12:     $\Delta t^l = \Delta t^{l_{\max}}$ for all $0 \le l < l_{\max}$

13: **end if**

14: **for** $n = 1, n_{\max}$ **do**                             ▷ $n_{\max}$ is the number of time steps in the simulation

15:     LEVELCYCLING($0$, $t_n^0$, $t_n^0 + \Delta t^0$, $\Delta t^0$)

16:     Apply the synchronization projection using Eqs. (5.41–5.42)

17:     Regrid the patch hierarchy and interpolate $\mathbf{u}$, $\psi$, $\phi$, and $p$ onto new patches

18: **end for**

19:

20: **procedure** LEVELCYCLING($l$, $t^l$, $t_{\max}^l$, $\Delta t^l$)

21:     **while** $t^l < t_{\max}^l$ **do**

22:         **if** $l < l_{\max}$ **then**

23:             Apply the force averaging algorithm on level $l$                 ▷ **Algorithm 5**

24:         **end if**

25:         Perform single-level advancement on level $l$ from $t^l$ to $t^l + \Delta t^l$

26:             • Advect the LS functions $\psi^{n,l}$ and $\phi^{n,l}$ and reinitialize them using Eqs. (5.11–5.19)

27:             • Solve momentum Eq. (5.24) to obtain $\tilde{\mathbf{u}}^{n+1,l}$

28:             • Calculate the Lagrangian correction velocity $\Delta \mathbf{U}_c$ based on a specific kinematic constraint

29:             • Correct the Eulerian velocity field $\mathbf{u}^{n+1,l}$ using Eq. (5.28)

30:         **if** $l < l_{\max}$ **then**

31:             LEVELCYCLING($l + 1$, $t^l$, $t^l + \Delta t^l$, $\Delta t^{l+1}$)

32:         **end if**

33:         $t^l \leftarrow t^l + \Delta t^l$

34:     **end while**

35:     **if** $l > 0$ **then**

36:         Average all data from finer levels to the coarser levels

37:     **end if**

38:     **if** $l < l_{\max}$ **then**

39:         Perform MAC synchronization and refluxing using Eqs. (2.6–2.26)

40:     **end if**

41: **end procedure**

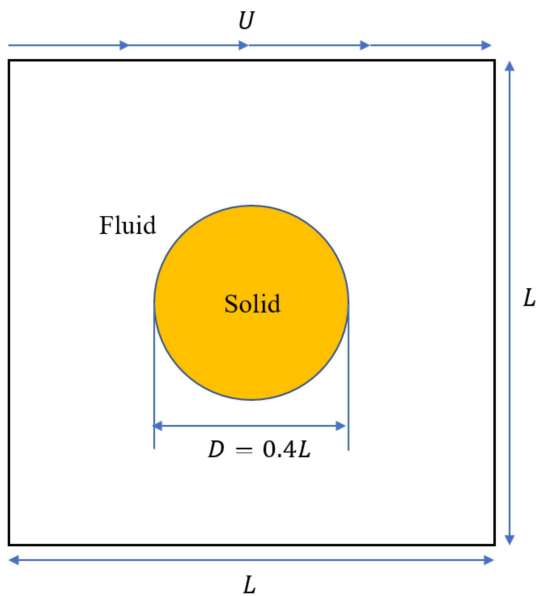

Figure 5.3: Schematic of a submerged cylinder in a lid-driven cavity flow.

## 5.4 Results

This section presents several canonical testing problems to validate the capabilities and robustness of the proposed AMR framework from different aspects. The following notations are used unless stated otherwise. For each case, $\Delta t_0$ denotes the time step on level 0, and $\Delta x_0$, $\Delta y_0$, and $\Delta z_0$ represent the grid spacings in the $x$-direction, $y$-direction, and $z$-direction, respectively, on level 0. For a multilevel grid, the grid spacings on the finer level $l$ satisfy $\Delta x_l = \Delta x_0/2^l$, $\Delta y_l = \Delta y_0/2^l$, and $\Delta z_l = \Delta z_0/2^l$ for all $0 \leq l \leq l_{max}$.

### 5.4.1 Lid-driven cavity with a submerged cylinder

We begin by considering the flow around a cylinder submerged in a lid-driven cavity. As shown in the left part of Fig. 5.3, a stationary cylinder with diameter $D = 0.4\,L$ is immersed at the center of a computational domain of extent $\Omega \in [0, L]^2$, with $L = 1$. The lid velocity at the domain top is set to $U = 1$. The Reynolds number of the flow is $Re = \rho_f U L/\mu_f = 1000$. No-slip boundary conditions are applied on all sides of the domain boundaries.

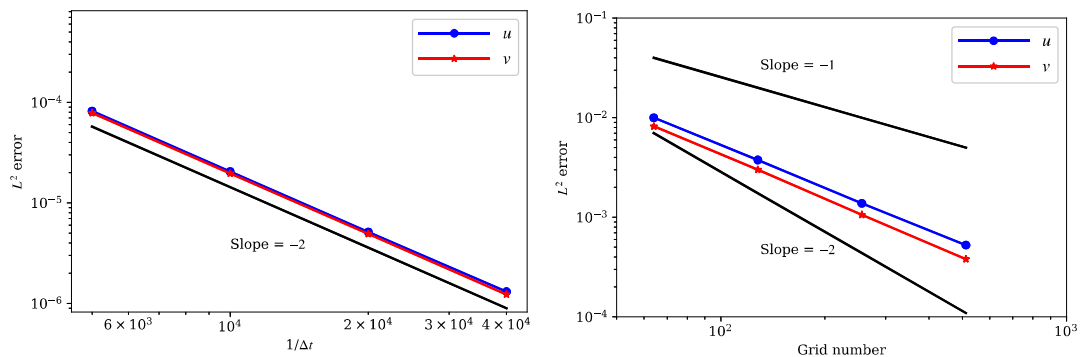

Figure 5.4: Temporal convergence (left) and spatial–temporal convergence (right) of $u$ and $v$ components of the flow velocity for the single-level case.

Both the single-level computational case and multilevel computational case are considered for this problem. For the single-level case, we test both the temporal convergence rate and the spatial-temporal convergence rate of the flow field. For the temporal convergence, it is defined as the rate of error reduction with decreasing time step size. When performing a temporal convergence study, it is necessary to design the tests such that the spatial error is minimal (Jacob and Ted, 2007; Almgren et al., 1998). We thus use a fixed grid with the smallest grid size and only change the time step size in different tests. For the spatial-temporal convergence, we consider grid numbers of $64 \times 64$, $128 \times 128$, $256 \times 256$, and $512 \times 512$. The CFL number is kept at a constant value 0.5 for all grids. Because no analytical solution is available for this problem, we compare the numerical solution against a reference solution obtained on a high-resolution $2048 \times 2048$ uniform grid. When the flow field reaches the steady state at around $tU/L = 30$, the $L^2$ norm of the velocity errors is computed (Balaras, 2004), and the temporal convergence rate and the spatial-temporal convergence rate of the solution are estimated, as shown in Fig. 5.4. In the left part of Fig. 5.4, it shows that our mid-point integration scheme can achieve the second order. The right part of Fig. 5.4 demonstrates that spatial-temporal convergence rate of the velocity is approximately 1.45 for the single-level case. This convergence rate is reasonable considering the smearing of the fluid-solid interface inherent in the DLM/IB method (Roma et al., 1999; Griffith et al., 2007). Also consistent with other IB papers (Bhalla et al., 2013; Patankar et al., 2000), the dominant errors are localized near the fluid-structure interface.

Table 5.1: Parameters of the lid-driven cavity with a submerged cylinder problem

| Case No. | Grid numbers on level 0 | $l_{\max}$ | Cycling methods | Force averaging used? |
|---|---|---|---|---|
| 1 | $2048 \times 2048$ | 0 | – | – |
| 2 | $64 \times 64$ | 2 | Subcycling | Yes |
| 3 | $64 \times 64$ | 2 | Subcycling | No |

For the multilevel grid case, we consider static mesh refinement. Grid cells in the rectangular region $(x, y) \in [0.2, 0.8]$ are refined to level 1, and grid cells in the region $(x, y) \in [0.25, 0.75]$ are further refined to $l_{\max} = 2$. The cylinder is kept on the finest level, and a refinement ratio of two is applied to both levels.

Before proceeding to test the convergence rate on a multilevel grid, we demonstrate the necessity of the force averaging algorithm (Section 5.3.4) for the level-by-level time advancement method. Three cases are considered, as listed in Table 5.1. Case 1 is the single-level case, which is employed as a reference case. Case 2 has the force averaging scheme, while Case 3 does not. Fig. 5.5 compares the time evolution of the maximum Eulerian IB forces among the different cases, where $fb_{x,max}$ and $fb_{y,max}$ represent the maximum IB force in the $x$ direction and $y$ direction, respectively. It is clearly seen that the time series of IB forces in Case 3 deviate from those in Case 1 and Case 2 because the IB forces on the finest level in Case 3 are not averaged to the coarser levels to obtain a proper estimation. By using the force averaging scheme, the time evolution of the IB forces in Case 2 shows agreement with that in Case 1, and their difference in the early stage of simulation can be explained by the fact that it takes some time for the simulations on the finer level to be tightly coupled with those on the coarser level before the data across multiple levels match. In addition, the effectiveness of the force averaging scheme is also demonstrated through the velocity fields. As shown in Fig. 5.6, the contours of the velocity magnitude in Cases 1 and 2 are almost identical, whereas in Case 3, a spurious vortex is generated on the upper right side of the finest level. These results show the necessity and efficacy of using the force averaging scheme when a multilevel grid is employed in the simulations. Although only the subcycling case results are shown here, the non-subcycling case also exhibits a similar behavior.

After validating the force averaging scheme, we now assess the temporal and spatial-temporal convergence rates of our numerical scheme on a multilevel grid. For assessing the temporal convergence rate, the grid size is kept fixed. For the spatial-temporal convergence rate test, the CFL number is fixed and the grid sizes on level 0 are taken to be $32 \times 32$, $64 \times 64$, and $128 \times 128$. Both the subcycling and non-subcycling methods are assessed. Similar to the single-level case, we compare the composite solution with the reference solution at $tU/L = 30$ using the $L^2$-norm of the error. In Fig. 5.7, it can be observed that we obtain second-order convergence rate in time for both the non-subcycling and subcycling methods; the latter uses an additional midpoint time interpolation scheme of variables defined on different levels (Section 5.3.4). As seen in Fig. 5.8, the $L^2$ errors of $\mathbf{u}$ decrease as the grid resolution increases. The $L^2$ errors at a given grid spacing for the subcycling method and non-subcycling methods are comparable. The two adaptive methods also achieve a spatial-temporal convergence rate of approximately 1.44, which is nearly the same as those observed for the single-level cases previously. These tests prove the efficacy of the force averaging scheme and demonstrate that our numerical schemes can achieve the expected order of accuracy on (static) multilevel mesh.

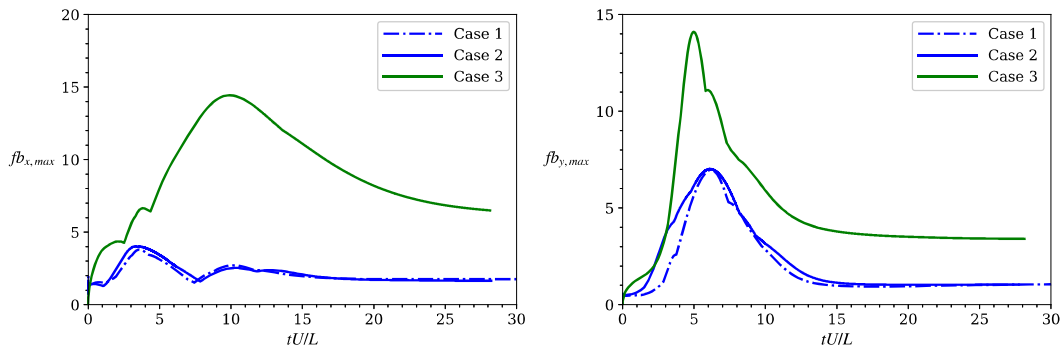

Figure 5.5: Comparison of the time evolution of the maximum Eulerian IB force among the different cases listed in Table 5.1. $fb_{x,max}$ (left) and $fb_{y,max}$ (right) represent the maximum IB force in the $x$ direction and $y$ direction, respectively.

### 5.4.2 Cylinder with prescribed motion

We first consider the flow past a cylinder oscillating about a mean position with a prescribed velocity. Specifically, the cylinder oscillates in the $x$ direction with the prescribed velocity $U_c = U_{max} \cos(2\pi t/T)$, where $U_{max} = 1.0$ is the maximum oscillating velocity and $T$ is the time period of the oscillation. To exclude wall effects, the size of the computational domain is set sufficiently large to $[-16D, 16D] \times [-8D, 8D]$. Here, $D$ is the diameter of the cylinder. The center of the cylinder is located at $(0, 0)$. To match the previous studies by Shen et al. (Shen et al., 2009) and Bhalla et al. (Bhalla et al., 2013), the Keulegan-Carpenter number is set to $KC = U_{\max}T/D = 5$, and the Reynolds number is $Re = \rho_f U_{\max} D/\mu_f = 100$. Three cases are considered, as listed in Table 5.2. The refinement criterion is the distance to the fluid-solid interface indicated by the LS function $\psi$. Specifically, the grid cells $(i, j)$ on level $l$ ($0 \le l < l_{\max}$) are refined to the finer level if $|\psi_{i,j}| < 4.0 \max(\Delta x_l, \Delta y_l)$, where $\Delta x_l$ and $\Delta y_l$ are the grid spacings in the $x$ direction and $y$ direction, respectively, on level $l$.

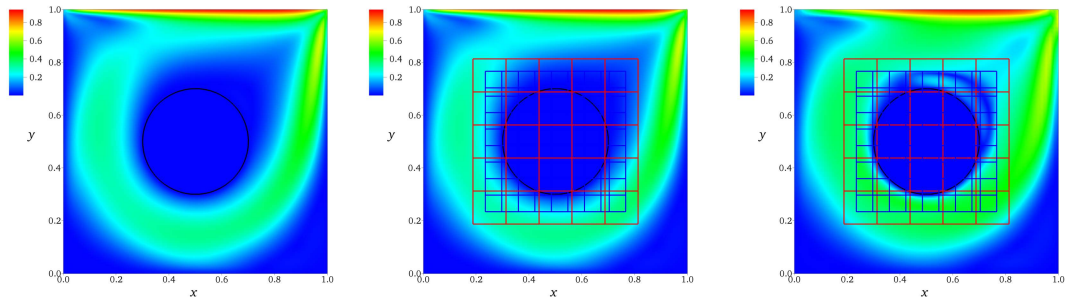

Figure 5.6: Comparison of the velocity magnitudes for the cylinder submerged within a lid-driven cavity problem at $tU/L = 30$. Left: single-level case (Case 1); middle: three-level subcycling case (Case 2) with the force averaging scheme; right: three-level subcycling case (Case 3) without the force averaging scheme. Black lines: fluid-solid interface; red lines: patches on level 1; blue lines: patches on level 2.

Table 5.2: Parameters of the cylinder with prescribed motion problem

| Case No. | Grid numbers on level 0 | $l_{\max}$ | $\Delta t_0$ | Cycling methods |
|----------|-------------------------|------------|--------------|-----------------|
| 1 | $1024 \times 512$ | 0 | $5 \times 10^{-4}$ | – |
| 2 | $256 \times 128$ | 2 | $2 \times 10^{-3}$ | Subcycling |
| 3 | $256 \times 128$ | 2 | $5 \times 10^{-4}$ | Non-subcycling |

The left part of Fig. 5.9 compares the drag coefficient in the single-level case, defined as $C_d = F_x/(0.5\rho_f U_{\max}^2 D)$, with the results of Shen et al. (Shen et al., 2009) and Bhalla et al. (Bhalla et al., 2013). We note that both Eq. (5.26) and Eq. (5.39) are used to calculate the drag coefficient in this test problem. Eq. (5.26) only considers the constraint force, while Eq. (5.39) includes both the constraint force and inertial force. The right part of Fig. 5.9 shows the comparison of the drag coefficients among the single-level case (Case 1), three-level subcycling case (Case 2), and three-level non-subcycling case (Case 3). The agreement indicates that when inertial effects are present, our FSI algorithm can accurately estimate the hydrodynamic force on the solid surface for both the single-level case (Case 1) and multilevel case (Cases 2 and 3).

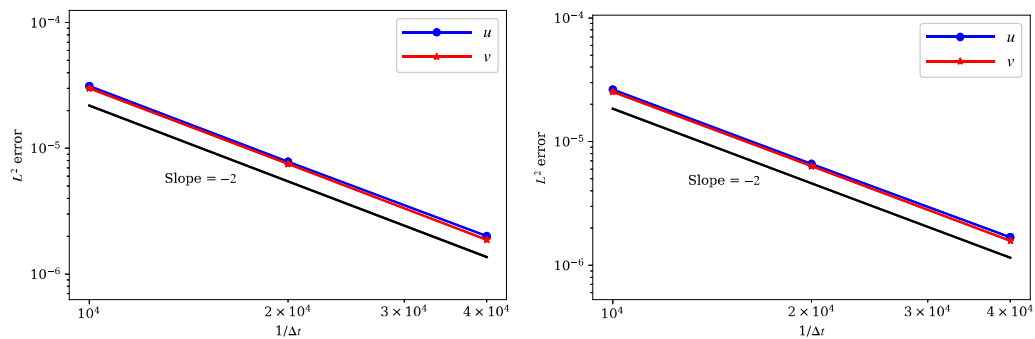

Figure 5.7: Temporal convergence of $u$ and $v$ for the problem of a cylinder submerged in a lid-driven cavity with mesh refinement. The finest level is $l_{\max} = 2$. Left: subcycling method. Right: non-subcycling method.

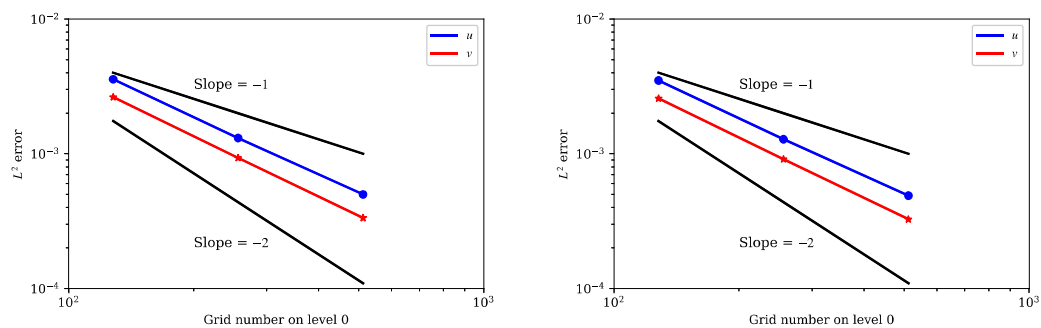

Figure 5.8: Spatial-temporal convergence of $u$ and $v$ for the problem of a cylinder submerged in a lid-driven cavity with mesh refinement. The finest level is $l_{\max} = 2$. Left: subcycling method. Right: non-subcycling method.

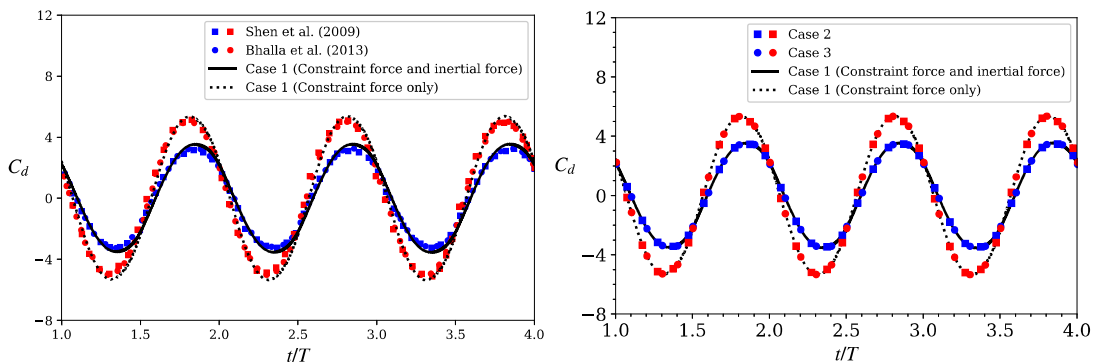

Figure 5.9: Time series of the drag coefficient of an oscillating cylinder at $Re = 100$ and $KC = 5$. Left: comparison between the single-level case (Case 1) and previous results. Right: comparison among the single-level case (Case 1), the three-level subcycling case (Case 2), and the three-level non-subcycling case (Case 3). The dotted line, red circle (●), and red square (■) represent the constraint force calculated from Eq. (5.26). The solid line, blue circle (●), and blue square (■) represent the sum of both the inertial term and the constraint force obtained from Eq. (5.39).

Table 5.3: Parameters of the falling sphere in quiescent flow problem

| Case No. | Grid numbers on level 0 | $l_{max}$ | $\Delta t_0$ | Cycling methods |
|----------|------------------------|-----------|--------------|-----------------|
| 1 | $64 \times 512 \times 64$ | 0 | $1 \times 10^{-3}$ | – |
| 2 | $16 \times 128 \times 16$ | 2 | $4 \times 10^{-3}$ | Subcycling |
| 3 | $16 \times 128 \times 16$ | 1 | $1 \times 10^{-3}$ | Non-subcycling |

### 5.4.3 Falling sphere in quiescent flow

In the test cases introduced above, the motion of the body was prescribed. In this section, the falling sphere problem is simulated to validate the DLM algorithm with a freely moving body. Here, we follow the same setup as (Bhalla et al., 2013). The computational domain is $[-1\,\text{m}, 1\,\text{m}] \times [0\,\text{m}, 8\,\text{m}] \times [-1\,\text{m}, 1\,\text{m}]$, and the sphere with diameter $D = 0.625\,\text{m}$ is initially centered at $(x, y, z) = (0\,\text{m}, 7\,\text{m}, 0\,\text{m})$. The density of the fluid and the density of the solid are set to $\rho_f = 2\,\text{kg/m}^3$ and $\rho_s = 3\,\text{kg/m}^3$, respectively. The fluid viscosity is $\mu_f = 0.05\,\text{Pa} \cdot \text{s}$, and the Reynolds number is $Re = \rho_f V D / \mu_f$, where $V$ is the terminal velocity of the falling sphere. No-slip boundary conditions are employed in all directions. Three cases are considered, as listed in Table 5.3.

When the sphere approaches the terminal state, its velocity in this case is $V =$

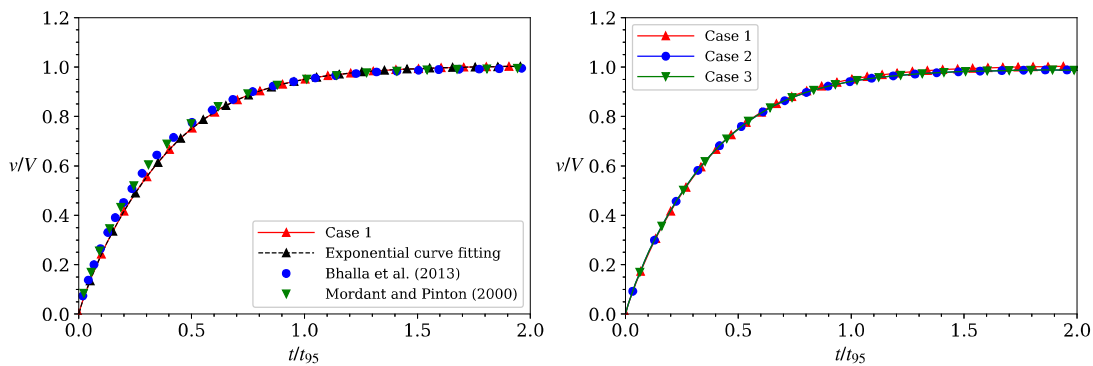

Figure 5.10: Time series of the vertical velocity of a falling sphere at $Re = 31$. Left: comparison among the single-level case (Case 1), the exponential curve fitting using Eq. (5.43), and previous results (Bhalla et al. (2013); Mordant and Pinton (2000)). Right: comparison among the single-level case (Case 1), three-level subcycling case (Case 2), and three-level non-subcycling case (Case 3).

$-1.24\,\mathrm{m/s}$, which agrees with prior studies (Bhalla et al., 2013; Sharma and Patankar, 2005). Previous studies have shown that the normalized vertical velocity can be fit using an exponential curve as

$$v/V = 1 - e^{-3t/t_{95}}, \tag{5.43}$$

where $t_{95}$ is the time required for the sphere to reach 95% of the terminal velocity $V$. The left part of Fig. 5.10 compares the time series of the vertical velocity among the single-level case (Case 1), exponential curve fitting results using Eq. (5.43), and previous numerical (Bhalla et al., 2013) and experimental (Mordant and Pinton, 2000) results. The right part of Fig. 5.10 shows that the results of the multilevel cases (Cases 2 and 3) agree well with the single-level results. These results indicate that our algorithms can accurately capture the transient velocity of the falling sphere when it freely interacts with the surrounding fluid.

### 5.4.4 Rotating cylinder in a shear flow

This section considers a rotating cylinder in a shear flow to validate the multilevel DLM algorithm for problems with a rotational degree of freedom (DOF). As shown in Fig. 5.11, a 2D cylinder is initially immersed in a shear-driven channel flow with its centroid at $(L/4, H/4)$. The flow is driven by two moving plates with velocity $U_w$ in opposite directions. No-slip and periodic boundary conditions are applied in the vertical direction and horizontal direction, respectively. Three non-subcycling cases are considered, as listed in Table 5.4, at Reynolds number $Re = 8\rho_f U_w D/\mu_f = 40$. The refinement criterion is based on the distance to the fluid-solid interface, i.e., the grid cells $(i, j)$ on level $l$ ($0 \leq l < l_{\max}$) are refined to the next finer level if $|\psi_{i,j}| < 4.0 \max(\Delta x_l, \Delta y_l)$.

Table 5.4: Parameters of the rotating cylinder in a shear flow problem

| Case No. | Grid numbers on level 0 | $l_{\max}$ | $\Delta t_0$ |
|----------|-------------------------|------------|--------------|
| 1 | $64 \times 512$ | 0 | $1 \times 10^{-3}$ |
| 2 | $16 \times 128$ | 1 | $1 \times 10^{-3}$ |
| 3 | $16 \times 128$ | 2 | $1 \times 10^{-3}$ |

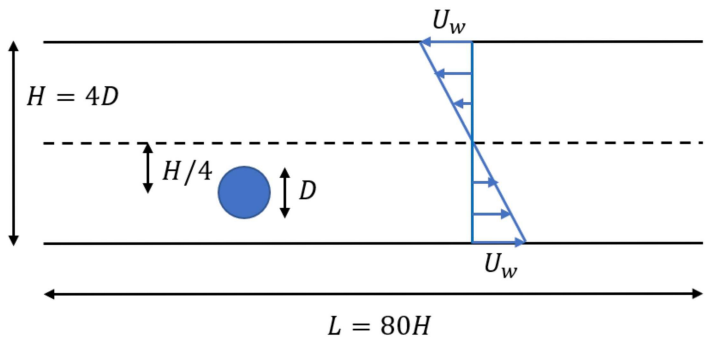

Figure 5.11: Sketch of a two-dimensional rotating cylinder in a shear flow.

Fig. 5.12 shows the time evolution of the angular velocity $\omega_z$ and the vertical position of the cylinder centroid $y_c$. The angular velocity of the cylinder increases rapidly and then reaches a steady state at approximately $2U_w t/H = 200$ at a value of 0.47. This value is consistent with the results of Yeo et al. (2010) and Bhalla et al. (2013). Similarly, the evolution of the vertical position of the cylinder $y_c$ in the single-level case (Case 1) also agrees with previous studies. Fig. 5.13 compares the time series of $\omega_z$ and $y_c$ between the single-level case (Case 1) and the multilevel non-subcycling cases (Cases 2 and 3). The agreement proves that the proposed multilevel algorithms can accurately simulate FSI problems with rotational degrees of freedom.

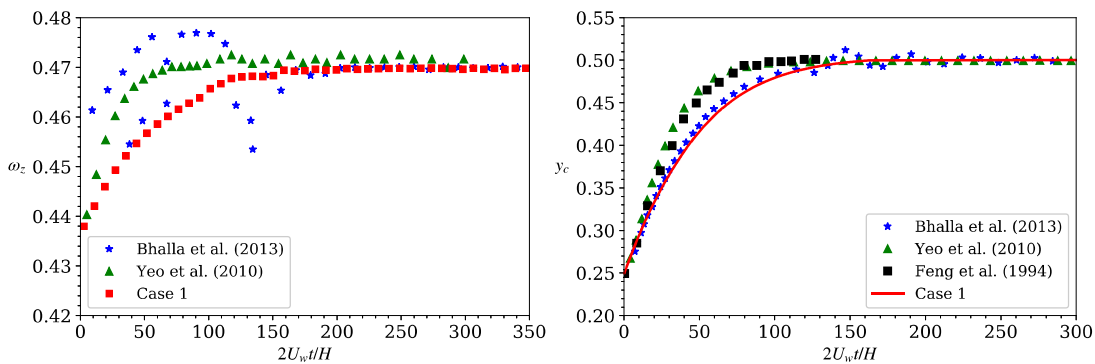

Figure 5.12: Comparison of the angular velocity $\omega_z$ (left) and the vertical position of the cylinder centroid $y_c$ (right) between the single-level case (Case 1) and previous results.

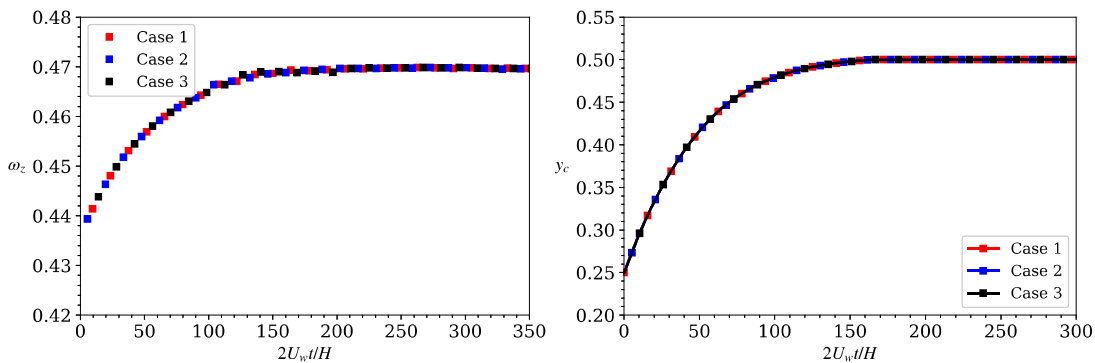

Figure 5.13: Comparison of the angular velocity $\omega_z$ (left) and vertical position of the cylinder centroid $y_c$ (right) between the single-level case (Case 1) and the multilevel non-subcycling cases (Cases 2 and 3).

### 5.4.5 Oscillating cylinder in a spring-mass-damper system

This section considers an oscillating cylinder with different densities in a spring-mass-damper system, aimed at testing the FSI algorithms when external restoring and damping forces are applied to the solid body at each time step. The schematic of the numerical configuration is shown on the left part of Fig. 5.14, in which the computational domain is $[0, 5D] \times [0, 10D]$ and no-slip boundary conditions are utilized on all sides except for at the top boundary (Zeng and Shen, 2020). The cylinder is attached to a spring-damper system with a damping constant of $b_s$ and a stiffness constant of $k_s$, and its initial centroid is located at $(X_0, Y_0) = (2.5D, 8D)$. The free length of the spring is set to $6.5D$, which means that the initial extension of the spring is $1.5D$. The density ratio between the solid and the fluid is $m^* = \rho_s/\rho_f$. Six cases are listed in Table 5.5, in which we choose two different density ratios, $m^* = 100$ and $0.8$, corresponding to a structure in a gas and a structure in a liquid, respectively. For all these cases, we fix the diameter of the cylinder $D = 0.2$ m and spring constant $k_s = 500$ N/m. We consider the damper with different values of $b_s/b_{\text{critical}} = \zeta$, where $b_{\text{critical}} = 2\sqrt{k_s M_b}$. We note that $M_b$ is the mass of the cylinder and $\zeta$ determines the behavior of the spring-damper system (Nayfeh and Pai, 2008): $\zeta < 1$ represents an underdamped system, $\zeta = 1$ leads to a critically damped system, and $\zeta > 1$ results in an overdamped system.

For the density ratio $m^* = 100$, the left part of Fig. 5.15 compares the time evolution of the vertical position of the cylinder centroid between the single-level case (Case 1) and the analytical solution. The analytical solution is obtained by disregarding the fluid forces on the solid body because these forces are small compared to the inertial force for this case with a large density ratio. Fig. 5.15 shows that our results are consistent with the analytical results. The right part of Fig. 5.15 compares Cases 1-3, which shows that both subcycling results and non-subcycling results match the single-level result. To visualize the flow field, the contours of the velocity vector at $t = 0.5$ and the patch hierarchy for Case 2 are plotted in the right part of Fig. 5.14.

For the low-density ratio $m^* = 0.8$, the fluid and solid have comparable densities, and the hydrodynamic forces cannot be disregarded, especially at low damping ratios. Although an analytical solution is not available for this density ratio, we compare our numerical results with those of Dafnakis et al. (2020), in which a Brinkman penalization (BP) method is used to simulate the oscillating cylinder and WEC problems. As shown on the left part of Fig. 5.16, the numerical results of our single-level case (Case 4) using the DLM method agree with those using the BP method in (Dafnakis et al., 2020), which proves the correctness and robustness of our algorithms at low-density ratios between the solid phase and fluid phase. The right part of Fig. 5.16 further compares the results between the single-level case (Case 4) and the multilevel cases (Cases 5 and 6), which shows the consistency in the numerical results regardless of the presence of AMR.

Table 5.5: Parameters of the oscillating cylinder in a spring-mass-damper system problem

| Case No. | Density ratio $m^*$ | Grid numbers on level 0 | $l_{max}$ | $\Delta t_0$ | Cycling methods |
|----------|---------------------|-------------------------|-----------|--------------|-----------------|
| 1 | 100 | $256 \times 512$ | 0 | $1 \times 10^{-3}$ | – |
| 2 | 100 | $64 \times 128$ | 2 | $4 \times 10^{-3}$ | Subcycling |
| 3 | 100 | $64 \times 128$ | 2 | $1 \times 10^{-3}$ | Non-subcycling |
| 4 | 0.8 | $256 \times 512$ | 0 | $1 \times 10^{-3}$ | – |
| 5 | 0.8 | $64 \times 128$ | 2 | $4 \times 10^{-3}$ | Subcycling |
| 6 | 0.8 | $64 \times 128$ | 2 | $1 \times 10^{-3}$ | Non-subcycling |

### 5.4.6 Wave energy converter (WEC)

In this section, a WEC simulation is conducted to evaluate the performance of our multilevel DLM algorithms in the presence of multiphase flows and external forces. Here, the LS-based two-phase flow solver (Sussman et al., 1999) is coupled to the multilevel DLM algorithm to simulate the WEC dynamics. A schematic representation of the WEC problem is shown on the left part of Fig. 5.17. A water wave of height $H = 0.01$ m and period $T = 0.8838$ s is generated from the left boundary using Stokes wave theory. Refer to Appendix A.3 and Dafnakis et al. (2020) for details. The wave then propagates to the right, transfers energy to the WEC, and is eventually dissipated in the damping zone at the right end of the domain. The wavelength is $\lambda = 1.216$ m, and the water depth is $d = 0.65$ m (Dafnakis et al., 2020; Zeng and Shen, 2020). A cylindrical-shaped WEC device with diameter $D = 0.16$ m and density ratio $m^* = \rho_s/\rho_f = 0.9$ is initially submerged to a depth of $d_s = 0.25$ m from the free water surface ( 5.17). The stiffness and damping constants of the power take-off (PTO) system are $k_s = 1995.2$ N/m and $b_s = 80.64$ Ns/m, respectively. Two multilevel computational cases are considered, as listed in Table 5.6. The air-water interface and water-solid interface are refined to the finest level (Fig. 5.17). In this work, the WEC device is only allowed to oscillate in the heave and surge directions, i.e., with two free DOFs, and the same $k_s$ and $b_s$ are applied to both directions.

Fig. 5.18 shows the heave and surge motions of the two-DOF WEC and its power generation during the steady state. Our multilevel results agree with the previous CFD simulation (Dafnakis et al., 2020), and the slow drift phenomenon appearing in the surge dynamics is also well captured. Compared with the one-DOF results (not shown here), the two-DOF WEC slightly increases the power absorption, which shows agreement with the theoretical analysis of (Falnes and Kurniawan, 2020). These results indicate that our multilevel DLM algorithm can accurately resolve the dynamics of a solid body when external forces are applied in a wave environment, for both the subcycling time advancement scheme and non-subcycling time advancement schemes.

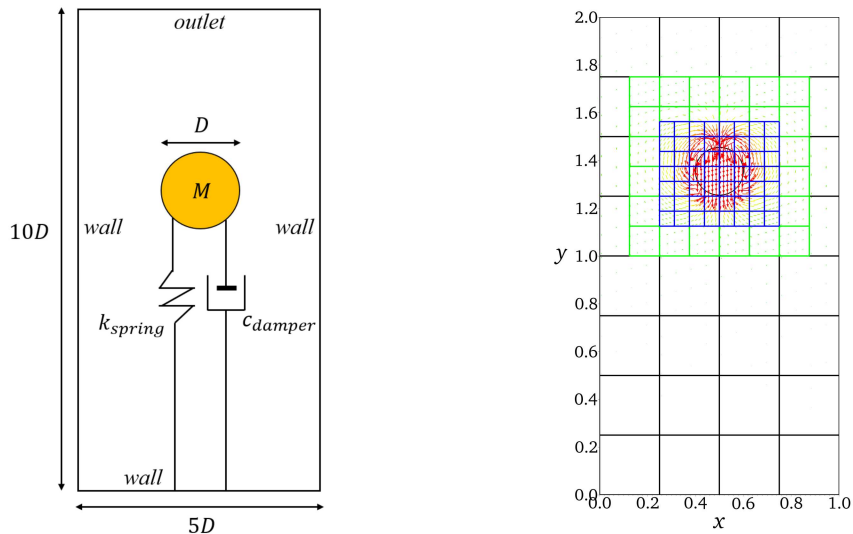

Figure 5.14: Left: schematic of an oscillating cylinder in a spring-damper system. Right: velocity vectors around the cylinder in an underdamped regime ($\zeta = 0.5$) for the three-level subcycling case (Case 2). Black lines: patches on level 0; green lines: patches on level 1; blue lines: patches on level 2.

Table 5.6: Parameters of the WEC problem

| Case No. | Grid numbers on level 0 | $l_{\max}$ | $\Delta t_0$ | Cycling methods |
|---|---|---|---|---|
| 1 | $800 \times 192$ | 2 | $8 \times 10^{-4}$ | Subcycling |
| 2 | $800 \times 192$ | 2 | $2 \times 10^{-4}$ | Non-subcycling |

### 5.4.7 Cylinder splashing onto a two-fluid interface

To further validate our multilevel algorithms in the multiphase flow scenario, we consider a cylinder splashing onto a two-fluid interface, which is different from the single-phase falling sphere problem (Section 5.4.3). The computational domain is $[8D, 48D]$, the upper half of which from $y = 12D$ to $y = 24D$ is filled with a lighter fluid of density $\rho_g = 1 \times 10^3 \, \text{kg/m}^3$ and the lower half of which from $y = 0$ to $y = 12D$ is filled with a heavier fluid of density $\rho_l = 1.25 \times 10^3 \, \text{kg/m}^3$. No-slip boundary conditions are imposed on all sides. A circular cylinder with diameter $D = 2.5 \times 10^{-3} \, \text{m}$ and density $\rho_s = 1.5 \times 10^3 \, \text{kg/m}^3$ is initially placed at location $[4D, 40D]$. For this FSI problem, two cases using either the subcycling or non-subcycling method are considered, as listed in Table 5.7.

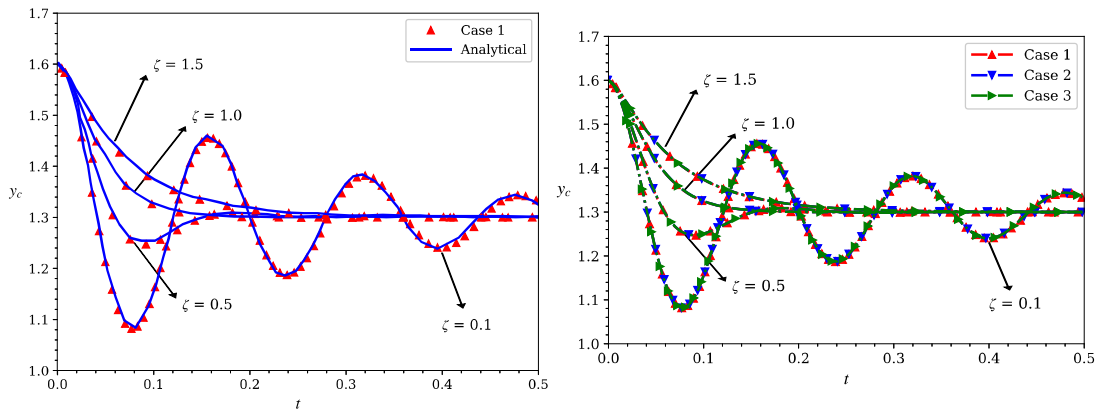

Figure 5.15: Left: comparison of the time series of the vertical position of the cylinder centroid at $m^* = 100$ between the single-level case (Case 1) and the analytical solution. Right: comparison of the time series of the vertical position of the cylinder centroid among the single-level case (Case 1), three-level subcycling case (Case 2), and three-level non-subcycling case (Case 3). Various values of damping ratio $\zeta$ are considered.

Table 5.7: Parameters of the cylinder splashing onto a two-fluid interface problem

| Case No. | Grid numbers on level 0 | $l_{\max}$ | $\Delta t_0$ | Cycling methods |
|---|---|---|---|---|
| 1 | $128 \times 768$ | 2 | $2 \times 10^{-3}$ | Subcycling |
| 2 | $128 \times 768$ | 2 | $5 \times 10^{-4}$ | Non-subcycling |

Fig. 5.19 shows the time evolution of the density contours of the three phases. The dimensionless time is defined as $T = t\sqrt{g/D}$. As plotted, a cavity forms in the wake of the cylinder as it penetrates the two-fluid interface. As the cavity collapses, a jet forms and breaks up into small droplets. We note that both the cavity and jet dynamics remain approximately symmetric in our simulations due to the initial symmetry of the problem setup. Symmetrical results are also observed in Nangia et al. (2019b).

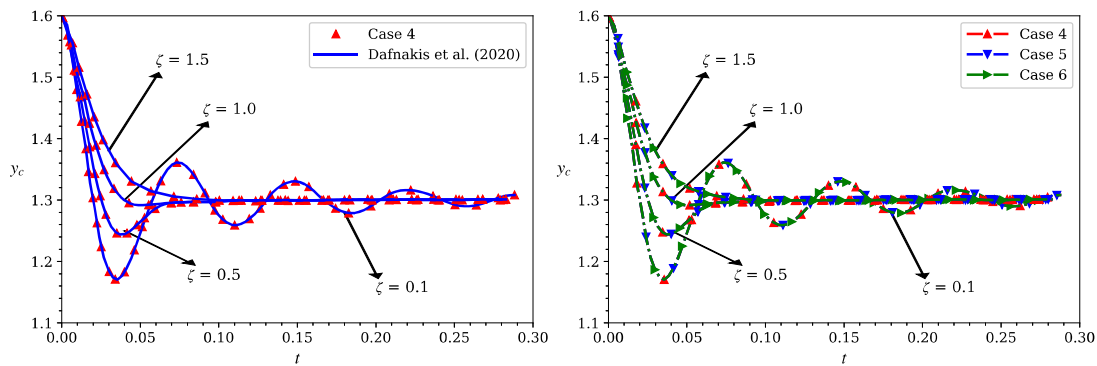

Figure 5.16: Left: comparison of the time series of the vertical position of the cylinder centroid at $m^* = 0.8$ between the single-level case (Case 4) and Dafnakis et al. (2020). Right: comparison of the time series of the vertical position of the cylinder centroid among the single-level case (Case 4), three-level subcycling case (Case 5), and three-level non-subcycling case (Case 6). Various values of damping ratio $\zeta$ are considered.

In chapter 2, we assess the conservation of a passive scalar. Here, instead, we assess the conservation of mass by our multiphase FSI solver. This is done by tracking the normalized fluid mass $m^\star(t) = |m(t)|/|m(0)|$ as a function of time for the cylinder splashing onto the two-fluid interface case. Here, $m(t) = \int \rho(\mathbf{x}, t)\, dV$ is the total fluid mass in the domain. Fig. 5.20 plots the time evolution of $m^\star$ over the course of the entire simulation. The results indicate that the total mass loss for Case 1 and 2 is less than 2%. We attribute the mass loss to the level set technique, which is known to be non-conservative in nature. Nevertheless, results of Fig. 5.20 indicate that our algorithm has a reasonably good performance in terms of conserving the total fluid mass for this complex problem including gas-liquid-solid interaction using both subcycling and non-subcycling methods.

To further quantitatively validate the accuracy of our algorithms, the time series of the vertical velocity and vertical position of the cylinder are plotted in Fig. 5.21. Our results show agreement with the studies of Nangia et al. (2019b), in which the same DLM approach was applied to capture the motion of the splashing cylinder. In summary, our simulations can accurately simulate the dynamics of the splashing cylinder and achieve good performance of the mass conservation in multiphase flow scenarios with both the subcycling time advancement scheme (Case 1) and the non-subcycling time advancement

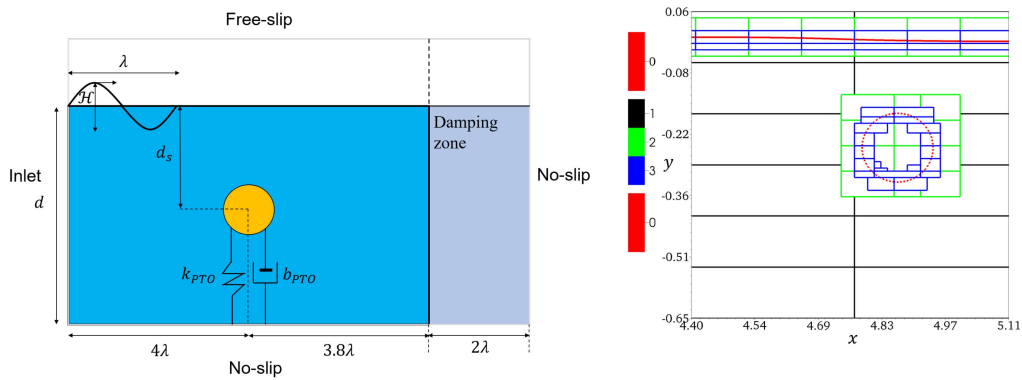

Figure 5.17: Left: schematic of the submerged cylindrical WEC device. Right: patch hierarchy and contours among different phases. Red solid line: air-water interface; red dashed line: solid-water interface; black lines: patches on level 0; green lines: patches on level 1; blue lines: patches on level 2.

scheme (Case 2).

### 5.4.8   Two-dimensional self-propelled eel

In this section, we consider the two-dimensional self-propelled eel problem to validate the proposed algorithms when the geometry and deformation of the body are prescribed (Section 5.3.2). Eels are elongated ray-finned fish that belong to the order *Anguilliformes*. Fig. 5.22 sketches the two-dimensional swimming eel considered in this study. The eel has translation and deformation motions at $t > 0$ while interacting with the surrounding fluid. For this problem, the eel is assumed to move forward in the negative $x$ direction, and thus, the local $h - \zeta$ coordinate, which is attached to the body of the eel, retains the same orientation as the fixed global $y - x$ coordinate. To describe the geometry of the fish, the vertical displacement of its middle line $h(\zeta, t)$ is given by Bhalla et al. (2013); Cui et al. (2018); Zeng et al. (2022a)

$$h(\zeta, t) = 0.125L\frac{\zeta/L + 0.03125}{1.03125}\sin\left[2\pi\left(\frac{\zeta}{L} - \frac{t}{T}\right)\right],\tag{5.44}$$

where $L$ is the projected body length and $T$ is the oscillating period of the tail tip. The cross-section of the fish body is centered about the middle line $h(\zeta, t)$, and its half width $r(\zeta)$ is given by

$$r(\zeta) = \begin{cases} \sqrt{2w_{\mathrm{H}}\zeta - \zeta^2} & \text{for } 0 \leqslant \zeta < \zeta_{\mathrm{H}}, \\ w_{\mathrm{H}}\dfrac{L-\zeta}{L-\zeta_{\mathrm{H}}} & \text{for } \zeta_{\mathrm{H}} \leqslant \zeta < L. \end{cases} \tag{5.45}$$

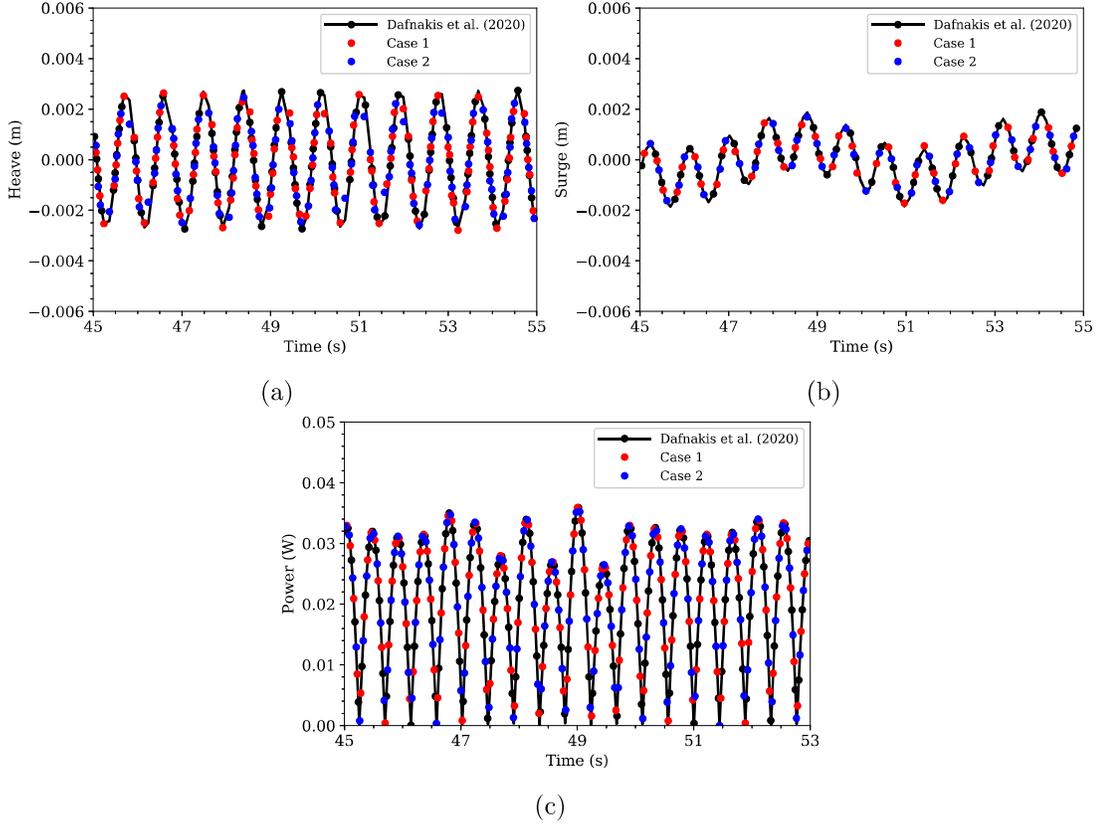

(a)                                         (b)

(c)

Figure 5.18: Comparison of the dynamics of the two-DOF WEC among Dafnakis et al. (2020), the three-level subcycling case (Case 1), and the three-level non-subcycling case (Case 2). (a) Heave, (b) surge, and (c) generated power.

Here, $w_{\mathrm{H}} = 0.04L$ is obtained from the observations reported in (Kern and Koumoutsakos, 2006; Bhalla et al., 2013). A periodic computational domain of size $8L \times 4L$ is employed for the two-dimensional eel simulation. The Reynolds number is $Re = V_{\max}L/\mu_{\mathrm{f}}$, where $V_{\max} = 0.785L/T$ is the maximum undulatory velocity of the tail tip. At $t = 0$, the head of the eel is centered at $(6L, 2L)$. The eel then moves forward along the negative $x$ direction when $t > 0$ because of self-propulsion. Other useful parameters for

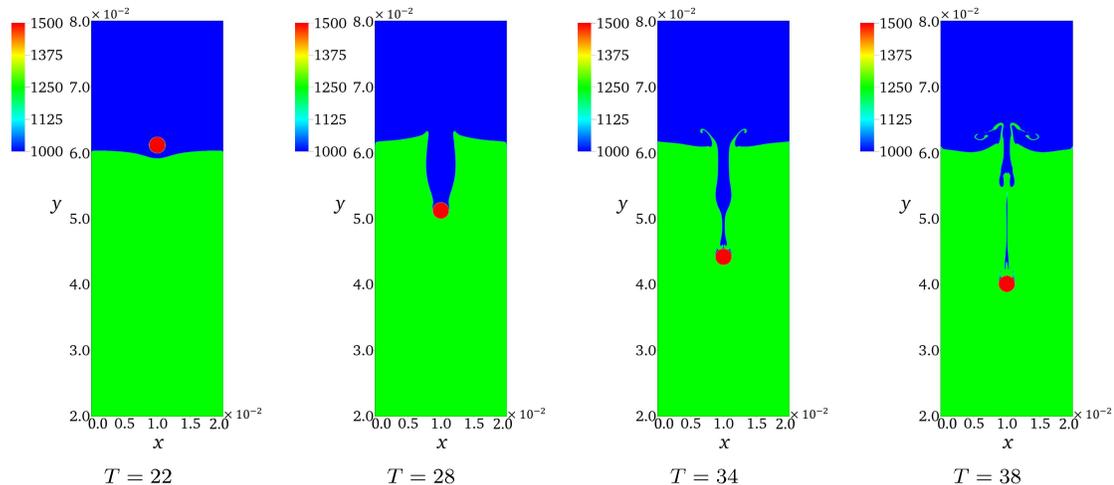

$T = 22$       $T = 28$       $T = 34$       $T = 38$

Figure 5.19: Evolution of the density field when a cylinder falls into a column containing two fluids at different time instances. Only part of the $y$ axis is plotted for better visualization.

Table 5.8: Parameters of the two-dimensional self-propelled eel problem

| Case No. | Grid numbers on level 0 | $l_{\max}$ | Cycling methods |
|----------|-------------------------|------------|-----------------|
| 1 | $4096 \times 2048$ | 0 | – |
| 2 | $512 \times 256$ | 3 | Subcycling |
| 3 | $512 \times 256$ | 3 | Non-subcycling |

different cases are listed in Table 5.8. There are two refinement criteria in the eel simulation problem. The first criterion is the distance to the interface, i.e., the grid cells $(i, j)$ on level $l$ ($0 \leq l < l_{\max}$) are refined to the finer level if $|\psi_{i,j}| < 4.0 \max(\Delta x^l, \Delta y^l)$, where $\Delta x^l$ and $\Delta y^l$ are the grid spacings in the $x$ direction and $y$ direction, respectively, on level $l$. The second criterion is based on the vorticity magnitude, in which grid cells are also tagged and refined to the next finer level if $|\omega| > 0.7|\omega^{\max}|$. The refinement stops when $l_{\max}$ is reached.

For the two-dimensional eel simulation, the left part of Fig. 5.23 shows the results of the evolution of the forward-moving velocity, which agrees with the results of Bhalla et al. (2013). Fig. 5.24 shows vorticity contours associated with the swimming eel and the grid hierarchy at different time instants. Vortices are generated on both sides of the eel and shed from the tail tip in the form of a reverse Von Kármán vortex sheet. The frequency of vortex shedding corresponds to the oscillating frequency of undulation. The eel moves forward as a result of self-propulsion. The adaptive grid helps accurately resolve the fluid-solid interface and the essential flow features at a reduced computational cost. The Strouhal number is defined as $St = 2A_{\text{tail}}/TU_{\text{steady}}$ (Triantafyllou et al., 1993), in which $U_{\text{steady}}$ is the average forward speed of the eel in the quasi-steady state and $A_{\text{tail}}$ is the amplitude of the tail tip flapping. In our simulations, we obtain $St = 0.41$, which is in the range of experimentally observed $St$ data for eels (Tytell, 2004).

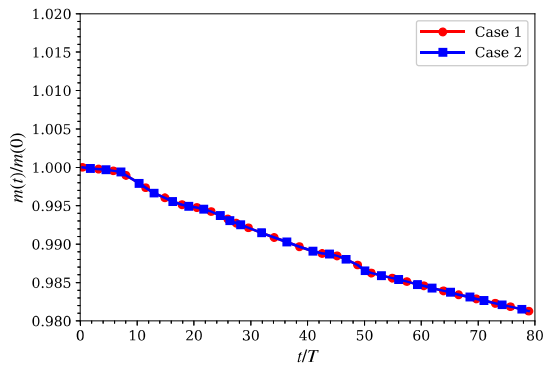

Figure 5.20: Comparison of the time evolution of the normalized fluid mass between the subcycling method (Case 1) and the non-subcycling method (Case 2) for a 2D cylinder free-falling into a column containing two fluids.

In summary, our adaptive algorithms can accurately capture the dynamics of swimming eels with prescribed geometry and deformation.

### 5.4.9 Three-dimensional self-propelled eel

This section investigates a three-dimensional, self-propelled swimming eel, a dynamic and complex problem that is considered to be computationally expensive. In addition to validating the adaptive DLM algorithms for 3D problems with prescribed deformations, another objective of this test is to compare the computational cost of the single-level, subcycling, and non-subcycling cases. The simulation parameters and refinement criteria for the three-dimensional eel problem are the same as those for the two-dimensional eel problem of Section 5.4.8, except that the geometry of its cross-section is an ellipsoid with semimajor and semiminor axis lengths of $a = 0.51L$ and $b = 0.08L$, respectively.

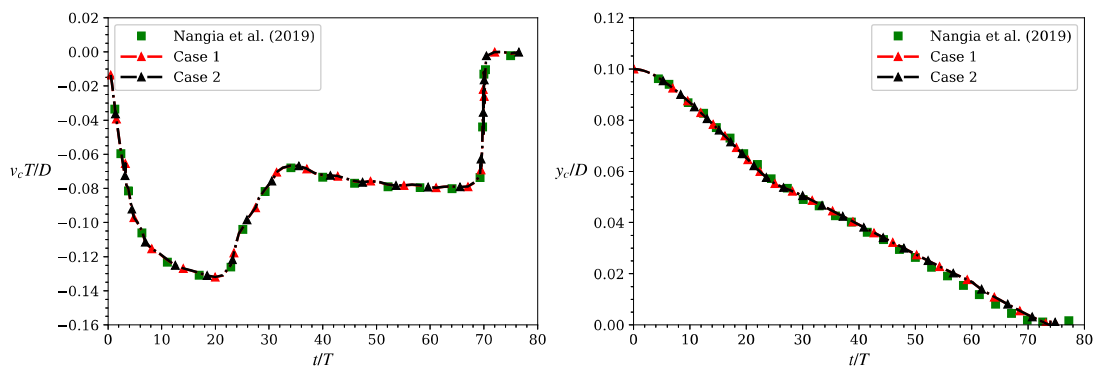

Figure 5.21: Time series of vertical velocity (left) and vertical position (right) for a 2D cylinder free-falling into a column containing two fluids. The simulation results are compared with those of (Nangia et al., 2019b).

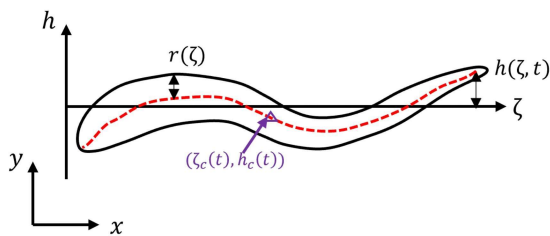

Figure 5.22: Sketch of a two-dimensional swimming eel. Here, $(\zeta_c(t), h_c(t))$ is the centroid of the eel, $h(\zeta, t)$ is the vertical displacement of its middle line (red dashed line), and $r(\zeta)$ is the half width of the cross-section centered at the middle line.

The height $h(s)$ is set to

$$h(\zeta) = b\sqrt{1 - \left(\frac{\zeta - a}{a}\right)^2}.$$

(5.46)

A periodic computational domain of size $8L \times 4L \times 4L$ is employed in the simulation. The head of the 3D eel at $t = 0$ is centered at $(6L, 2L, 2L)$. Three cases listed in Table 5.9 are considered. As seen from the right part of Fig. 5.23, we obtain agreement with the literature on the forward swimming velocity of the eel. The left part of Fig. 5.25 shows the vorticity contour and grid hierarchy of the four-level subcycling case (Case 2) at $Re = 5609$. The flow features are consistent with those reported in (Kern and Koumoutsakos, 2006).

To compare the computational cost for different cases, we profile each case for $t/T =$

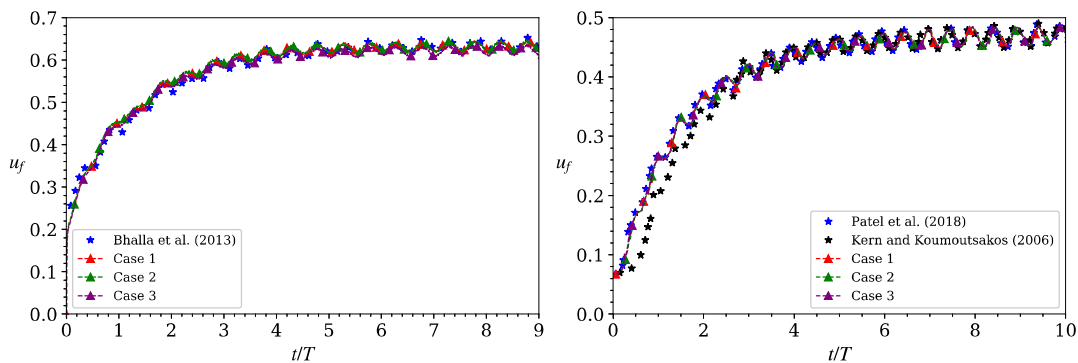

Figure 5.23: Comparison of the evolution of the forward velocity $u_f$ of the cases for the two-dimensional eel problem (left) and the three-dimensional eel problem (right). Case 1: single-level case; Case 2: four-level subcycling case; Case 3: four-level non-subcycling case.

$0 - 0.1$ on the Cray XC40/50 (Onyx) system at the U.S. Army Engineer Research and Development Center, excluding the I/O costs. Table 5.10 shows the total number of grid cells for different cases at $t/T = 0.1$. Compared with the adaptive cases with $l_{max} = 3$ (Cases 2 and 3), the single-level case (Case 1) has nearly 11.42 times more cells, i.e., the adaptive refinement considerably reduces the total number of grid cells.

The right part of Fig. 5.25 compares the wall clock time between the single-level case and the multilevel case for the time range $t/T = 0 - 0.1$. Compared with the single-level case (Case 1), the four-level subcycling case (Case 2) achieves more than a $20\times$ speedup in terms of the wall clock time, which significantly saves the computational cost of the 3D simulation. By comparing the non-subcycling case (Case 3) with the subcycling case (Case 2), we find that the subcycling case further lowers the computational cost by a factor of 1.5. The reason is that, compared to the non-subcycling method, the subcycling method uses a larger time step size for the coarser levels.

In addition to the total wall clock time, the wall clock time spent on some key parts of the algorithm is also documented, including the MAC projection, viscous solver, level projection, synchronization, and DLM algorithm. Among them, the DLM algorithm, which includes the initialization and redistribution of the Lagrangian markers and the time advancement of the solid structure, is the most time-consuming operation due to the suboptimal parallelization of the Lagrangian markers while using the domain decomposition technique for the background (Cartesian) grid. The second most time-consuming part and third most time-consuming part are the level projection and MAC projection, respectively, in which the Poisson equation needs to be solved on the multi-level grid. The optimization of these three parts is beyond the scope of this work and deferred to future studies.

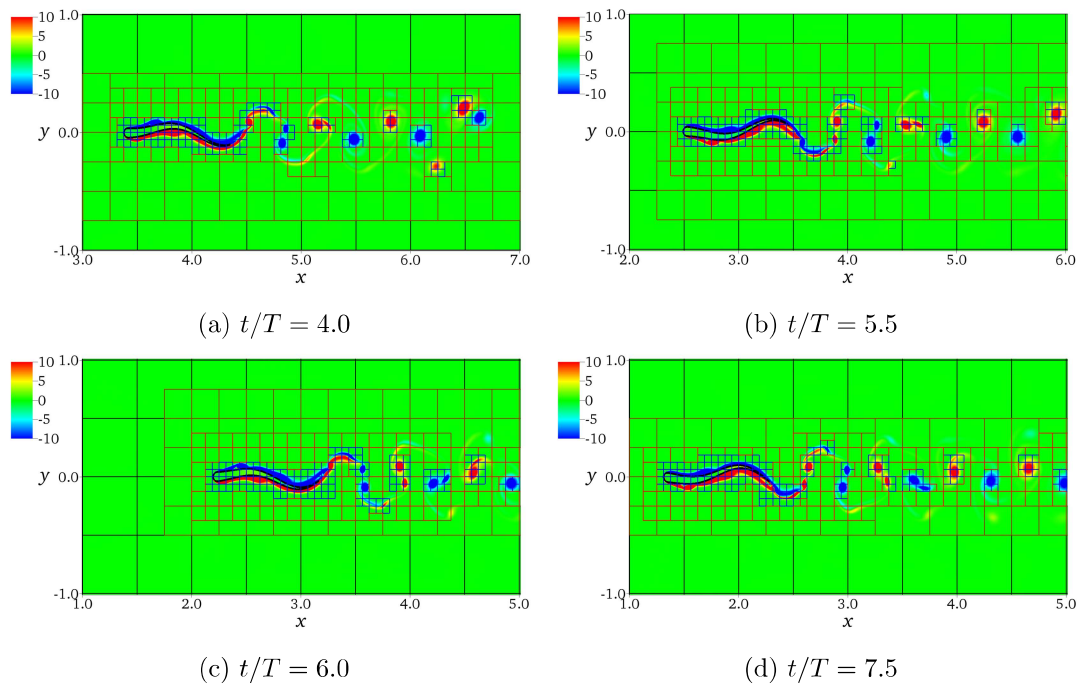

(a) $t/T = 4.0$                          (b) $t/T = 5.5$

(c) $t/T = 6.0$                          (d) $t/T = 7.5$

Figure 5.24: Vorticity contours of the four-level subcycling case (Case 2) at $Re = 5609$ for the two-dimensional eel problem. Black patches: level 0; brown patches: level 1; red patches: level 2; blue patches: level 3.

## 5.5 Concluding Remarks

In this chapter, we have established a novel adaptive distributed Lagrangian multiplier (DLM) framework with subcycling and non-subcycling time advancement methods for simulating fluid-structure interaction (FSI) problems on adaptively refined grids. We note that the proposed multilevel advancement algorithm uses the level-by-level advancement technique for time-marching the variables in valid and invalid regions of the adaptive mesh refinement (AMR) hierarchy and decouples the time advancement at different levels. Because of this decoupling, the time step constraint on the coarser levels is relaxed compared to the finer levels when the subcycling method is applied. On the other hand, the non-subcycling method avoids the time interpolation process across different levels because data on all levels are located at the same time instant during the simulation.

Table 5.9: Parameters of the three-dimensional eel problem

| Case No. | Grid numbers on level 0 | $l_{\max}$ | Cycling methods |
|:---:|:---:|:---:|:---:|
| 1 | $1536 \times 512 \times 256$ | 0 | – |
| 2 | $192 \times 64 \times 32$ | 3 | Subcycling |
| 3 | $192 \times 64 \times 32$ | 3 | Non-subcycling |

Table 5.10: Number of grid cells for the 3D self-propelled eel problem at $t/T = 0.1$

| Case No. | Level 0 cells | Level 1 cells | Level 2 cells | Level 3 cells | Total cells |
|:---:|:---:|:---:|:---:|:---:|:---:|
| 1 | 201,326,592 | – | – | – | 201,326,592 |
| 2 | 393,216 | 1,081,344 | 3,932,160 | 12,222,464 | 17,629,184 |
| 3 | 393,216 | 1,081,344 | 3,932,160 | 12,222,464 | 17,629,184 |

We also developed a force-averaging algorithm to maintain the consistency of Eulerian immersed boundary (IB) forces across multiple levels. The efficacy of the force averaging algorithm is validated using the lid-driven cavity with a submerged cylinder problem, in which the expected order of the convergence rate is obtained for the multilevel cases. When a fine level catches up with a coarse level, synchronization operations are applied to represent the composite solution during the level-by-level advancement. The MAC synchronization and refluxing operation ensures mass and momentum conservation of the entire flow field by correcting the multilevel solution using velocity and flux registers.

The accuracy and robustness of the computational framework were validated using several canonical test problems. The results show that our multilevel numerical schemes can simulate various FSI problems with different types of solid constraints, including prescribed motion, free motion, and prescribed shape change. The subcycling and non-subcycling methods produced consistent and accurate results for all of these problems. We also combined the DLM algorithm with our previous two-phase flow solver and incorporated external spring-damper forces into the simulation. This approach enabled us to simulate the wave energy converter (WEC) problem.

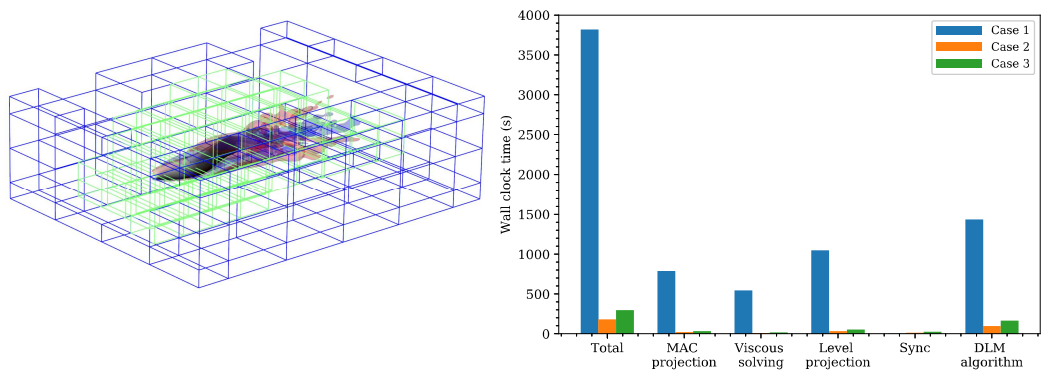

Figure 5.25: Left: vorticity contour ($|\omega| = 1.8$) and grid hierarchy of the four-level subcycling case (Case 2) at $Re = 5609$ for the three-dimensional eel problem. Blue lines: patches on level 2; green lines: patches on level 3. Right: comparison of the wall clock time of key advancing steps among the single-level case (Case 1), four-level subcycling case (Case 2), and four-level non-subcycling case (Case 3).

Finally, we demonstrated that multilevel simulation can achieve the same level of accuracy with substantially fewer grid cells compared to the single-level fine-grid simulation. In particular, for the 3D swimming eel problem, the multilevel simulation is able to accurately capture the forward velocity with a nearly $20\times$ speedup compared to the single-level simulation. In summary, we conclude that our proposed BSAMR framework is promising for high-fidelity and computationally demanding single-phase and multiphase fluid-structure interaction problems.

# Chapter 6

# Summary and discussions

## 6.1   Contributions of this thesis

The Adaptive Mesh Refinement (AMR) is a response to the need to resolve small-scale flow structures with a hierarchy of refined grids only at the locations where high resolution is required, without resorting to a uniformly fine-resolution grid across the entire computational domain, thus greatly reducing the overall computational cost and memory requirement. In this thesis, we have developed a novel block structured adaptive mesh refinement (BSAMR) method for simulations of two-phase flows and fluid-structure interactions (Zeng and Shen, 2019, 2020; Zeng et al., 2022b).

For the two-phase flow part, we first devise an inconsistent scheme and combine it with AMR for the simulations of wave breaking and bubbly flow problems (Zeng et al., 2022b). Compared with the other AMR tools in the CFD community, our AMR code has the following advanced features: (1) Distinct from previous work using staggered grids (Nangia et al., 2019a; Patel and Natarajan, 2017), a purely collocated grid is employed in the present framework, in which all variables are defined at the center of the grid cells. This design eases the implementation because only one set of interpolation and averaging schemes is needed when the multilevel grid is involved. (2) A synchronization algorithm is proposed in this work, which is the key to maintain the consistency of variables across multiple levels so that the variables can better represent the composite solution during the level-by-level advancement. When a fine level catches up with a coarse level, the cell-centered averaging sub-step first replaces the variables on the coarser levels with the more accurate solution on the finer levels. Then, the MAC synchronization and refluxing sub-step provide corrections to the multi-level solution from the velocity and flux registers to ensure the momentum conservation on the entire flow field. (3) Both subcycling and non-subcycling advancement methods are embedded in this unified framework. One can choose either of these two methods or even combine them for time integration. Note that multilevel advancement decouples the time advancement for different levels, which relaxes the time step constraint on the coarser levels if the subcycling method is applied. On the other hand, the non-subcycling method avoids time interpolation across the different levels because data on all levels are synchronized to the same time instant during the simulation. Numerical simulations demonstrate that the subcycling and non-subcycling methods can produce consistent results and accurately capture the complex dynamics of the two-phase flow.

We then embed the consistent scheme into the BSAMR framework for two-phase flow simulations. In the consistent scheme, we use the advanced Cubic Upwind Interpolation (CUI) scheme for consistent mass and momentum transport, which greatly improves the accuracy and robustness for simulating two-phase flows with a high density ratio and high Reynolds number. The interface capturing level set method is coupled with the conservative form of the Navier–Stokes equations, and the multilevel reinitialization technique is implemented on the multilevel grid to improve the mass conservation of the two-phase flow. Various numerical tests demonstrate that the consistent scheme results in a numerically stable solution in flows with high density ratios (up to $10^6$) and high Reynolds numbers (up to $10^6$). Therefore, we conclude that the proposed AMR framework is promising for high-fidelity simulations of complex two-phase flows.

For tackling single- and multiphase fluid-structure interaction (FSI) problems, we provide an adaptive implementation of the distributed Lagrange multiplier (DLM) immersed boundary (IB) approach on multilevel collocated grids. Both a non-subcycling and a subcycling time advancement scheme are described, which are used to time-march the composite grid variables on a level-by-level basis; these schemes employ the same time step size on different levels and a different time step size on different levels, respectively. This is in contrast to previous adaptive variants of the IB technique in the literature, which solve and advance coarse- and fine-level variables in a coupled fashion.

## 6.2  Future studies

The AMR is a powerful technique that can be combined with different physical modules in the scientific computing areas. In this thesis, we build a unified adaptive framework for the two-phase flow and FSI problems with AMR. Based on this framework, here are some future work about the algorithms and the applications.

We build our parallel algorithms based on the Message Passing Interface (MPI). As the *AMReX* package is moving forward, we can re-implement our algorithms using GPU (Zhang et al., 2020), which will provide an optimal combination of portability, readability, and performance for the simulations on supercomputers. Besides, we only use the LS method to capture the interface of the two-phase flow. In the future, more interface-capturing methods, such as the VOF (Natarajan et al., 2019a), CLSVOF (Sussman and Puckett, 2000), and ghost fluid methods, can be involved. In addition, we only focus on the level-by-level advancement methods (e.g. the subcycling and non-subcycling methods) in this work. We plan to compare the computational efficiency between the composite advancement method and the level-by-level advancement method. We can also combine these two advancement methods to find an optimal hybrid to improve the efficiency for different physical simulations.

In future studies, it is also worthwhile to combine AMR with the particle collision models. The reason is that the time-step size of the collision is around 50 times smaller than that in the background flow (Uhlmann, 2005). The gap of the temporal scales can be properly dealt with when the AMR, especially the subcycling method, is utilized. Besides, we now utilize the MLMG solver for solving the PDEs on all levels. We can leverage the deep learning (DL) technique to accelerate the solving process on the multi-level grid (Margenberg et al., 2022). Finally, most of the refinement criteria in the validation cases are manually set up. It is also interesting to investigate how we can improve the efficiency of the refinement and de-refinement process when the reinforcement learning method is utilized (Yang et al., 2021a).

Most of the current work focus on the development of the two-phase flow and FSI algorithms in this adaptive framework. In the future, we plan to apply these algorithms to practical applications, including the bubbles (Gao et al., 2021b), droplets (Liu et al., 2018), and sea spray (Wang et al., 2019) under breaking waves. With the help of AMR, we can put more computational resources around these small flow structures in simulations. Another good topic is to apply this adaptive framework to simulate dynamics of renewable energy devices, such as wind turbines (Calderer et al., 2014), hydro-power turbines (Gao et al., 2022), and WECs (Yu and Li, 2013; Zeng and Shen, 2020). Finally, we can extend our AMR framework to simulate turbulent flows (Shen et al., 1999), non-Newtonian fluids (Sverdrup et al., 2018), electromagnetic fluids (Bhalla et al., 2014), and micro-electronic devices (Yao et al., 2021).

# Appendix A

# Appendix

## A.1   Inconsistent scheme

For the inconsistent scheme described in chapter 3, the discretizations are applied to the nonconservative forms of the Navier–Stokes equations (Eqs. (4.4) and (4.5)) (Zeng et al., 2022b). Due to the same advection and reinitialization schemes of the LS function $\phi$, **Step 1** and **Step 3** in the inconsistent scheme are the same as those in the consistent scheme (Section 4.2.1). **Step 5**, which is used to obtain the updated velocity $\widetilde{\mathbf{u}}^{n+1}$ and pressure $p^{n+1/2}$, is also the same as **Step 5** in the consistent scheme, assuming the intermediate velocity $\widetilde{\mathbf{u}}^{*,n+1}$ has been calculated. Thus, we write only the details of **Step 3** and **Step 4** for the inconsistent scheme.

**Step 3**: Both the viscosity $\mu^{n+1}$ and the density $\rho^{n+1}$ fields are reset through the Heaviside function as

$$\mu^{n+1} = \mu_{\mathrm{g}} + (\mu_{\mathrm{l}} - \mu_{\mathrm{g}})\,\widetilde{H}\left(\phi^{n+1}\right), \tag{A.1}$$

$$\rho^{n+1} = \rho_{\mathrm{g}} + (\rho_{\mathrm{l}} - \rho_{\mathrm{g}})\,\widetilde{H}\left(\phi^{n+1}\right), \tag{A.2}$$

where $\mu_{\mathrm{l}}$ and $\rho_{\mathrm{l}}$ are the viscosity and density of the liquid phase and $\mu_{\mathrm{g}}$ and $\rho_{\mathrm{g}}$ are those for the gas phase, respectively. The smoothed Heaviside function, which is regularized over $n_{\text{cells}}$ grid cells on either side of the two-phase interface, is defined as

$$\widetilde{H}\left(\phi^{n+1}\right) = \begin{cases} 0, & \phi^{n+1} < -n_{\text{cells}}\Delta x \\ \frac{1}{2}\left(1 + \frac{1}{n_{\text{cells}}\Delta x}\phi^{n+1} + \frac{1}{\pi}\sin\left(\frac{\pi}{n_{\text{cells}}\Delta x}\phi^{n+1}\right)\right), & |\phi^{n+1}| \leq n_{\text{cells}}\Delta x \\ 1, & \text{otherwise} \end{cases}$$
(A.3)

where the uniform grid spacing $\Delta x = \Delta y$ is assumed (Nangia et al., 2019a) and $n_{cells} = 1$ or 2 is applied for all testing cases. The midpoint values of $\rho$ and $\mu$ are then calculated as $\rho^{n+\frac{1}{2}} = (\rho^{n+1} + \rho^n)/2$ and $\mu^{n+\frac{1}{2}} = (\mu^{n+1} + \mu^n)/2$, respectively.

**Step 4**: The intermediate velocity $\widetilde{\mathbf{u}}^{*,n+1}$ is solved semi-implicitly as

$$\rho^{n+\frac{1}{2}}\left(\frac{\widetilde{\mathbf{u}}^{*,n+1} - \mathbf{u}^n}{\Delta t} + \boldsymbol{\nabla}\cdot(\mathbf{uu})^{n+\frac{1}{2}}\right) = -\boldsymbol{\nabla}p^{n-\frac{1}{2}} + \frac{1}{2}\left(\boldsymbol{\nabla}\cdot\mu(\phi^{n+1})\boldsymbol{\nabla}\widetilde{\mathbf{u}}^{*,n+1} + \boldsymbol{\nabla}\cdot\mu(\phi^n)\boldsymbol{\nabla}\mathbf{u}^n\right) \\ + \rho^{n+\frac{1}{2}}\mathbf{g} + \mathbf{f}_{\text{s}}^{n+\frac{1}{2}},$$
(A.4)

where the convective term $\boldsymbol{\nabla}\cdot(\mathbf{uu})^{n+\frac{1}{2}}$ is calculated using the Godunov scheme (Almgren et al., 1998; Sussman et al., 1999).

## A.2 Discretization of the advection terms

We use the Godunov scheme (Colella, 1990, 1985) for discretization of the advection terms $[\boldsymbol{\nabla}\cdot(\mathbf{u}\psi)]^{n+1/2}$ in Eq. (5.11) and $\boldsymbol{\nabla}\cdot(\mathbf{uu})^{n+1/2}$ in Eq. (A.4). This scheme is robust for a wide range of Reynolds numbers. There are four substeps to calculate the advection terms:

1. The unsplit Godunov approach is utilized to estimate the edge-centered velocity $(\mathbf{u}^{n+1/2,L}, \mathbf{u}^{n+1/2,R})$ and edge-centered LS function $(\psi^{n+1/2,L}, \psi^{n+1/2,R})$ at the middle time step $t^{n+1/2}$ on the edges perpendicular to the $x$ direction. The superscripts $L$ and $R$ indicate that the edge-centered values are approximated on the left edge and right edge, respectively. The edge-centered velocity $(\mathbf{u}^{n+1/2,U}, \mathbf{u}^{n+1/2,D})$ and edge-centered LS function $(\psi^{n+1/2,U}, \psi^{n+1/2,D})$ can be calculated on the edges perpendicular to the $y$ direction at the middle time step $t^{n+1/2}$, where the superscripts $U$ and $D$ denote that the edge-centered values are calculated from the up edge and down edge, respectively, of the computational cell.

Table A.1: Parameters of the wave generation and wave absorption problem

| Case No. | Grid numbers on level 0 | $l_{\max}$ | $\Delta t_0$ | Cycling methods |
|----------|-------------------------|------------|--------------|-----------------|
| 1 | $1280 \times 128$ | 0 | 0.0002 | – |
| 2 | $2560 \times 256$ | 0 | 0.0001 | – |
| 3 | $5120 \times 512$ | 0 | 0.00005 | – |
| 4 | $640 \times 64$ | 2 | 0.0004 | Subcycling |
| 5 | $640 \times 64$ | 2 | 0.0001 | Non-subcycling |

2. The MAC projection (Bell et al., 1989; Almgren et al., 1998; Sussman et al., 1999) is then applied to obtain the divergence-free edge-centered advection velocity $\mathbf{u}^{\mathrm{adv}}$.

3. The advection velocity $\mathbf{u}^{\mathrm{adv}}$ is then used to calculate the edge-centered approximate state $\mathbf{u}^{n+1/2}$ and $\psi^{n+1/2}$ based on $\mathbf{u}^{n+1/2,L}$, $\mathbf{u}^{n+1/2,R}$, $\mathbf{u}^{n+1/2,U}$, $\mathbf{u}^{n+1/2,D}$, $\psi^{n+1/2,L}$, $\psi^{n+1/2,R}$, $\psi^{n+1/2,U}$, and $\psi^{n+1/2,D}$.

4. The advection velocity $\mathbf{u}^{\mathrm{adv}}$ is then used to advect the approximate state $\mathbf{u}^{n+1/2}$ and $\psi^{n+1/2}$. The advection terms $\boldsymbol{\nabla} \cdot (\mathbf{uu})^{n+1/2}$ and $[\boldsymbol{\nabla} \cdot (\mathbf{u}\psi)]^{n+1/2}$ are calculated as $\boldsymbol{\nabla} \cdot (\mathbf{u}^{\mathrm{adv}}\mathbf{u}^{n+1/2})$ and $\boldsymbol{\nabla} \cdot (\mathbf{u}^{\mathrm{adv}}\psi^{n+1/2})$.

We note that the calculation of the advection term $[\boldsymbol{\nabla} \cdot (\mathbf{u}\phi)]^{n+1/2}$ in Eq. (5.12) is the same as that for $[\boldsymbol{\nabla} \cdot (\mathbf{u}\psi)]^{n+1/2}$.

## A.3    Wave generation and wave absorption

This section presents the validation of the wave generation and wave absorption algorithms utilized for simulating the WEC problem in Section 5.4.6. These algorithms are only briefly introduced here; their numerical details can be referenced in (Zeng and Shen, 2020; Nangia et al., 2019b; Dafnakis et al., 2020). In short, the horizontal and vertical velocity components are prescribed at the left boundary based on wave theory such that Stokes waves can be generated and propagate towards the right side. To mitigate the reflection of waves from the right boundary, a damping zone is placed at a downstream location to smoothly relax the velocities and LS function $\phi$.

A 2D example is presented here for validating the wave generation and wave absorption algorithms. This example has the same computational parameters as the WEC problem of Section 5.4.6, except that the WEC device is not included in the simulation. Five cases are considered, as listed in Table A.1. The first three single-level cases have different grid sizes and are employed to show the numerical convergence of the wave generation algorithms. As shown in the left part of Fig. A.1, the simulations converge to the theoretical wave elevation using Stokes wave theory (Lamb, 1993) as the grid resolution increases. We find that the resolution of Case 2 is sufficient to accurately capture the wave profile, in which there are approximately ten grid cells per wave height. Case 4 and Case 5 have the same resolution on their finest level as Case 2. From the right part of Fig. A.1, it is seen that the results of these three cases are consistent with each other and agree with the theoretical result.

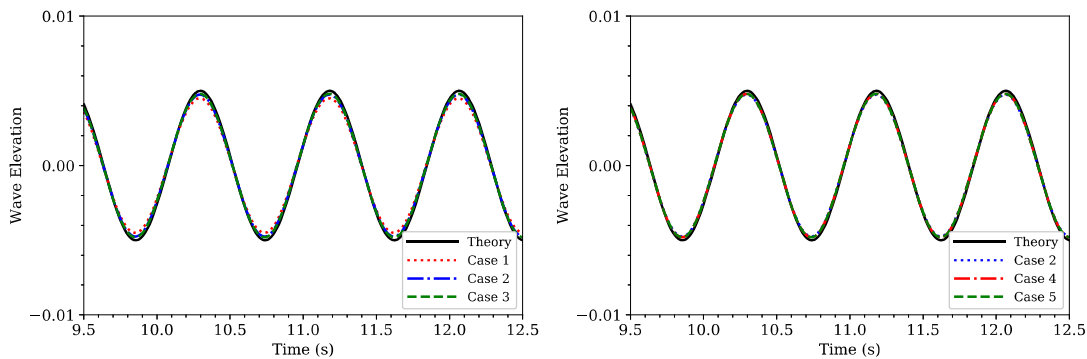

Figure A.1: Left: comparison of time series of the wave elevation at $x = 2.87\lambda$ among the cases with different grid sizes for the convergence study. Right: comparison of time series of the wave elevation at $x = 2.87\lambda$ among the single-level case (Case 2), the three-level subcycling case (Case 4), the three-level non-subcycling case (Case 5), and the theory.